\documentclass[aip,jcp,amsmath,amssymb,floatfix,reprint,citeautoscript,articletitle=false,noeprint]{revtex4-2}

\usepackage{amssymb}
\usepackage{amsmath}
\usepackage{graphicx}
\usepackage{dcolumn}
\usepackage{xcolor}
\usepackage{fancyhdr}
\usepackage[colorlinks,allcolors=black,citecolor=blue,urlcolor=blue]{hyperref}

\newcommand{\tochange}[1]{\textcolor{blue}{#1}}

\begin{document}

\title{How reactive is water at the nanoscale and how to control it?}

\author{Xavier R. Advincula}
\affiliation{Yusuf Hamied Department of Chemistry, University of Cambridge, Lensfield Road, Cambridge, CB2 1EW, UK}
\affiliation{Cavendish Laboratory, Department of Physics, University of Cambridge, Cambridge, CB3 0HE, UK}
\affiliation{Lennard-Jones Centre, University of Cambridge, Trinity Ln, Cambridge, CB2 1TN, UK}

\author{Yair Litman}
\affiliation{Yusuf Hamied Department of Chemistry, University of Cambridge, Lensfield Road, Cambridge, CB2 1EW, UK}
\affiliation{Lennard-Jones Centre, University of Cambridge, Trinity Ln, Cambridge, CB2 1TN, UK}
\affiliation{Max Planck Institute for Polymer Research, Ackermannweg 10, 55128 Mainz, Germany}

\author{Kara D. Fong}
\affiliation{Yusuf Hamied Department of Chemistry, University of Cambridge, Lensfield Road, Cambridge, CB2 1EW, UK}
\affiliation{Lennard-Jones Centre, University of Cambridge, Trinity Ln, Cambridge, CB2 1TN, UK}
\affiliation{Division of Chemistry and Chemical Engineering, California Institute of Technology, Pasadena, California 91125, USA}

\author{William C. Witt}
\affiliation{Harvard John A. Paulson School of Engineering and Applied Sciences, Harvard University, Cambridge, MA, USA}

\author{Christoph Schran}
\email{cs2121@cam.ac.uk}
\affiliation{Cavendish Laboratory, Department of Physics, University of Cambridge, Cambridge, CB3 0HE, UK}
\affiliation{Lennard-Jones Centre, University of Cambridge, Trinity Ln, Cambridge, CB2 1TN, UK}

\author{Angelos Michaelides}
\email{am452@cam.ac.uk}
\affiliation{Yusuf Hamied Department of Chemistry, University of Cambridge, Lensfield Road, Cambridge, CB2 1EW, UK}
\affiliation{Lennard-Jones Centre, University of Cambridge, Trinity Ln, Cambridge, CB2 1TN, UK}

\begin{abstract} 

Nanoconfined water plays a key role in nanofluidics, electrochemistry, and catalysis, yet its reactivity remains a matter of debate.
Prior studies have reported both enhanced and suppressed water self-dissociation relative to the bulk, but without a consistent explanation.
Here, using enhanced sampling molecular dynamics with machine-learned potentials trained at first-principles accuracy, we investigate dissociation behavior in water confined within 2D slit pores and nanodroplets, using graphene and hexagonal boron nitride as model materials.
We find that reactivity is extremely sensitive to water density, confinement width, geometry, material flexibility, and surface chemistry.
Despite this complexity, we show that chemical potential—together with interfacial interactions—governs dissociation trends and explains the variability observed in prior studies.
This thermodynamic perspective reconciles previous contradictions and reveals how nanoscale environments can drastically shift water reactivity.
Our findings provide molecular-level insight and offer a design lever for modulating water chemistry at the nanoscale.
\end{abstract}
\maketitle

\section*{Introduction}

Water is central to chemical and biological function, not only as a medium but also as an active participant in countless processes.
Among its most fundamental properties is the spontaneous self-dissociation into hydronium (H$_3$O$^+$) and hydroxide (OH$^-$) ions, a reaction that defines pH and drives acid–base chemistry, proton transport, and catalytic behavior across a vast range of systems \cite{geissler_2001, reac_int_2020, dissoc_membranes_2020, nat_cat_2022, dissoc_solar_2023, laage_pt_natchem_2024}.
The equilibrium constant for this process, $\textrm{K}_{\textrm{w}}$, governs the balance between neutral and ionized species in aqueous environments.
Although well-characterized in bulk water \cite{geissler_2001, ali_2011, dissoc_efield_prl_2012, review_h3o_oh_2016, van_erp_2018, joutsuka_dissoc_2022, selloni_2023}, many other real-world environments such as biological membranes, mineral interfaces, nanopores, and catalytic surfaces, feature water confined to nanometer-scale dimensions, where 
water's self-dissociation is much less well explored.

At the nanoscale, confinement can drastically reshape water’s structural \cite{algara-siller_square_2015, vk_chris_2022, quasionedimensional}, dynamical \cite{radha_b_2016, dripplon_2018, robin_2023, Jiang2024}, and dielectric properties \cite{fumagalli_dielectric_2018, fumagalli_2024}, often leading to behavior that departs markedly from that of bulk water.
These effects arise across a wide range of systems that vary in the nature and rigidity of their confining environments.
For example, in soft environments such as biomolecular cavities or atmospheric aerosols, interfaces are typically flexible and chemically heterogeneous.
By contrast, rigid confinements like carbon nanotubes and 2D slit pores impose well-defined geometric constraints.
The nature of confinement has a direct impact on solvation structure, interfacial interactions, and ultimately, chemical reactivity \cite{micelles_2005, marx_chem_2017}.
While reactivity in soft environments has been widely explored \cite{micelles_2005, reac_micelles_2020, annurev_2020_droplet, reac_int_2020, miguel_ph_2024}, much less is known about fundamental proton-transfer processes such as water self-dissociation in rigid confinement.
In rigid media, geometric constraints can disrupt the hydrogen-bond network, alter solvation, and shift the equilibrium between neutral and dissociated species \cite{marx_chem_2017, marx_conf_rev_2021, Stolte2022}.
These effects offer the possibility of modulating water reactivity through confinement alone.
This behavior is especially relevant in nanofluidic channels \cite{lyderic_2010_rev, lyderic_crossroads_2023}, electrochemical systems \cite{Lozada-Hidalgo2018, electrochem_conf, Gomes2024}, and surface catalysis on 2D materials \cite{cat_2d_2016, nat_cat_2022}.
Even modest deviations from bulk-like behavior in these settings can lead to emergent reactivity patterns\cite{marx_chem_2017}, highlighting the need to understand water’s behavior under nanoscale confinement for the design of functional interfaces and reaction environments.

\begin{figure*}
    \centering
    \includegraphics[width=\textwidth]{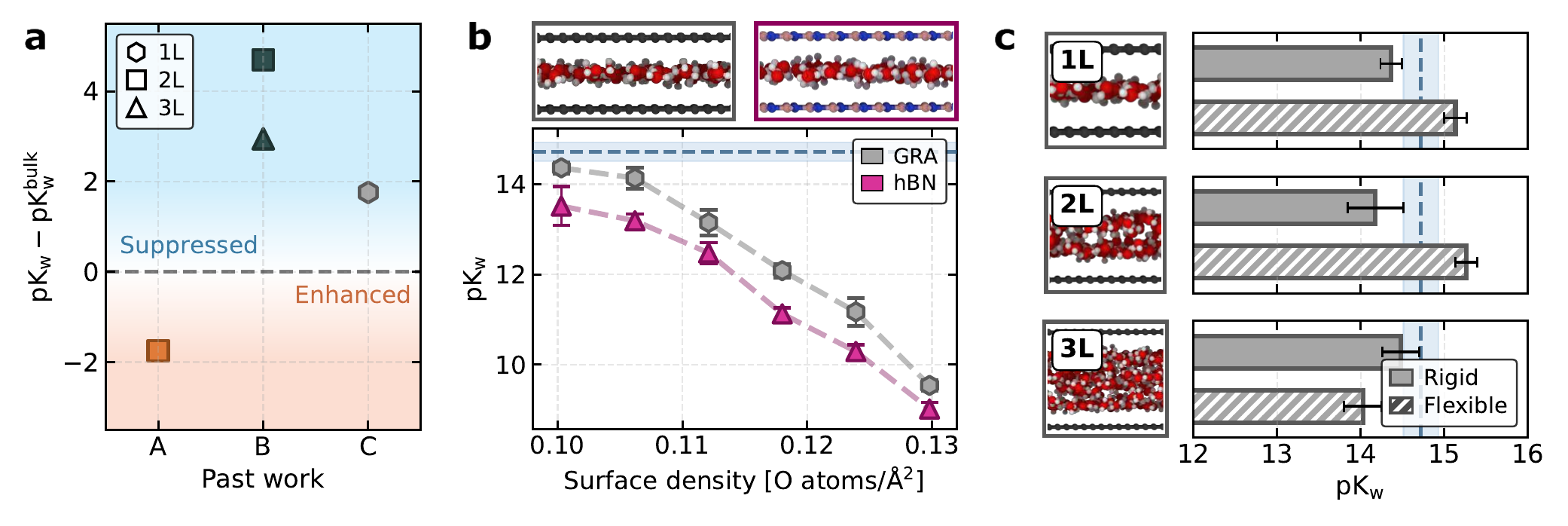}
    \caption{\textbf{The reactivity of nanoconfined water depends sensitively on the nature of the confinement.}  
(a) Reported $p\textrm{K}_{\textrm{w}}$ values for nanoconfined water from prior studies, illustrating both suppression and enhancement of self-dissociation relative to bulk water.
References: (A) Muñoz-Santiburcio and Marx \cite{marx_dissoc_conf}, (B) Di Pino \textit{et al.} \cite{hassanali_dissoc_ange}, and (C) Dasgupta \textit{et al.} \cite{paesani2025_dissoc}
These studies involve different confining environments: A corresponds to layered minerals, while B and C examine graphene-based systems using QM/MM and MLP approaches, respectively.
1L, 2L, and 3L indicate the number of confined water layers formed within the slit pores, corresponding to monolayer, bilayer, and trilayer water, respectively.
(b) Dependence of $p\textrm{K}_{\textrm{w}}$ on the surface density for monolayer water confined in rigid graphene (GRA) and hexagonal boron nitride (hBN) slit pores, highlighting the influence of both material type and density.
(c) $p\textrm{K}_{\textrm{w}}$ for 1L, 2L, and 3L graphene slit pores, showing the influence of pore width and flexibility on dissociation behavior under confinement.
The rigid 1L, 2L, and 3L setups correspond to slit widths of 6.70, 10.05, and 13.40 \AA, respectively. These values are commensurate with pore sizes that can be experimentally realized using van der Waals (vdW) assembly \cite{radha_b_2016, kara_pairing_2024}.
The horizontal and vertical dashed blue lines in panels (b) and (c) indicate our bulk reference estimate at 1 bar with its errors.
}
    \label{fig:fig1}
\end{figure*}

Although it is established that confinement can significantly alter the structural and dynamical properties of water, its influence on self-dissociation remains an open question.
First-principles molecular dynamics (MD) simulations have provided important insights, particularly in geometrically well-defined environments such as carbon nanotubes \cite{pccp_2013_ang, cnt_dissoc_2018} and 2D slit pores composed of chemically inert materials such as mackinawite or graphene \cite{marx_dissoc_conf, hassanali_dissoc_ange, paesani2025_dissoc}.
Yet, these studies have reported markedly different outcomes. Muñoz-Santiburcio and Marx \cite{marx_dissoc_conf} found a substantial, 55-fold enhancement in the rate of water self-dissociation when confined as a bilayer between mackinawite sheets, attributing this acceleration directly to nanoconfinement.
This corresponds to a pronounced decrease in $p\textrm{K}_{\textrm{w}}$, which reflects a greater degree of ionization in water.
In contrast, Di Pino \textit{et al.} \cite{hassanali_dissoc_ange} reported no enhancement, and even suppression, when water was confined between graphene layers, arguing that earlier results may have been influenced by overpressurization rather than confinement \textit{per se}.
More recently, Dasgupta \textit{et al.} \cite{paesani2025_dissoc} arrived at a similar conclusion, showing that water confined to a sub-nanometer monolayer exhibits significantly reduced dissociation.
This corresponds to an increase in $p\textrm{K}_{\textrm{w}}$, indicating lower concentrations of dissociated ionic species under such extreme confinement.
These contrasting findings, illustrated in Figure~\ref{fig:fig1}a, emphasize the complexity of confined aqueous systems, where dissociation behavior arises from a subtle interplay of environmental and thermodynamic factors.
Disentangling these contributions and isolating the specific role of confinement remains a central challenge.
Adding to this uncertainty, indirect experimental studies have also yielded contrasting interpretations \cite{c60_diss_2018, sauer_2021, fumagalli_2024}.
Due to the indirect nature of these measurements and their sensitivity to local conditions, these findings remain difficult to reconcile, contributing to a fragmented picture of how nanoconfinement influences water reactivity.
Altogether, these discrepancies highlight the need for a consistent and well-controlled framework to determine how confinement affects water self-dissociation: one capable of capturing both molecular-scale structure and thermodynamic factors that govern dissociation equilibria across a range of conditions.

In response to these challenges, we investigate the key factors that govern the self-dissociation of water under nanoconfinement.
To navigate this complex landscape, we employ carefully developed and validated machine learning potentials (MLPs), trained on density functional theory reference data (see \tochange{Methods}), which enable large-scale MD simulations with near first-principles accuracy. 
We focus on two archetypal materials for studying nanoconfined water: graphene and hexagonal boron nitride (hBN), as their similar interface structures contrast with the different behavior that water exhibits near their surfaces \cite{Siria2013, tocci_friction, mlb_oh}.
We first examine the key factors that complicate direct comparisons between confined and bulk water, including variations in density, confining material, pore width, and pore flexibility. 
From this extensive set of simulations, a strong sensitivity of water to the nature of the confinement is highlighted. 
By analyzing dissociation behavior as a function of chemical potential in monolayer graphene slit pores, we show that confinement alone does not inherently alter water’s acid–base chemistry.
Instead, we find that, when compared at equivalent chemical potentials, confined water exhibits dissociation behavior that closely resembles that of the bulk.
We then extend our study to more complex geometries, including material-encapsulated nanodroplets, which introduce interfacial curvature, spatial heterogeneity, and edge reactivity, and represent another class of experimentally realizable nanoconfinement systems.
In these droplets, we find that variations in surface chemistry and local structure can lead to significant departures from bulk-like behavior.
Notably, at hBN interfaces, we identify an alternative dissociative pathway in which hydroxide ions produced during self-dissociation are stabilized through chemisorption at droplet edges.
This finding highlights that interfacial chemistry, rather than geometric confinement alone, offers a powerful means of tuning water reactivity at the nanoscale\cite{yongkang_natcoms_2025}.
Building on these insights, we illustrate how interfacial geometry, chemistry, and local structure influence dissociation equilibria in confined systems.
Altogether, our findings reveal when confined water behaves like the bulk and when it departs from it, offering molecular-scale principles for understanding and controlling acid–base behavior under confinement.

\section*{Results}

\subsection*{Strong sensitivity of nanoconfined water to the confinement conditions}

The question ``How reactive is water at the nanoscale?" is not a simple one to answer because many factors can, in principle, influence the tendency of water to dissociate. 
To tackle this question, we started by systematically exploring slit pore geometries with water at different densities, in different confining materials (graphene and hBN), and with rigid and flexible confining materials at different confinement widths. 
Since water self-dissociation is a rare event on accessible simulation timescales, we employ umbrella sampling to enhance sampling along a reaction coordinate associated with the dissociation process.
Specifically, we bias the coordination number of a water oxygen, $n_{\text{H}}$, which tracks the number of hydrogen atoms covalently bonded to it \cite{sprik_pk_coord_2000}.
This variable captures the transition from a neutral H$_2$O molecule to the ionized species H$_3$O$^+$ and OH$^-$.
By sampling along $n_{\text{H}}$, we reconstruct the dissociation pathway and compute the corresponding free energy profiles.
From these, we extract the dissociation constant, $p\textrm{K}_{\textrm{w}}$, using the relation $p\textrm{K}_{\textrm{w}} = \Delta F^{\ddagger} / (RT \ln 10)$, where $\Delta F^{\ddagger}$ is the free energy barrier between reactant and product states\cite{joutsuka_dissoc_2022, litman_efield_2025} (see \tochange{Methods} for details).
We note that our computational approach shows excellent agreement with experiment for the bulk dissociation constant as a function of temperature, as discussed in detail in \tochange{Section S2}.

The results of our systematic analysis of dissociation constants are summarized in Figure~\ref{fig:fig1}.
Let us start by considering how density influences water self-dissociation of monolayer-confined water in rigid graphene slit pores (see \tochange{Section S1} for setup details).
Density is difficult to define in nanoconfined systems, and consequently, we explored a range of water densities across our simulations. 
Indeed, even in experiments, a range of densities can be expected depending upon the conditions used to create the systems of interest. 
To facilitate an initial comparison of trends, we considered bulk-like density variations ranging from approximately 0.9--1.2 g/cm$^{3}$.
However, in our analysis, we focus specifically on variations in surface density, defined as the number of water molecules per unit area of the confining surface (see \tochange{Section S1}), as this is a more well-defined quantity under confinement than 3D density.
Figure~\ref{fig:fig1}b shows that as the surface density (i.e., the 2D density of the O atoms in the water molecules) increases, the $p\textrm{K}{\textrm{w}}$ decreases.
Higher densities are related to higher effective pressures. 
Thus, this trend suggests that pressure plays a key role in enhancing self-dissociation under confinement and mirrors the known behavior of bulk water, where elevated pressure lowers $p\textrm{K}_{\textrm{w}}$ and promotes dissociation \cite{pkw_press_experiments_bulk_2005}.
However, as we will explore in more detail later, drawing such parallels requires a careful comparison.

To examine the influence of the confining material, we also consider hBN slit pores. 
The same qualitative trend is observed: higher density leads to a lower $p\textrm{K}_{\textrm{w}}$. 
However, consistently lower $p\textrm{K}_{\textrm{w}}$ values are found compared to graphene.
This difference highlights the role of surface chemistry, such as hydrophobicity or hydrogen-bonding characteristics, in shaping the local reactivity environment.
Figure~\ref{fig:fig1}c extends this analysis to different pore widths, comparing monolayer (1L), bilayer (2L), and trilayer (3L) water confined in rigid graphene slit pores.
In the systems confined by rigid materials, $p\textrm{K}_{\textrm{w}}$ increases with the number of layers, with 1L showing the lowest values, suggesting that stronger confinement enhances dissociation.
We then assess the role of pore flexibility. In contrast to the rigid cases, flexible slit pores show the opposite trend: thinner pores exhibit reduced dissociation.
However, direct comparisons between rigid and flexible systems are complicated by the fact that flexible pores, even when exhibiting a similar average interlayer spacing, can accommodate a wider range of densities and layering motifs \cite{teresa_flexible_2018}.
The situation is further complicated by distinct phase behavior.
For example, bilayer water has been shown to exhibit ice-like structural characteristics, driven by an anomalously high melting temperature that exceeds that of bulk ice \cite{valeria_bilayer_2010, artacho_bilayer_2016}.

Overall, these findings reveal the intricate interplay of density, pressure, surface chemistry, confinement geometry, and phase behavior.
This complexity not only makes direct comparisons with bulk water challenging but also helps explain the conflicting results reported in the literature, highlighting the importance of evaluating confined systems under thermodynamically consistent conditions.

\subsection*{Confined and bulk water show similar dissociation behavior}

\begin{figure}
    \centering
    \includegraphics[width=0.43\textwidth]{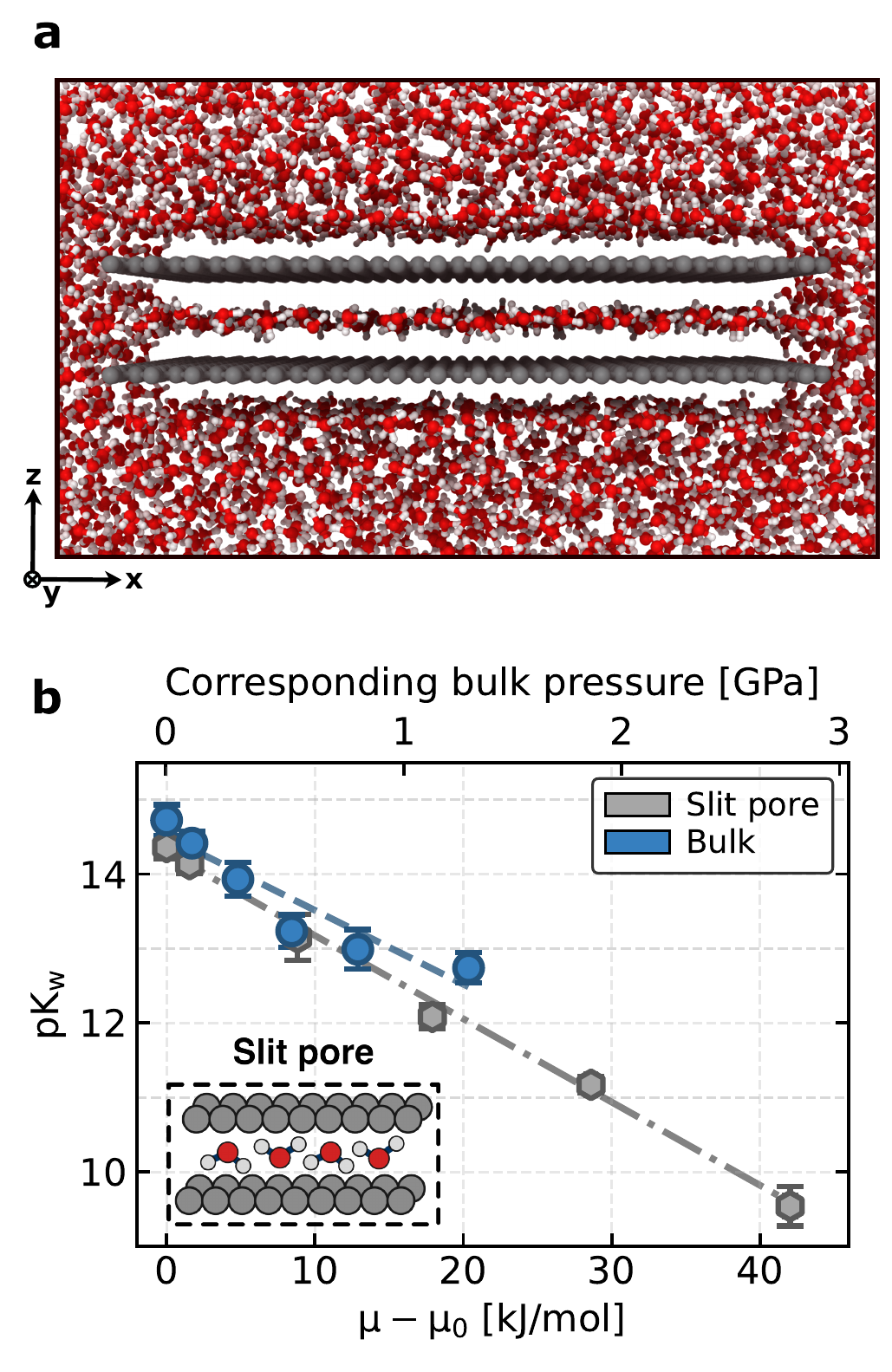}
    \caption{\textbf{Comparison of water self-dissociation in bulk and nanoconfined environments under equivalent thermodynamic conditions.}
    (a) Representative simulation snapshot showing the type of system used to equilibrate confined water: a rigid graphene slit pore with armchair edges immersed in a bulk water reservoir in the $NPT$ ensemble (see \tochange{Section S1} and \tochange{Section S4} for further details).
    This approach enables consistent determination of a chemical potential reference for the confined system, allowing direct comparison with bulk water (see \tochange{Methods}).
    (b) $pK_{\mathrm{w}}$ as a function of the chemical potential difference relative to the corresponding reference, shown for both bulk water and graphene slit pores.
    The accompanying schematic shows the graphene slit pore systems simulated for comparison with bulk water, consisting of monolayer water confined between rigid, parallel graphene sheets (see \tochange{Section S1}).
    The corresponding bulk pressure is indicated on the secondary (top) axis.
    The dashed lines represent linear fits to guide the eye.
}
    \label{fig:fig2}
\end{figure}

Because water self-dissociation is an equilibrium process, meaningful comparisons between bulk and confined environments require a consistent thermodynamic basis.
To this end, we focus on the chemical potential, $\mu$, which governs the equilibrium distribution of molecular and ionic species and offers a natural reference point for comparing dissociation behavior across different conditions \cite{nitzan_book}.
Rather than attempting to match pressure, which becomes ill-defined in nanoconfined systems due to the ambiguity in assigning volume and slit width, as discussed above, we analyze how $p\textrm{K}_{\textrm{w}}$ varies relative to a reference chemical potential, $\mu - \mu_0$\cite{scox_chempot2022}.
We focus specifically on monolayer water confined between rigid graphene sheets, which offers a controlled environment for isolating the thermodynamic factors that influence dissociation.
For bulk water, $\mu_0$ corresponds to the chemical potential of liquid water at 1 bar, a well-established reference in both experiments and simulations.
For confined water, $\mu_0$ corresponds to the chemical potential at which the surface density in the central region of the slit pore matches its equilibrium value when the system is in contact with a bulk water reservoir at 1 bar.
This ensures that both systems are compared under equivalent chemical potential, thereby enabling a thermodynamically consistent comparison.
To determine the equilibrium density in confined water, we simulate a series of systems consisting of two parallel graphene layers, which are periodic along the $y$-axis and immersed in a bulk water reservoir, using the $NPT$ ensemble with pressure maintained at 1 bar (Figure~\ref{fig:fig2}a).
Due to volume fluctuations inherent to the $NPT$ ensemble, the total simulation box dimensions vary, reaching up to approximately 105.742\;\AA\;$\times$\;77.004\;\AA\;$\times$\;36.000\;\AA\;($\approx$30,000 atoms).
This highlights the key role of MLPs in this work, especially recent advances in large-scale models used here (see \tochange{Methods}), in enabling simulations at this scale.
By computing the surface density at the center of the slit and extrapolating to the infinite-size limit, we obtain a thermodynamically consistent reference state for the confined system.

With these reference points established, we systematically vary the chemical potential away from $\mu_0$ in both bulk and confined setups, and compute the corresponding $p\textrm{K}_{\textrm{w}}$ values (see \tochange{Section S4} for details).
For the confined cases, we focus on monolayer water situated between rigid, parallel graphene sheets in a periodic slit pore geometry (similar to the setups introduced in Figure~\ref{fig:fig1}; see \tochange{Section S1}).
This controlled setup enables a direct comparison of dissociation behavior across environments under matched thermodynamic conditions, avoiding the ambiguities associated with pressure- or density-based comparisons.
As shown in Figure~\ref{fig:fig2}b, once framed in terms of $\mu - \mu_0$, the dissociation behavior of confined water is similar to that of the bulk.
This finding indicates that confinement alone does not inherently enhance or suppress water self-dissociation.
Rather, the differences observed across environments primarily reflect shifts in the underlying thermodynamic state imposed by confinement.

\begin{figure*}
    \centering
    \includegraphics[width=0.88\textwidth]{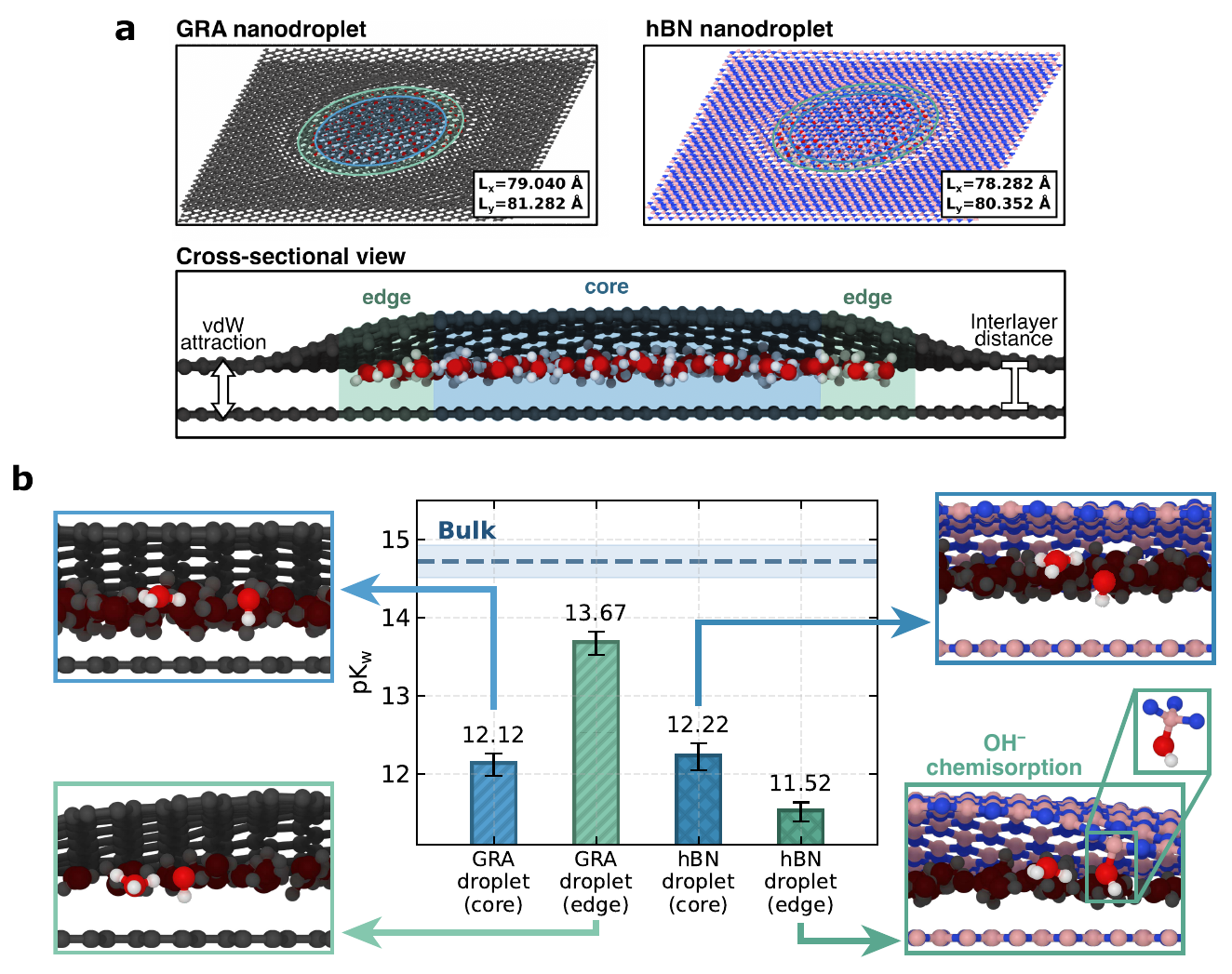}
    \caption{\textbf{Control of water dissociation in nanodroplets.}
    (a) Representative snapshots of water confined within grapehne (GRA) and hBN nanodroplets.
    The full system and cross-sectional views illustrate the droplet geometry, including lateral dimensions ($L_x$, $L_y$).
    These nanodroplets represent a distinct and experimentally realizable form of confinement, complementary to slit pores.
    Core and edge regions of the droplets are indicated by their respective color coding.
    (b) Comparison of the $p\textrm{K}_{\textrm{w}}$ for bulk water and for water confined in GRA and hBN nanodroplets.
    The horizontal dashed blue line marks our bulk reference estimate at 1 bar (corresponding to the equilibrium bulk density), along with its associated uncertainty.
    In the nanodroplet systems, dissociation is induced either at the core or edge to probe spatial variations in water self-dissociation (see \tochange{Section S6}).
    Snapshots for each case are shown, highlighting that edge dissociation in hBN leads to OH$^-$ chemisorption on the surface.
    }
    \label{fig:fig3}
\end{figure*}

\subsection*{Controlling dissociation equilibria in experimentally realizable systems}

Having established a thermodynamically consistent framework to interpret water dissociation across bulk and confined systems, we now pose a broader question: Can this understanding be used to modulate reactivity at the nanoscale?

To explore this, we complement our analysis of slit pores by investigating material-encapsulated nanodroplets, a distinct form of nanoconfinement that introduces interfacial curvature, vdW pressure, and spatial heterogeneity.
These characteristics provide additional degrees of control over the local environment and offer a useful comparison to planar confinement.
In addition to their relevance as experimentally realizable systems (e.g., nanocapillaries) \cite{algara-siller_square_2015}, nanodroplets also serve as representative models for intercalated water layers found in nanofluidic and nanoelectronic devices, as well as in energy storage systems such as batteries and supercapacitors, where confined water plays a key role in mediating wetting, transport, and chemical reactivity \cite{dewet_2012, 2d_diff_2014, clare_2016, forse_nanop_2024}.

Figure~\ref{fig:fig3}a shows representative configurations of the graphene and hBN nanodroplets studied in this work.
Each system consists of a rigid lower substrate, representing a supported 2D material, and a flexible top sheet.
In the absence of water, the vdW attraction brings the sheets into close contact. 
When water is present, however, the interlayer spacing expands locally to accommodate the fluid, forming a droplet-like structure.
These setups are constructed to ensure that the density in the central region of the droplet converges with respect to droplet size, reaching a well-defined equilibrium density that can be meaningfully compared to the equilibrium density of bulk water (see \tochange{Section S5} for details).

We now compare the dissociation behavior in these nanodroplet systems to that of bulk water.
Within each droplet, dissociation events can occur either in the core or near the edge, where the local solvation environments differ substantially.
We begin our discussion with the droplet core.
As shown in Figure~\ref{fig:fig3}b, for both graphene and hBN systems, we observe a decrease in $p\textrm{K}_{\textrm{w}}$ of roughly 2.5 units compared to bulk.
Given the logarithmic scale of $p\textrm{K}_{\textrm{w}}$, this reflects a significant enhancement in water self-dissociation within the interior of the droplet.
At the droplet edge, however, the behavior shifts.
In graphene nanodroplets, moving from the core to the edge leads to a $p\textrm{K}_{\textrm{w}}$ increase of approximately 1.5 units, indicating suppressed dissociation near the droplet edge.
This trend reflects changes in the local solvation environment, underlining the influence of interfacial conditions on water self-dissociation.
A similar shift has been reported at the air–water interface, where moving from bulk to interface incurs a dissociation free energy penalty of around 2 kcal/mol (or approx. 1.5~$p\textrm{K}_{\textrm{w}}$ units) \cite{miguel_ph_2024}.
Given the structural similarities between the graphene–water and air–water interfaces \cite{gra_wat_acid_2025}, these results highlight how comparable interfacial environments can give rise to similar dissociation trends.

In hBN nanodroplets, we observe a starkly contrasting trend: moving from the core to the edge leads to a decrease in $p\textrm{K}_{\textrm{w}}$, indicating enhanced dissociation near the interface.
This behavior arises from a distinct dissociative pathway in which the OH$^-$ ion produced during water dissociation becomes covalently bound to a boron atom at the droplet edge via chemisorption \cite{grosjean_2016, mlb_oh}, as shown in the simulation snapshot in Figure~\ref{fig:fig3}b.
This binding lowers the free energy of the dissociated state and reduces the likelihood of recombination, thereby shifting the dissociation equilibrium, as evidenced by the plateau in the free energy profile (Figure~S12).
It occurs preferentially at the droplet edge, where local curvature and strain alter boron hybridization and enhance surface reactivity.
These observations align with prior studies of hydroxide adsorption on hBN \cite{wang2025}.

\subsection*{Unifying dissociation thermodynamics and routes to modulation}

Until now, our analysis has primarily focused on how interfacial conditions and surface chemistry influence water self-dissociation.
While these factors clearly affect reactivity, the consistently reduced $p\textrm{K}_{\textrm{w}}$ values observed in nanodroplet systems raise a broader question: to what extent are these changes driven by confinement itself?

\begin{figure}
    \centering
    \includegraphics[width=0.48\textwidth]{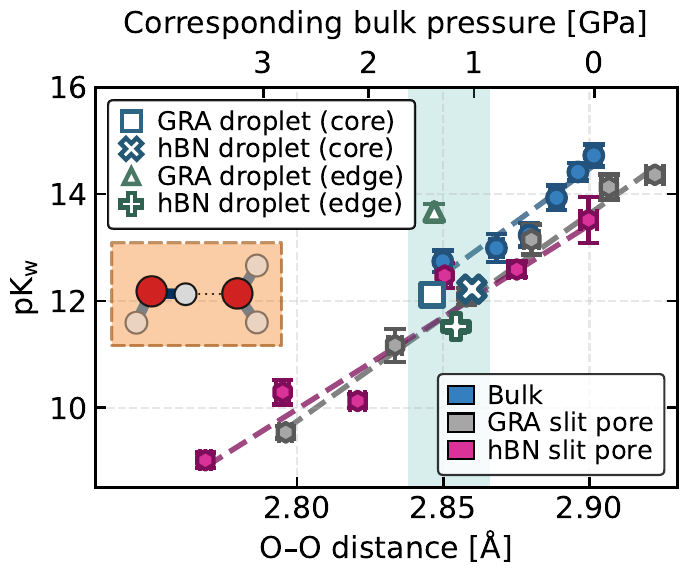}
    \caption{\textbf{Unifying bulk and confined dissociation thermodynamics and outlining routes to modulation.}
    Relationship between $p\textrm{K}_{\textrm{w}}$ and the average O-O distances in O–H$\cdots$O pairs, which promote water self-dissociation.
    The O–O distances are calculated either at the droplet core or near the edge, depending on where the dissociated species are located in each case (see \tochange{Section S6}).
    The corresponding bulk pressure is indicated on the secondary (top) axis.
    The teal-shaded region highlights results for nanodroplets across different scenarios and indicates a corresponding bulk pressure of $\sim$1 GPa.
    The accompanying schematic illustrates a representative O–H$\cdots$O pair.
    }
    \label{fig:fig4}
\end{figure}

To explore this, we turn to a microscopic structural descriptor that reflects a system’s ability to support proton transfer, a key step in self-dissociation.
Specifically, we examine the average O–O distance in hydrogen-bonded O–H$\cdots$O pairs, which couples strongly with the proton transfer barrier modulating how readily protons can be transferred.
Figure~\ref{fig:fig4} presents these distances across all systems studied and offers a unified view of how structural changes relate to dissociation behavior under both bulk and confined conditions.

As shown in Figure~\ref{fig:fig4}, in most cases—whether in bulk water, slit pores, or nanodroplets—the variation in $p\textrm{K}{\textrm{w}}$ is well captured by changes in O–H$\cdots$O distances.
Although absolute O–O values vary slightly between systems, they exhibit parallel trends, indicating a common structural response to compression.
This behavior mirrors the well-known effect of pressure in bulk water, where shorter O–O distances are associated with enhanced dissociation \cite{pkw_press_experiments_bulk_2005}.
Moreover, it aligns with the thermodynamic trends in Figure~\ref{fig:fig2}b, where comparable shifts in chemical potential led to similar reductions in $p\textrm{K}{\textrm{w}}$ across environments.
Notably, the degree of structural compression observed in nanodroplet cores corresponds to effective pressures approaching 1~GPa, consistent with earlier estimates \cite{algara-siller_square_2015, encaps_wat_2017}, further reinforcing this pressure-like interpretation of confinement.
Together, these observations suggest that confinement enhances dissociation through a shared mechanism: compression-induced shortening of hydrogen bonds that facilitates proton transfer.
In addition to their mechanistic relevance, the O–O distances also serve as a practical descriptor for reactivity.
Unlike full thermodynamic treatments such as the chemical potential-controlled comparisons presented earlier, O–O distances can be readily extracted from both simulations and experiments, offering a physically interpretable and broadly applicable proxy for dissociation behavior across environments.
Crucially, however, the relationship between compression and dissociation is not universal.
While shorter O–O distances generally correlate with increased proton transfer; this trend can be modulated by local interfacial properties, which alter the hydrogen-bonding environment and the stability of dissociated species.
For example, at the edges of graphene and hBN nanodroplets, dissociation behavior may begin to diverge from bulk-like trends, influenced by localized interfacial effects such as strain or reactive surface sites that can stabilize dissociated species.
In such cases, even subtle changes to the local environment can shift dissociation equilibria, highlighting how interfacial design can offer a route for tuning acid–base behavior at the nanoscale.

These findings reveal that confinement alone does not inherently alter water’s dissociation chemistry.
Instead, the differences often arise from changes in the underlying thermodynamic conditions between confined and bulk systems.
Notable deviations emerge when chemical or interfacial complexity becomes significant, marking the breakdown of bulk-like behavior and underscoring how surface chemistry and interface design can be used as effective levers for tuning acid–base equilibria at the nanoscale.

\section*{Discussion}
Our results reframe how nanoscale aqueous reactivity should be interpreted.
Rather than treating confinement as a standalone variable, we show that water reactivity depends critically on the thermodynamic and interfacial context, a perspective enabled by adopting chemical potential as the central lens for comparison.
This shift in framing not only reconciles previous contradictions but also lays the groundwork for more predictive control of confined water chemistry.

In practical terms, this means that observed reactivity in experimental systems such as carbon membranes, nanocapillaries, or layered 2D materials may vary significantly depending on local curvature, hydration level, or surface chemistry.
Such sensitivity offers both a challenge and an opportunity: a challenge in interpreting inconsistent results, and an opportunity to deliberately tune reactivity through structural and environmental design.

This work addresses a long-standing ambiguity, where water self-dissociation under nanoconfinement has been reported as both enhanced and suppressed relative to the bulk.
Using machine-learned molecular dynamics simulations trained at first-principles accuracy, we systematically investigated dissociation equilibria across a diverse range of nanoconfined systems.
By varying the chemical potential to enable thermodynamically consistent comparisons, we demonstrated that water confined within rigid graphene slit pores behaves similarly to bulk water under equivalent conditions.
This finding demystifies many prior discrepancies, where confinement was assumed to alter reactivity directly.
Instead, we show that geometric confinement alone does not intrinsically promote or inhibit self-dissociation. Apparent differences often stem from inconsistent comparisons, particularly in pressure or density, among others.

At the same time, our results reveal clear, physically motivated pathways through which confinement can modulate reactivity, especially when interfacial chemistry and structural heterogeneity are introduced.
In encapsulated nanodroplets, a complex class of experimentally realizable confinement that introduces interfacial curvature and spatial heterogeneity, we identify a dissociative pathway at hBN interfaces where hydroxide ions are stabilized via chemisorption at droplet edges.
This alternative mechanism lowers the free energy cost of dissociation, unlocking reactivity inaccessible in bulk water or chemically inert environments.

Rather than offering a single answer to how reactive water is at the nanoscale, here we provide a unifying perspective that explains when and why confinement affects self-dissociation.
This molecular-level insight opens new avenues for controlling aqueous reactivity in applied settings ranging from ion transport in energy storage to proton-mediated processes in catalysis and sensing.
Understanding how to selectively stabilize reactive species in confined environments lays the foundation for interface design strategies that combine geometric control with chemical specificity to tune water chemistry at the nanoscale.

\section*{Methods}
\label{sec:methods}

\textbf{Machine Learning Potentials.} 
The MLPs used in this work were developed using the MACE architecture \cite{mace_canonical}, employing 128 invariant channels, two layers, and a cutoff distance of 6\;\AA.
Each model captures semi-local interactions through an effective receptive field, which extends to 12\;\AA, corresponding to the product of the number of layers and the cutoff distance per layer. 
All models were trained to reproduce the revPBE-D3 reference potential energy surface with high fidelity\cite{litman_efield_2025} (see \tochange{Section S2}).
We developed two separate MLPs: one for graphene–water systems and one for hBN–water systems.

The graphene–water MLP is based on training data from Ref.~\citenum{gra_wat_acid_2025}, which includes water self-dissociated configurations across various density regimes within graphene confinement, including the ultra-confined limit.
It also incorporates data from Ref.\citenum{flt_nanotubes}, covering interfaces from planar graphene to highly confined water in carbon nanotubes of varying radii.
Together, these datasets allow the model to accurately capture nanoconfined water behavior in slit pores while also describing graphene’s bending rigidity, which is a key property for modeling nanodroplet formation (see \tochange{Section S2}).
To improve the description of graphene–graphene interactions, we included structures at equilibrium and slightly perturbed interlayer distances, for both AA and AB stacking.
This ensures that the model properly describes the vdW-driven closure of the nanodroplet, which is governed by the attractive forces between the graphene sheets.
Given the broad density range explored in this work, we incorporated representative high- and low-density configurations into the training set, including corresponding structures under rigid graphene confinement to ensure transferability across relevant thermodynamic conditions.
We also varied the graphene layer separations to capture water behavior across distinct interlayer environments.
Finally, to reflect the large-scale systems considered here, we extended the training set with configurations from larger confined water systems. 
The final training set comprised 5,845 structures, yielding root-mean-square errors of 0.9\;meV/atom for energies and 26.3\;meV/\AA\; for forces.

The hBN–water MLP was developed following the same procedure, with additional training data from Ref.~\citenum{wang2025} to accurately describe OH$^{–}$–hBN interactions.
As with the graphene model, this MLP includes configurations from both slit pore and nanodroplet geometries, ensuring broad coverage of structural and thermodynamic conditions relevant to this study.
The final training set comprised 5,395 structures, yielding root-mean-square errors of 0.7\;meV/atom for energies and 21.1\;meV/\AA\; for forces.

\textbf{Electronic Structure.} All electronic structure calculations to train the MLP were carried out using the CP2K/Quickstep code \cite{cp2k_2020}. 
The revPBE-D3 functional \cite{revpbed3_1, revpbed3_2} was chosen for its reliable performance in capturing the structure and dynamics of liquid water \cite{angelos_dft_water_2016, tobias_vdw_2016, ondrej_revpbe_2017} and its ionized products \cite{marsalek_mlp_pt, gra_wat_acid_2025} as well as the interaction between water and graphene \cite{Brandenburg2019}.
Atomic cores were represented using dual-space Goedecker-Teter-Hutter pseudopotentials \cite{gth_1996}.
In the Gaussian and plane waves method, the Kohn-Sham orbitals of oxygen and hydrogen atoms were represented using the TZV2P basis set, whereas carbon, nitrogen, and boron atoms were described using the DZVP basis set. 
The electron density was represented using an auxiliary plane-wave basis with a cutoff energy of 1050~Ry. 
See \tochange{Section S2} for further details.

\textbf{Molecular Dynamics Simulations.} All simulations were performed using machine-learned potentials. Except for the large-scale simulations described at the end of this section, all simulations were carried out using the ASE software package \cite{ase_2017}, with enhanced sampling implemented via PLUMED \cite{plumed}.
Dynamics were propagated at a temperature of 300 K in the $NVT$ ensemble, with a 0.5 fs time step and a Langevin thermostat with a friction coefficient of 2.5 ps$^{-1}$.
All systems were modeled within orthorhombic simulation cells, applying periodic boundary conditions along all three spatial dimensions.
All simulations used hydrogen atom masses.
To avoid interactions between periodic images, a 15\;\AA\;vacuum, greater than the model's receptive field, was introduced along the $z$-axis.

Each of the $p\textrm{K}_{\textrm{w}}$ values reported in this work was obtained from the free energy profiles for the water self-dissociation reaction via umbrella sampling (see \tochange{Section S3}).
In each umbrella window, the coordination number of a selected oxygen atom (O$^{*}$) with all hydrogen atoms in the system was defined as:
\begin{equation}
n_{\textrm{H}} = \sum^{N}_{i=1} \frac{1 - \left( r_i / R_0 \right)^{12} }{1 - \left( r_i / R_0 \right)^{24}}
\end{equation}
where $i$ iterates over all the hydrogens in the simulation box, $r_{i}$ is the distance between hydrogen $i$ and O$^{*}$, and $R_{0}$ is a switching distance set to 1.38\;\AA\;from Ref.~\citenum{sprik_pk_coord_2000}.
To enforce sampling along the reaction coordinate, $n_{\text{H}}$ was restrained around a target value $n_{\textrm{H}}^{\prime}$ using a harmonic potential with a force constant of 200 kcal/mol per coordination unit squared.
A total of 31 windows with $n_{\textrm{H}}^{\prime} = 1.00, 1.04, \ldots, 2.20$ were considered for each energy profile.
For each window, a simulation of 100 ps was performed.
To reconstruct the free energy profile, thermodynamic integration \cite{umbrella_int_2005, umbrella_error_2006} was used.
In total, 27 $p\textrm{K}_{\textrm{w}}$ estimates are reported in this work, each based on 3.1~ns of simulation time, amounting to over 80~ns of total simulation.
This highlights the critical role of machine learning-based MD in making such extensive sampling computationally feasible.
Because our goal is to understand how confinement modulates water dissociation, we emphasize relative differences in $p\textrm{K}_{\textrm{w}}$ across systems rather than precise absolute values.
This strategy ensures that our conclusions remain robust and transferable, as they are less sensitive to the choice of electronic structure method, neglect of nuclear quantum effects, or the specific sampling protocol.

To establish a chemical potential reference for nanoconfined water, allowing direct comparison with bulk water, we determined the equilibrium density of water confined within a rigid graphene slit pore immersed in an aqueous liquid reservoir.
These simulations were performed in the $NPT$ ensemble, maintaining a pressure of 1 bar.
To preserve the structural rigidity of the graphene sheets, the barostat was applied only to the water molecules, and only along the $x$-direction (perpendicular to the slit pore walls).
No pressure coupling was applied along the $y$-axis, which remained fully periodic.
The system consisted of two parallel graphene layers of equal length (see Figure~\ref{fig:fig2}a), and the surface density was computed within the central region of the slit pore to characterize the confined water.
To obtain the equilibrium confined density, we systematically varied the size of the slit pores.
For each pore size, the graphene sheets were immersed in a sufficiently large water reservoir, with the amount of surrounding water adjusted to ensure bulk-like behavior at the lateral edges of the graphene layers.
The equilibrium density in the confined system was determined by extrapolating the densities obtained from these simulations (see \tochange{Section S4}).
To overcome the computational cost associated with these large-scale simulations, ranging from 15,000 to 30,000 atoms, we employed the Symmetrix library \cite{mace_off, symmetrix}, an optimized C++ and Kokkos implementation that accelerates machine-learned potentials for efficient large-scale inference.
Symmetrix interfaces directly with LAMMPS \cite{lammps_2022}, providing efficient large-scale inference.

\begin{acknowledgements}
We thank Stephen J. Cox for fruitful discussions.
X.R.A., Y.L., and A.M. acknowledge support from the European Union under the “n-AQUA” European Research Council project (Grant No. 101071937). 
K.D.F. acknowledges support from Schmidt Science Fellows, in partnership with the Rhodes Trust, and Trinity College, Cambridge.
W.C.W. acknowledges support from the EPSRC (Grant EP/V062654/1).
C.S. acknowledges financial support from the Deutsche Forschungsgemeinschaft (DFG, German Research Foundation) project number 500244608, as well as from the Royal Society grant number RGS/R2/242614.
This work used the ARCHER2 UK National Supercomputing Service via the UK’s HEC Materials Chemistry Consortium, funded by EPSRC (EP/F067496).
We also utilized resources from the Cambridge Service for Data Driven Discovery (CSD3), supported by EPSRC (EP/T022159/1) and DiRAC funding, with additional access through a University of Cambridge EPSRC Core Equipment Award (EP/X034712/1).
The Cirrus UK National Tier-2 HPC Service at EPCC, funded by the University of Edinburgh and the EPSRC (EP/P020267/1), also provided computational support.
We further acknowledge the EuroHPC Joint Undertaking for awarding this project access to the EuroHPC supercomputer LEONARDO, hosted by CINECA (Italy) and the LEONARDO consortium through an EuroHPC Regular Access call.
\end{acknowledgements}

\section*{Data Availability}
All data required to reproduce the findings of this work will be made openly available on GitHub upon acceptance of this manuscript.

\section*{References}
%aipnum4-2.bst 2019-01-14 (MD) hand-edited version of apsrev4-1.bst
%Control: production of eprint (1) enabled
%

%

\end{document}

\def\mytitle{How reactive is water at the nanoscale and how to control it?}
\title{Supporting Information for: \mytitle}

\author{Xavier R. Advincula}
\affiliation{Yusuf Hamied Department of Chemistry, University of Cambridge, Lensfield Road, Cambridge, CB2 1EW, UK}
\affiliation{Cavendish Laboratory, Department of Physics, University of Cambridge, Cambridge, CB3 0HE, UK}
\affiliation{Lennard-Jones Centre, University of Cambridge, Trinity Ln, Cambridge, CB2 1TN, UK}

\author{Yair Litman}
\affiliation{Yusuf Hamied Department of Chemistry, University of Cambridge, Lensfield Road, Cambridge, CB2 1EW, UK}
\affiliation{Lennard-Jones Centre, University of Cambridge, Trinity Ln, Cambridge, CB2 1TN, UK}
\affiliation{Max Planck Institute for Polymer Research, Ackermannweg 10, 55128 Mainz, Germany}

\author{Kara D. Fong}
\affiliation{Yusuf Hamied Department of Chemistry, University of Cambridge, Lensfield Road, Cambridge, CB2 1EW, UK}
\affiliation{Lennard-Jones Centre, University of Cambridge, Trinity Ln, Cambridge, CB2 1TN, UK}
\affiliation{Division of Chemistry and Chemical Engineering, California Institute of Technology, Pasadena, California 91125, USA}

\author{William C. Witt}
\affiliation{Harvard John A. Paulson School of Engineering and Applied Sciences, Harvard University, Cambridge, MA, USA}

\author{Christoph Schran}
\email{cs2121@cam.ac.uk}
\affiliation{Cavendish Laboratory, Department of Physics, University of Cambridge, Cambridge, CB3 0HE, UK}
\affiliation{Lennard-Jones Centre, University of Cambridge, Trinity Ln, Cambridge, CB2 1TN, UK}

\author{Angelos Michaelides}
\email{am452@cam.ac.uk}
\affiliation{Yusuf Hamied Department of Chemistry, University of Cambridge, Lensfield Road, Cambridge, CB2 1EW, UK}
\affiliation{Lennard-Jones Centre, University of Cambridge, Trinity Ln, Cambridge, CB2 1TN, UK}

\keywords{}

{\maketitle}
\tableofcontents

\onecolumngrid

\FloatBarrier

\newpage
\section{Molecular dynamics simulations} \label{sec:md_sims}
\subsection{System setup} 

The systems investigated in this work span both bulk water and a range of nanoconfined environments.
These include water confined between parallel rigid graphene (GRA) sheets, water confined between hexagonal boron nitride (hBN) sheets, and water encapsulated within graphene and hBN nanodroplets.
In addition, we considered graphene slit pores immersed in an aqueous liquid reservoir, used to establish a chemical potential reference for confined water.

All simulations used orthorhombic cells with periodic boundary conditions in all three dimensions.
In graphene-containing systems, the graphene sheets were constructed by repeating the unit cell dimensions $a = \sqrt{3}d_c$ and $b = 3d_c$ along the $x$ and $y$ directions, respectively, where $d_c = 1.42$\;\AA\; is the carbon–carbon bond length\cite{RevModPhys.81.109}.
For example, the graphene dimensions of $L_x = 44.460$\;\AA\; and $L_y = 47.058$\;\AA\; were obtained by repeating the unit cell 18 times along $x$ and 11 times along $y$.
This tiling of the unit cell ensures that the graphene sheet maintains its characteristic hexagonal lattice structure with a consistent carbon-carbon bond distance throughout the extended sheet.
An analogous approach was applied to construct the hBN sheets.

A vacuum space of 15\;\AA\; was added in the $z$ direction to prevent interactions between the periodic images in the confined systems, as this exceeds the model’s receptive field.
An overview of the systems investigated in this work is presented in Tables \ref{tab:syst_bulk}--\ref{tab:syst_liq_reserv}, which contains all the relevant information of these systems as well as illustrative schematics. \\

\newpage

\begin{longtable*}{@{}c >{\centering\arraybackslash}p{6cm} >{\centering\arraybackslash}p{5cm}}
\toprule
\toprule
\begin{tabular}[c]{@{}c@{}}\textbf{System}\\\textbf{(dimensions)}\end{tabular} & \textbf{Simulation details} & \textbf{Illustration} \\* \midrule
\endfirsthead
\endhead

\begin{tabular}[c]{@{}c@{}}Bulk water\\ (19.86\;\AA\;$\times$\;19.86\;\AA\;$\times$\;19.86\;\AA)\end{tabular}  & 
\begin{tabular}[c]{@{}c@{}} $N_{\textrm{atoms}}=768$\\ $N_{\textrm{H}_2\textrm{O}}=256$\\ $N_{\textrm{umbrellas}}=31$\\ $t_{\textrm{eq}}=50$ ps\\ $t_{\textrm{sim/umbrella}}=100$ ps \\ $\rho_{\textrm{O}}=0.9777$ g/cm$^{3}$\\ $p\textrm{K}_{\textrm{w}}=14.72 \pm 0.21$\\ $\mu-\mu_0=0$ kJ/mol \end{tabular}  & 
\begin{minipage}{0.2\textwidth}
\vspace{0.4cm}
      \includegraphics[width=\linewidth]{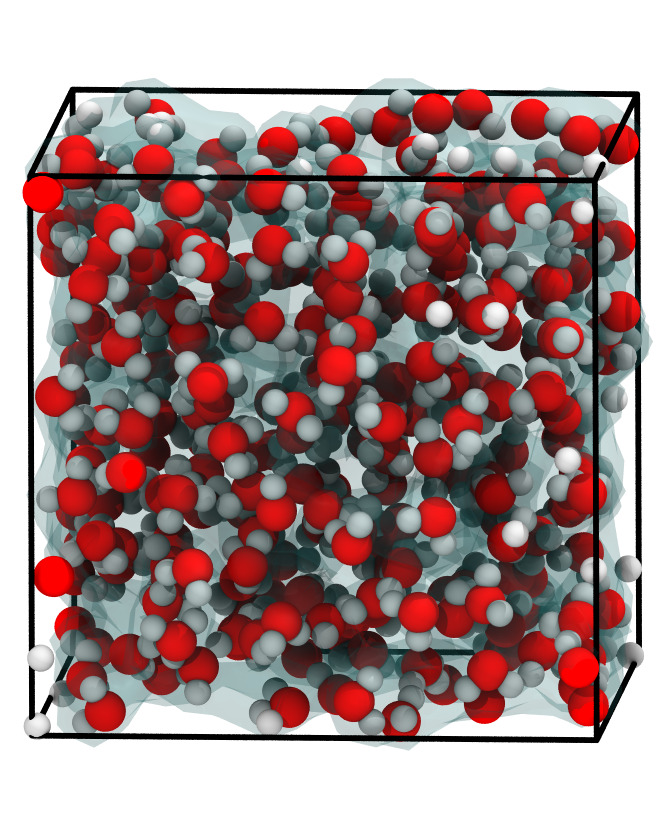} \\
\end{minipage}   \\* \midrule

\begin{tabular}[c]{@{}c@{}}Bulk water\\ (19.60\;\AA\;$\times$\;19.60\;\AA\;$\times$\;19.60\;\AA)\end{tabular}  & 
\begin{tabular}[c]{@{}c@{}} $N_{\textrm{atoms}}=768$\\ $N_{\textrm{H}_2\textrm{O}}=256$\\ $N_{\textrm{umbrellas}}=31$\\ $t_{\textrm{eq}}=50$ ps\\ $t_{\textrm{sim/umbrella}}=100$ ps \\ $\rho_{\textrm{O}}=1.0171$ g/cm$^{3}$\\ $p\textrm{K}_{\textrm{w}}=14.42 \pm 0.16$\\ $\mu-\mu_0=1.71$ kJ/mol \end{tabular}  & 
\begin{minipage}{0.2\textwidth}
\vspace{0.4cm}
      \includegraphics[width=0.9603\linewidth]{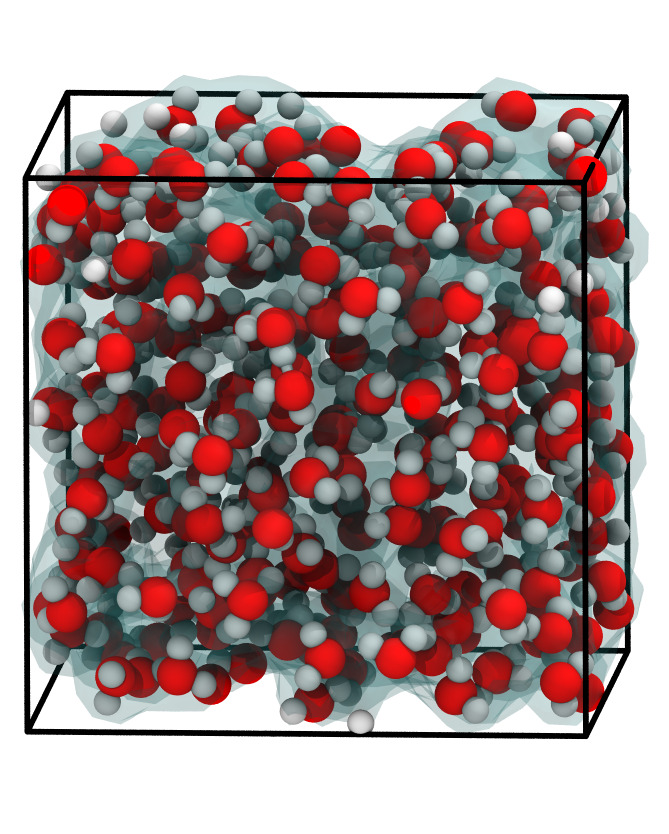} \\
\end{minipage}   \\* \midrule

\begin{tabular}[c]{@{}c@{}}Bulk water\\ (19.24\;\AA\;$\times$\;19.24\;\AA\;$\times$\;19.24\;\AA)\end{tabular}  & 
\begin{tabular}[c]{@{}c@{}} $N_{\textrm{atoms}}=768$\\ $N_{\textrm{H}_2\textrm{O}}=256$\\ $N_{\textrm{umbrellas}}=31$\\ $t_{\textrm{eq}}=50$ ps\\ $t_{\textrm{sim/umbrella}}=100$ ps \\ $\rho_{\textrm{O}}=1.0753$ g/cm$^{3}$\\ $p\textrm{K}_{\textrm{w}}=13.93\pm 0.23
$\\ $\mu-\mu_0=4.81$ kJ/mol \end{tabular}  & 
\begin{minipage}{0.2\textwidth}
\vspace{0.4cm}
      \includegraphics[width=0.9079\linewidth]{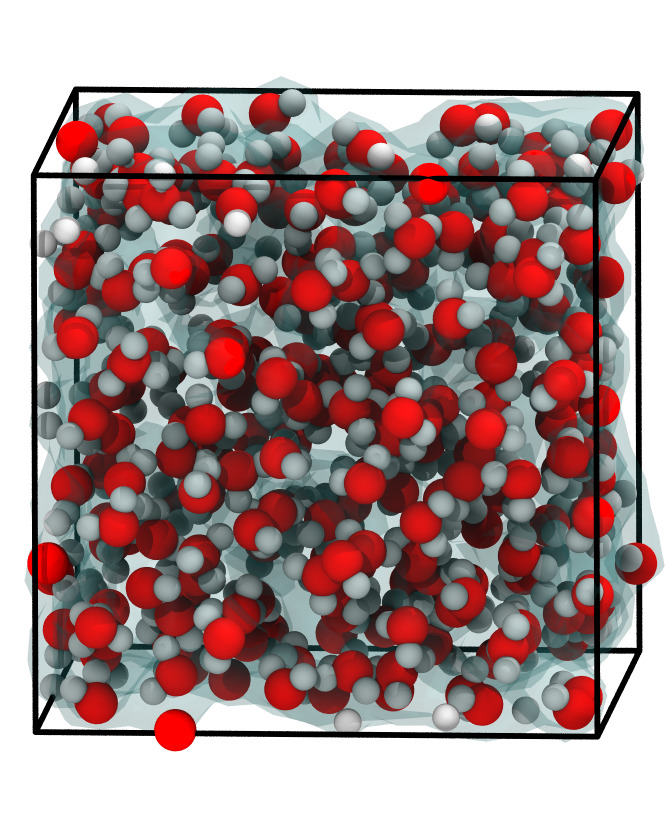} \\
\end{minipage}   \\* \midrule

\begin{tabular}[c]{@{}c@{}}Bulk water\\ (18.93\;\AA\;$\times$\;18.93\;\AA\;$\times$\;18.93\;\AA)\end{tabular}  & 
\begin{tabular}[c]{@{}c@{}} $N_{\textrm{atoms}}=768$\\ $N_{\textrm{H}_2\textrm{O}}=256$\\ $N_{\textrm{umbrellas}}=31$\\ $t_{\textrm{eq}}=50$ ps\\ $t_{\textrm{sim/umbrella}}=100$ ps \\ $\rho=1.1290$ g/cm$^{3}$\\ $p\textrm{K}_{\textrm{w}}=13.24 \pm 0.22
$\\ $\mu-\mu_0=8.43$ kJ/mol \end{tabular}  & 
\begin{minipage}{0.2\textwidth}
\vspace{0.4cm}
      \includegraphics[width=0.8650\linewidth]{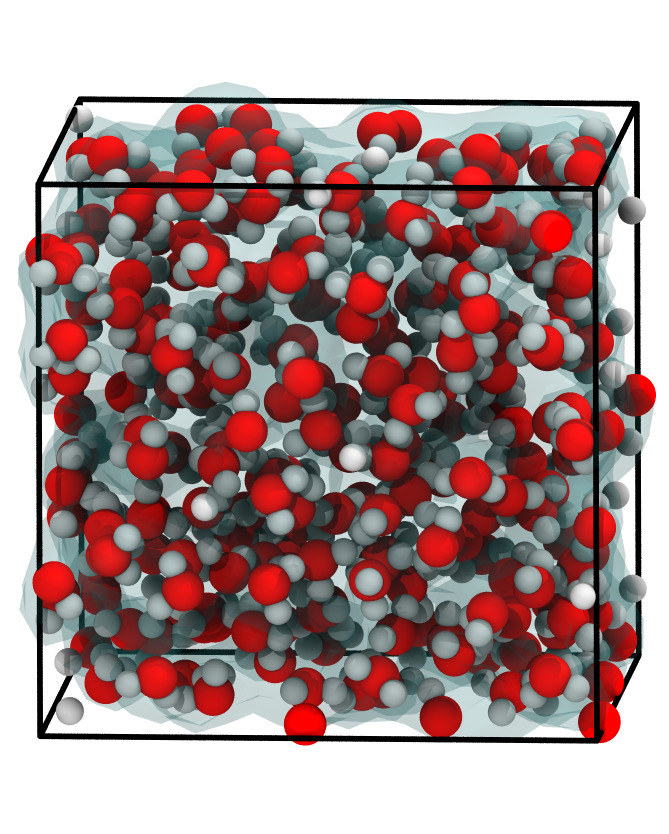} \\
\end{minipage}   \\* \midrule

\begin{tabular}[c]{@{}c@{}}Bulk water\\ (18.61\;\AA\;$\times$\;18.61\;\AA\;$\times$\;18.61\;\AA)\end{tabular}  & 
\begin{tabular}[c]{@{}c@{}} $N_{\textrm{atoms}}=768$\\ $N_{\textrm{H}_2\textrm{O}}=256$\\ $N_{\textrm{umbrellas}}=31$\\ $t_{\textrm{eq}}=50$ ps\\ $t_{\textrm{sim/umbrella}}=100$ ps \\ $\rho_{\textrm{O}}=1.1882$ g/cm$^{3}$\\ $p\textrm{K}_{\textrm{w}}=12.99 \pm 0.26
$\\ $\mu-\mu_0=12.89$ kJ/mol \end{tabular}  & 
\begin{minipage}{0.2\textwidth}
\vspace{0.4cm}
      \includegraphics[width=0.8224\linewidth]{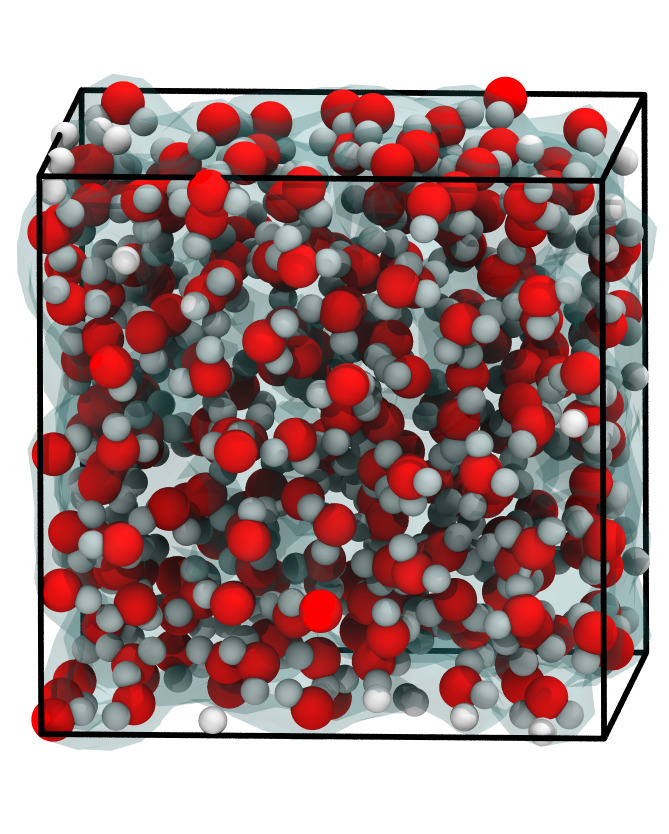} \\
\end{minipage}   \\* \midrule

\begin{tabular}[c]{@{}c@{}}Bulk water\\ (18.20\;\AA\;$\times$\;18.20\;\AA\;$\times$\;18.20\;\AA)\end{tabular}  & 
\begin{tabular}[c]{@{}c@{}} $N_{\textrm{atoms}}=768$\\ $N_{\textrm{H}_2\textrm{O}}=256$\\ $N_{\textrm{umbrellas}}=31$\\ $t_{\textrm{eq}}=50$ ps\\ $t_{\textrm{sim/umbrella}}=100$ ps \\ $\rho_{\textrm{O}}=1.2704$ g/cm$^{3}$\\ $p\textrm{K}_{\textrm{w}}=12.74 \pm 0.20
$\\ $\mu-\mu_0=20.36$ kJ/mol \end{tabular}  & 
\begin{minipage}{0.2\textwidth}
\vspace{0.4cm}
      \includegraphics[width=0.7679\linewidth]{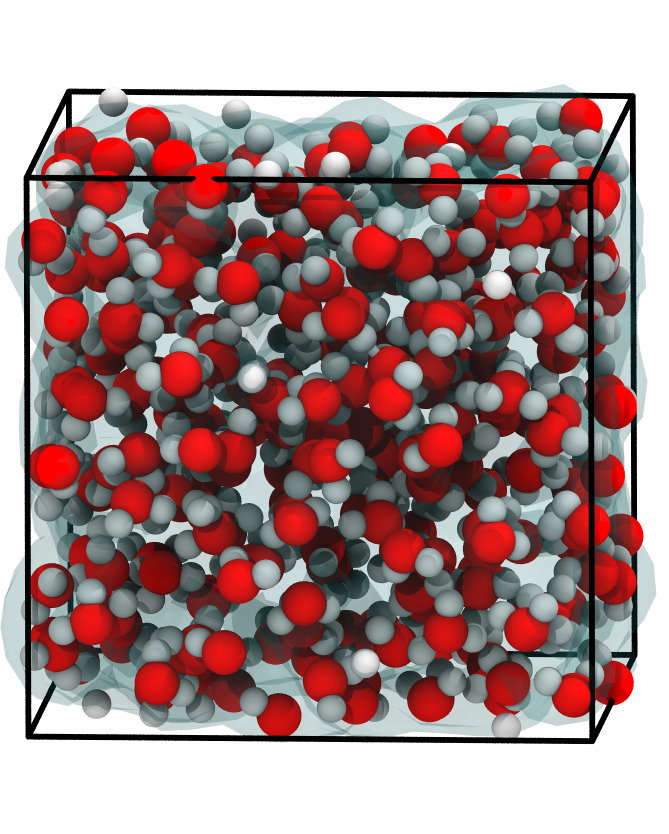} \\
\end{minipage}   \\* \bottomrule

\bottomrule

\multicolumn{3}{c}{
\parbox{0.95\textwidth}{
  \vspace{0.6cm}
  Table S1: Detailed overview of the bulk systems considered in this work.
  For each system, we report the total number of atoms, $N_{\textrm{atoms}}$; the corresponding number of water molecules, $N_{\textrm{H}_2\textrm{O}}$; the number of umbrellas sampled $N_{\textrm{umbrellas}}$; the equilibration time, $t_{\textrm{eq}}$; the simulation production time per umbrella, $t_{\textrm{sim/umbrella}}$; its density, $\rho_{\textrm{O}}$; its $p\textrm{K}_{\textrm{w}}$ value; and the chemical potential difference relative to the reference, $\mu - \mu_0$.}}
  \label{tab:syst_bulk}
\end{longtable*}

\begin{longtable}[c]{@{}c >{\centering\arraybackslash}p{6cm} >{\centering\arraybackslash}p{5cm}}
\toprule \toprule
\begin{tabular}[c]{@{}c@{}}\textbf{System}\\\textbf{(dimensions)}\end{tabular} & \textbf{Simulation details} & \textbf{Illustration} \\* \midrule
\endfirsthead
\endhead
\begin{tabular}[c]{@{}c@{}}Monolayer water confined between\\ parallel rigid GRA sheets\\ (44.460\;\AA\;$\times$\;47.058\;\AA\;$\times$\;21.700\;\AA)\end{tabular}  & 
\begin{tabular}[c]{@{}c@{}} $N_{\textrm{atoms}}=2214$\\ $N_{\textrm{C}}=1584$\\ $N_{\textrm{H}_2\textrm{O}}=210$\\ $N_{\textrm{umbrellas}}=31$\\ $t_{\textrm{eq}}=50$ ps\\ $t_{\textrm{sim/umbrella}}=100$ ps \\ $\rho_{\textrm{O}}^ {\textrm{2D}}=0.1003$ \#O/\AA$^{2}$\\ $p\textrm{K}_{\textrm{w}}=14.36 \pm 0.13	
$\\ $\mu-\mu_0=0.00$ kJ/mol\end{tabular}  &
\begin{minipage}{0.25\textwidth}
      \includegraphics[width=\linewidth]{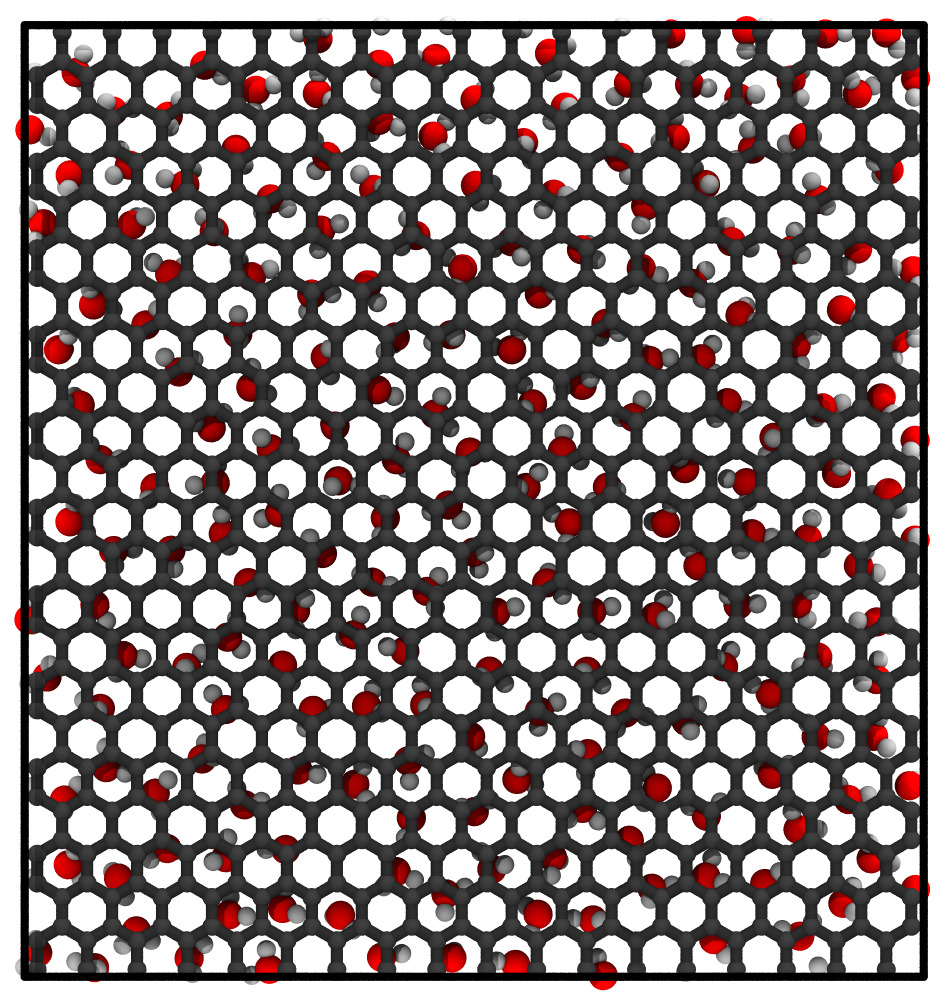}
      \includegraphics[width=\linewidth]{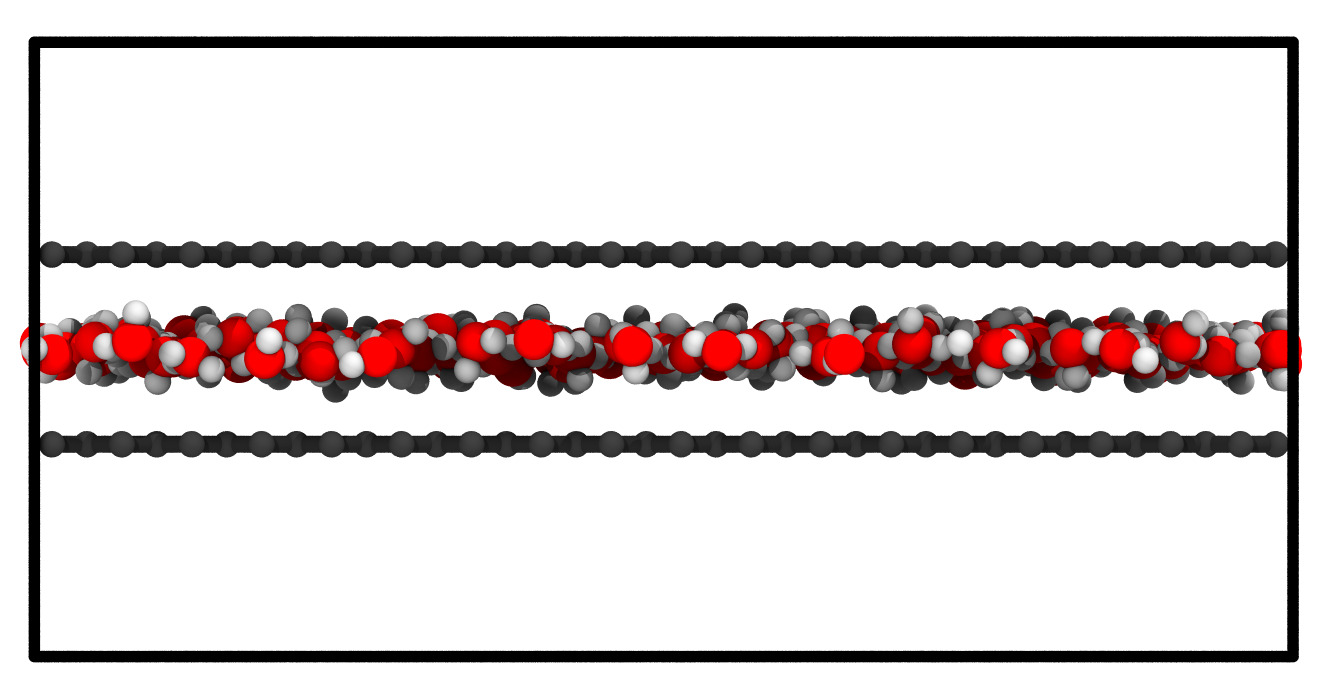}
\end{minipage}   \\* \midrule
\begin{tabular}[c]{@{}c@{}}Monolayer water confined between\\ parallel rigid GRA sheets\\  (44.460\;\AA\;$\times$\;47.058\;\AA\;$\times$\;21.700\;\AA)\end{tabular}  & 
\begin{tabular}[c]{@{}c@{}} $N_{\textrm{atoms}}=2250$\\ $N_{\textrm{C}}=1584$\\ $N_{\textrm{H}_2\textrm{O}}=222$\\ $N_{\textrm{umbrellas}}=31$\\ $t_{\textrm{eq}}=50$ ps\\ $t_{\textrm{sim/umbrella}}=100$ ps \\ $\rho_{\textrm{O}}^ {\textrm{2D}}=0.1062$ \#O/\AA$^{2}$\\ $p\textrm{K}_{\textrm{w}}=14.13 \pm 0.23	
$\\ $\mu-\mu_0=1.55$ kJ/mol\end{tabular}  &
\begin{minipage}{0.25\textwidth}
      \includegraphics[width=\linewidth]{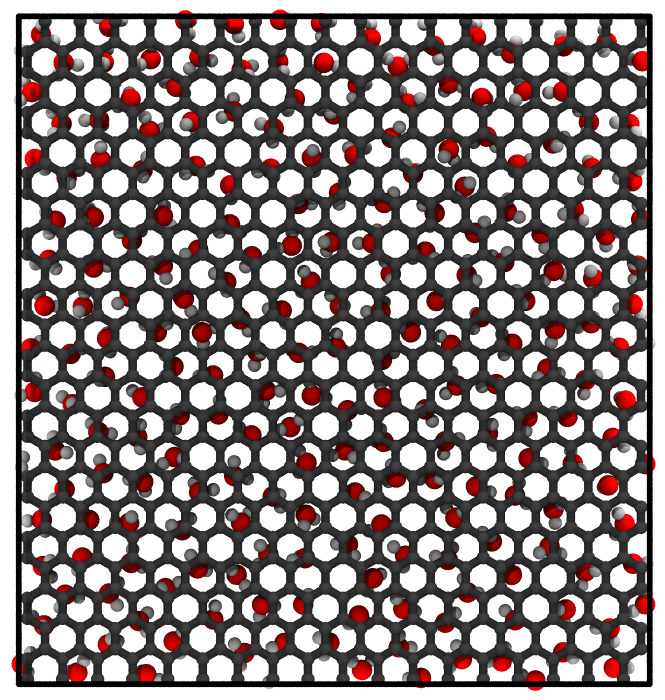}
      \includegraphics[width=\linewidth]{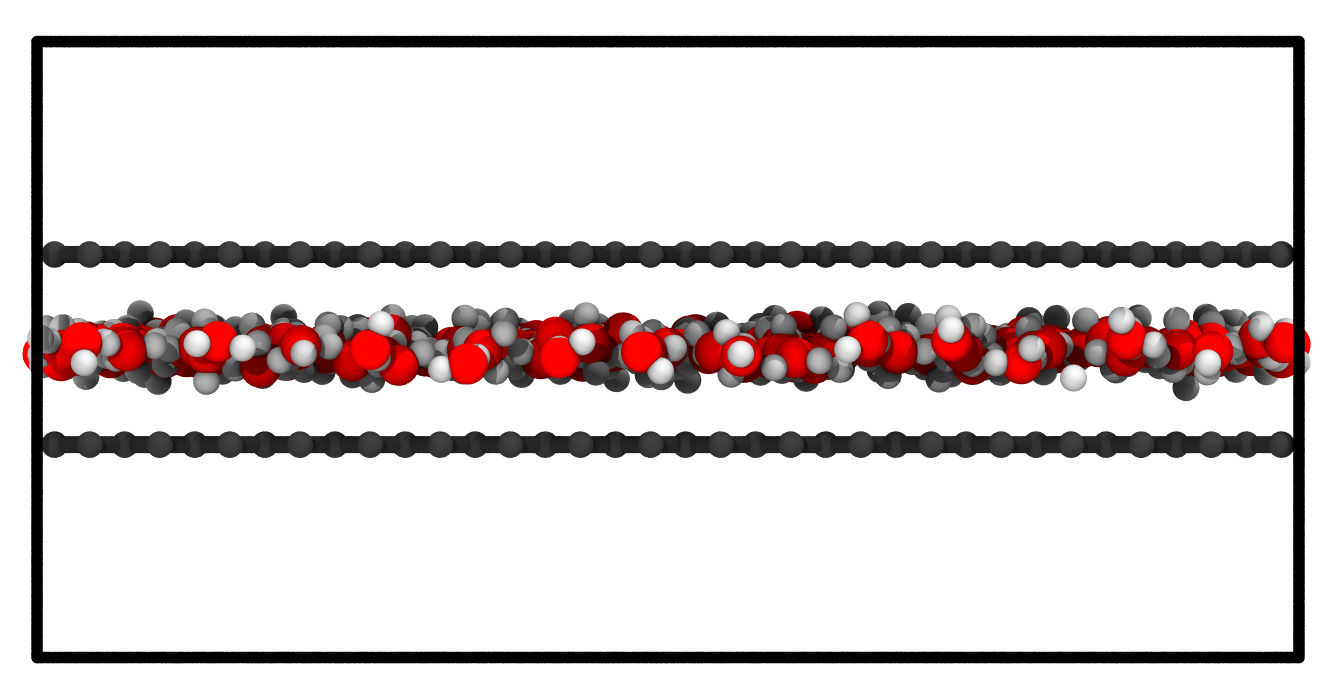}
\end{minipage}   \\* \midrule
\begin{tabular}[c]{@{}c@{}}Monolayer water confined between\\ parallel rigid GRA sheets\\ (44.460\;\AA\;$\times$\;47.058\;\AA\;$\times$\;21.700\;\AA)\end{tabular}  & 
\begin{tabular}[c]{@{}c@{}} $N_{\textrm{atoms}}=2289$\\ $N_{\textrm{C}}=1584$\\ $N_{\textrm{H}_2\textrm{O}}=235$\\ $N_{\textrm{umbrellas}}=31$\\ $t_{\textrm{eq}}=50$ ps\\ $t_{\textrm{sim/umbrella}}=100$ ps \\ $\rho_{\textrm{O}}^ {\textrm{2D}}=0.1121$ \#O/\AA$^{2}$\\ $p\textrm{K}_{\textrm{w}}=13.15 \pm 0.28	
$\\ $\mu-\mu_0=8.83$ kJ/mol\end{tabular}  &
\begin{minipage}{0.25\textwidth}
      \includegraphics[width=\linewidth]{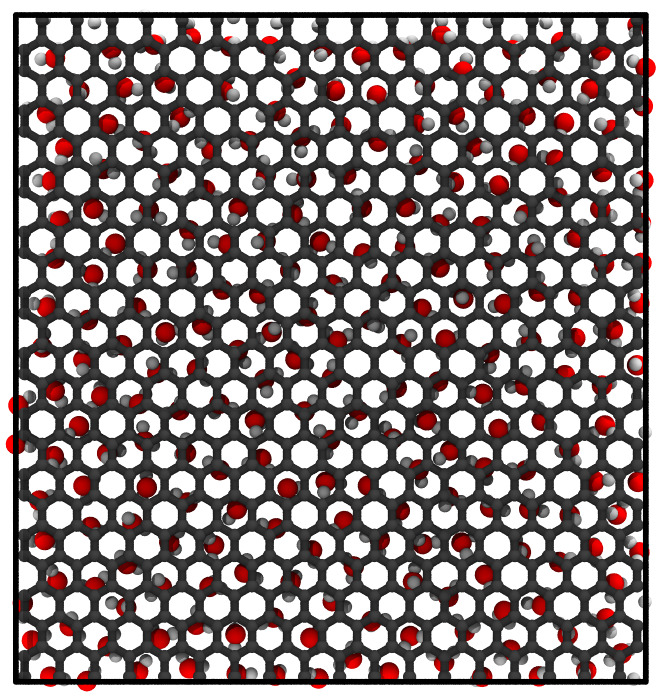}
      \includegraphics[width=\linewidth]{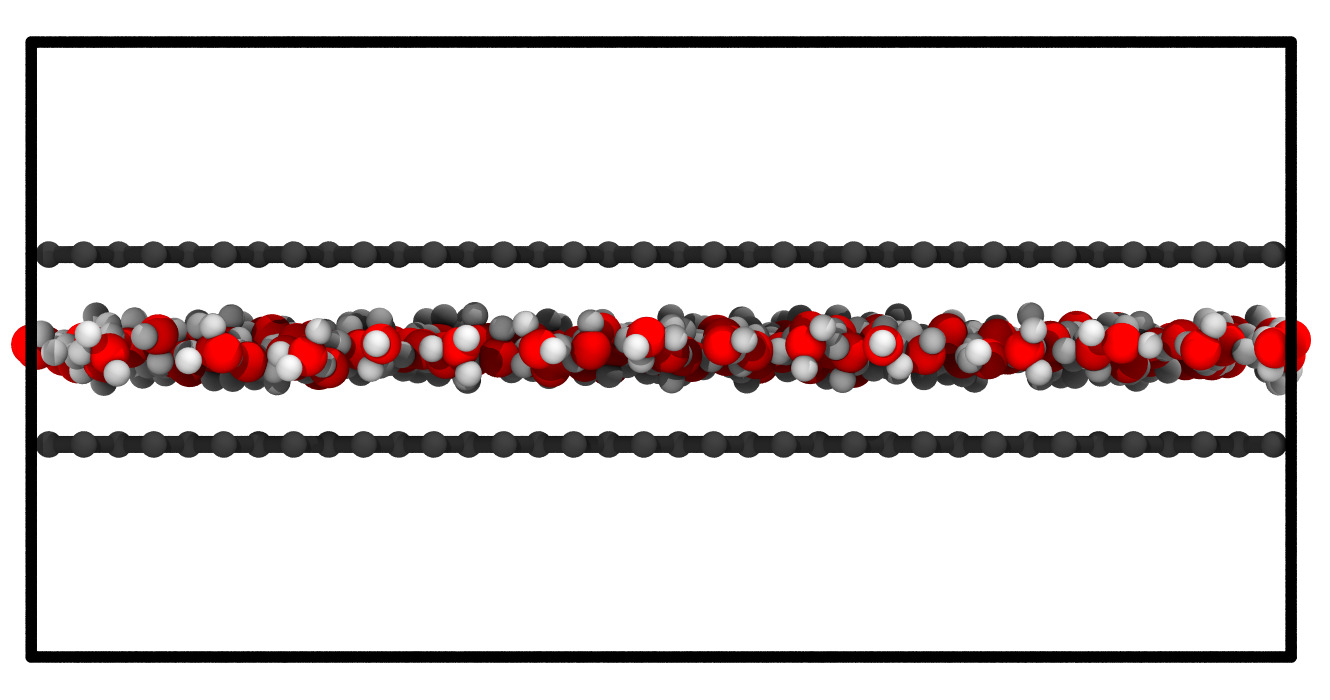}
\end{minipage}   \\* \midrule
\begin{tabular}[c]{@{}c@{}}Monolayer water confined between\\ parallel rigid GRA sheets\\ (44.460\;\AA\;$\times$\;47.058\;\AA\;$\times$\;21.700\;\AA)\end{tabular}  & 
\begin{tabular}[c]{@{}c@{}} $N_{\textrm{atoms}}=2325$\\ $N_{\textrm{C}}=1584$\\ $N_{\textrm{H}_2\textrm{O}}=247$\\ $N_{\textrm{umbrellas}}=31$\\ $t_{\textrm{eq}}=50$ ps\\ $t_{\textrm{sim/umbrella}}=100$ ps \\ $\rho_{\textrm{O}}^ {\textrm{2D}}=0.1180$ \#O/\AA$^{2}$\\ $p\textrm{K}_{\textrm{w}}=12.08 \pm 0.16	
$\\ $\mu-\mu_0=17.91$ kJ/mol\end{tabular}  &
\begin{minipage}{0.25\textwidth}
      \includegraphics[width=\linewidth]{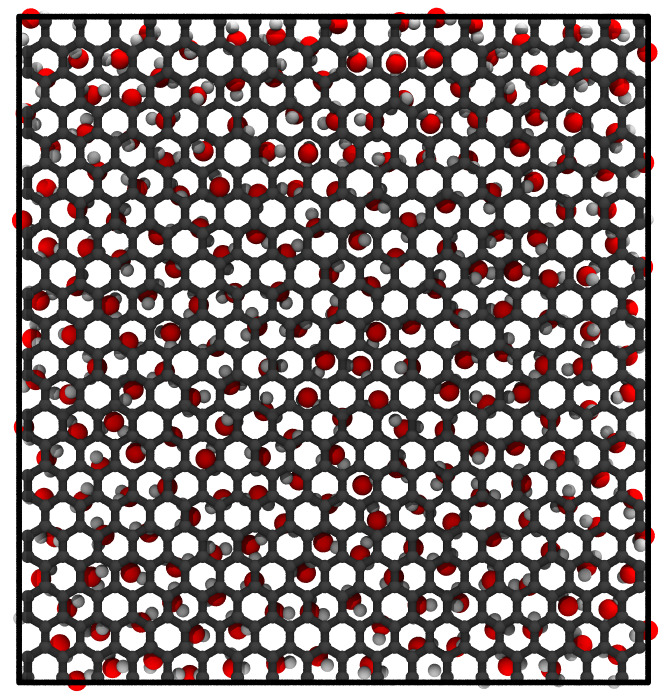}
      \includegraphics[width=\linewidth]{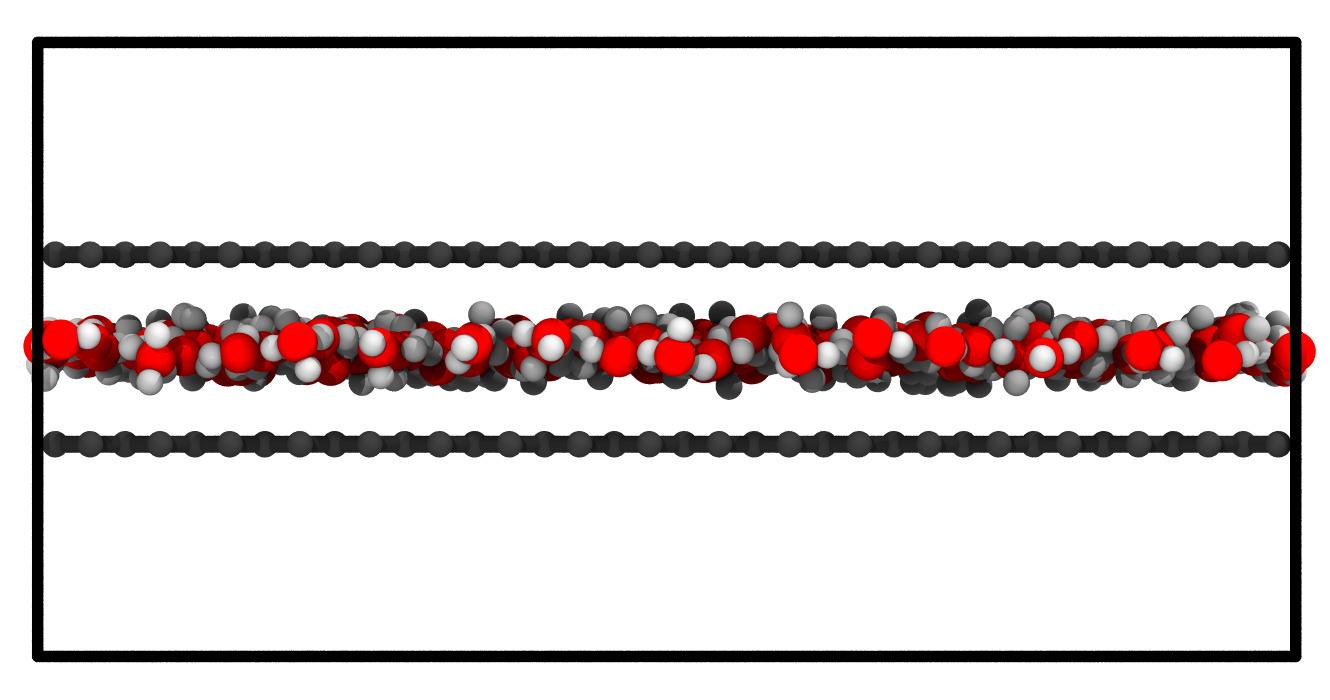}
\end{minipage}   \\* \midrule
\begin{tabular}[c]{@{}c@{}}Monolayer water confined between\\ parallel rigid GRA sheets\\  (44.460\;\AA\;$\times$\;47.058\;\AA\;$\times$\;21.700\;\AA)\end{tabular}  & 
\begin{tabular}[c]{@{}c@{}} $N_{\textrm{atoms}}=2361$\\ $N_{\textrm{C}}=1584$\\ $N_{\textrm{H}_2\textrm{O}}=259$\\ $N_{\textrm{umbrellas}}=31$\\ $t_{\textrm{eq}}=50$ ps\\ $t_{\textrm{sim/umbrella}}=100$ ps \\ $\rho_{\textrm{O}}^ {\textrm{2D}}=0.1239$ \#O/\AA$^{2}$\\ $p\textrm{K}_{\textrm{w}}=11.17 \pm 0.31	
$\\ $\mu-\mu_0=28.61$ kJ/mol\end{tabular}  & 
\begin{minipage}{0.25\textwidth}
      \includegraphics[width=\linewidth]{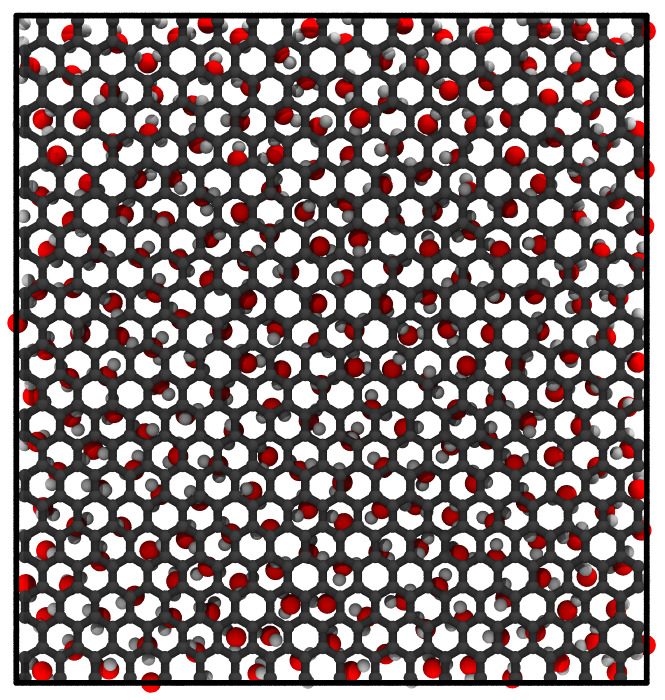}
      \includegraphics[width=\linewidth]{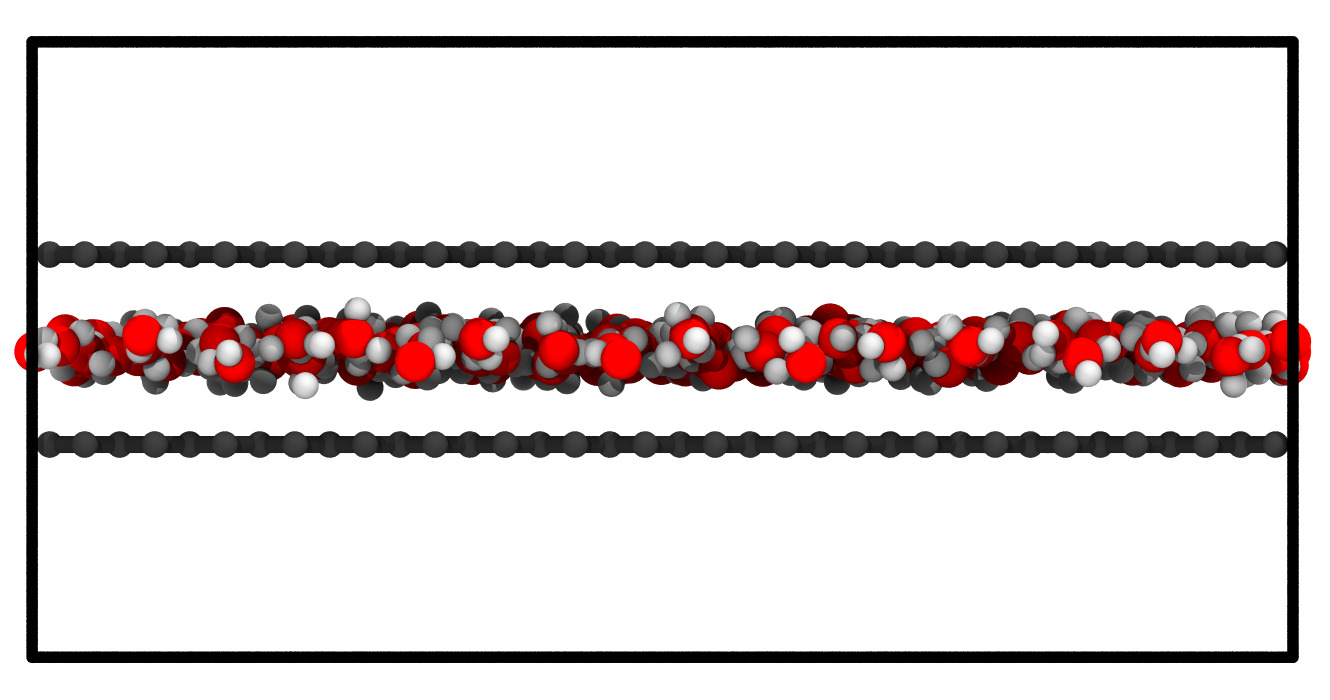}
\end{minipage}   \\* \midrule
\begin{tabular}[c]{@{}c@{}}Monolayer water confined between\\ parallel rigid GRA sheets\\  (44.460\;\AA\;$\times$\;47.058\;\AA\;$\times$\;21.700\;\AA)\end{tabular}  & 
\begin{tabular}[c]{@{}c@{}} $N_{\textrm{atoms}}=2400$\\ $N_{\textrm{C}}=1584$\\ $N_{\textrm{H}_2\textrm{O}}=272$\\ $N_{\textrm{umbrellas}}=31$\\ $t_{\textrm{eq}}=50$ ps\\ $t_{\textrm{sim/umbrella}}=100$ ps \\ $\rho_{\textrm{O}}^ {\textrm{2D}}= 0.1298$ \#O/\AA$^{2}$\\ $p\textrm{K}_{\textrm{w}}=9.54 \pm 0.11	
$\\ $\mu-\mu_0=42.03$ kJ/mol\end{tabular}  & 
\begin{minipage}{0.25\textwidth}
      \includegraphics[width=\linewidth]{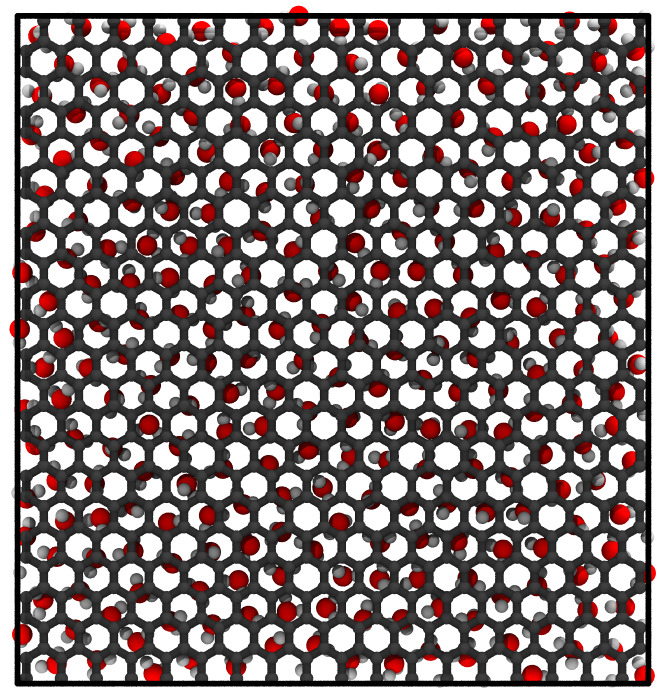}
      \includegraphics[width=\linewidth]{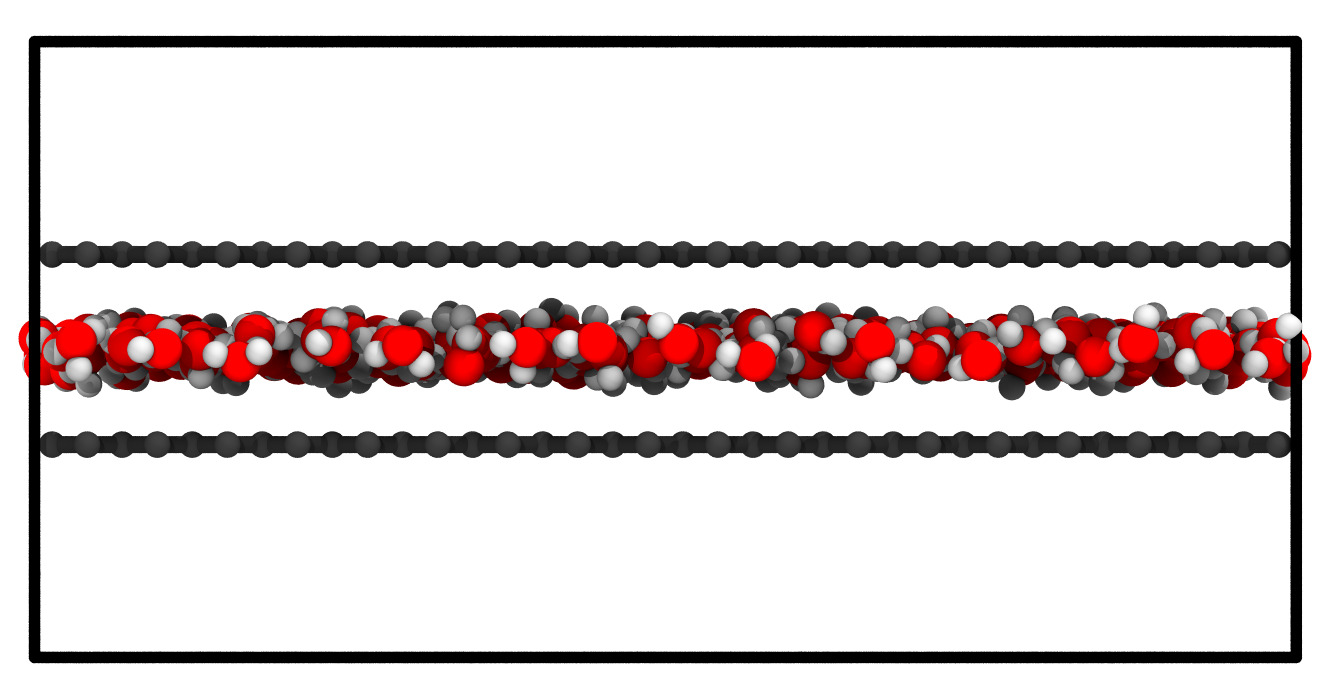}
\end{minipage}   \\* \bottomrule

\bottomrule
\multicolumn{3}{c}{
\parbox{0.95\textwidth}{
  \vspace{0.6cm}
  Table S2: Detailed overview of the monolayer confined water systems between parallel graphene layers considered in this work.
  For each system, we report the total number of atoms, $N_{\textrm{atoms}}$; the number of carbon atoms, $N_{\textrm{C}}$; the corresponding number of water molecules, $N_{\textrm{H}_2\textrm{O}}$; the number of umbrellas sampled $N_{\textrm{umbrellas}}$; the equilibration time, $t_{\textrm{eq}}$; the simulation production time per umbrella, $t_{\textrm{sim/umbrella}}$; its surface density, $\rho_{\textrm{O}}^ {\textrm{2D}}$; its $p\textrm{K}_{\textrm{w}}$ value; and the chemical potential difference relative to the reference, $\mu - \mu_0$.}}
  \label{tab:syst_gra}
\end{longtable}

\newpage

\begin{longtable}[c]{@{}c >{\centering\arraybackslash}p{6cm} >{\centering\arraybackslash}p{5cm}}
\toprule \toprule
\begin{tabular}[c]{@{}c@{}}\textbf{System}\\\textbf{(dimensions)}\end{tabular} & \textbf{Simulation details} & \textbf{Illustration} \\* \midrule
\endfirsthead
\endhead
\begin{tabular}[c]{@{}c@{}}Monolayer water confined between\\ parallel rigid GRA sheets\\ (44.460\;\AA\;$\times$\;47.058\;\AA\;$\times$\;21.700\;\AA)\end{tabular}  & 
\begin{tabular}[c]{@{}c@{}} $N_{\textrm{atoms}}=2214$\\ $N_{\textrm{C}}=1584$\\ $N_{\textrm{H}_2\textrm{O}}=210$\\ $N_{\textrm{umbrellas}}=31$\\ $t_{\textrm{eq}}=50$ ps\\ $t_{\textrm{sim/umbrella}}=100$ ps \\ $\rho_{\textrm{O}}^ {\textrm{2D}}=0.1003$ \#O/\AA$^{2}$\\ $p\textrm{K}_{\textrm{w}}=14.36 \pm 0.13	
$\end{tabular}  &
\begin{minipage}{0.25\textwidth}
      \includegraphics[width=\linewidth]{figs/supplementary/vmd_slit_210_top.jpg}
      \includegraphics[width=\linewidth]{figs/supplementary/vmd_slit_210_mid.jpg}
\end{minipage}   \\* \midrule
\begin{tabular}[c]{@{}c@{}}Bilayer water confined between\\ parallel rigid GRA sheets\\ (44.460\;\AA\;$\times$\;47.058\;\AA\;$\times$\;25.050\;\AA)\end{tabular}  & 
\begin{tabular}[c]{@{}c@{}} $N_{\textrm{atoms}}=2844$\\ $N_{\textrm{C}}=1584$\\ $N_{\textrm{H}_2\textrm{O}}=420$\\ $N_{\textrm{umbrellas}}=31$\\ $t_{\textrm{eq}}=50$ ps\\ $t_{\textrm{sim/umbrella}}=100$ ps \\ $\rho_{\textrm{O}}^ {\textrm{2D}}=0.1003$ \#O/\AA$^{2}$\\ $p\textrm{K}_{\textrm{w}}=14.18 \pm 0.33	
$\end{tabular}  &
\begin{minipage}{0.25\textwidth}
      \includegraphics[width=\linewidth]{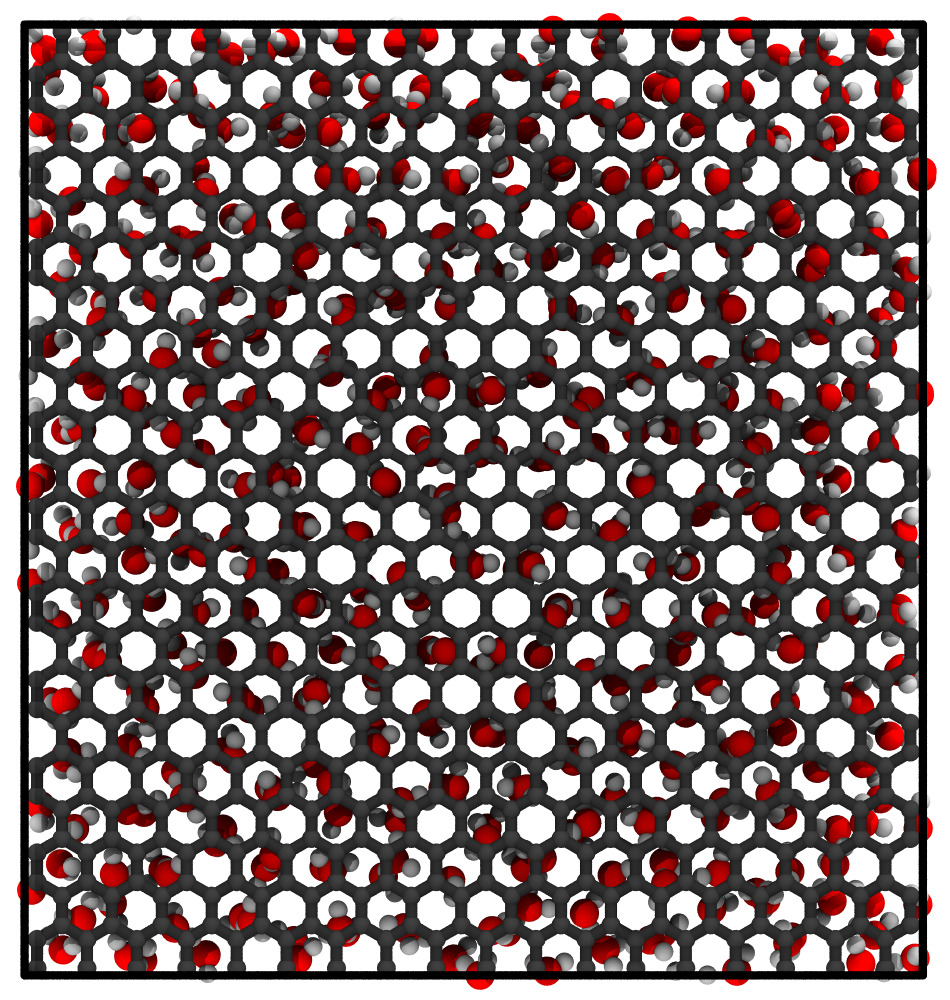}
      \includegraphics[width=\linewidth]{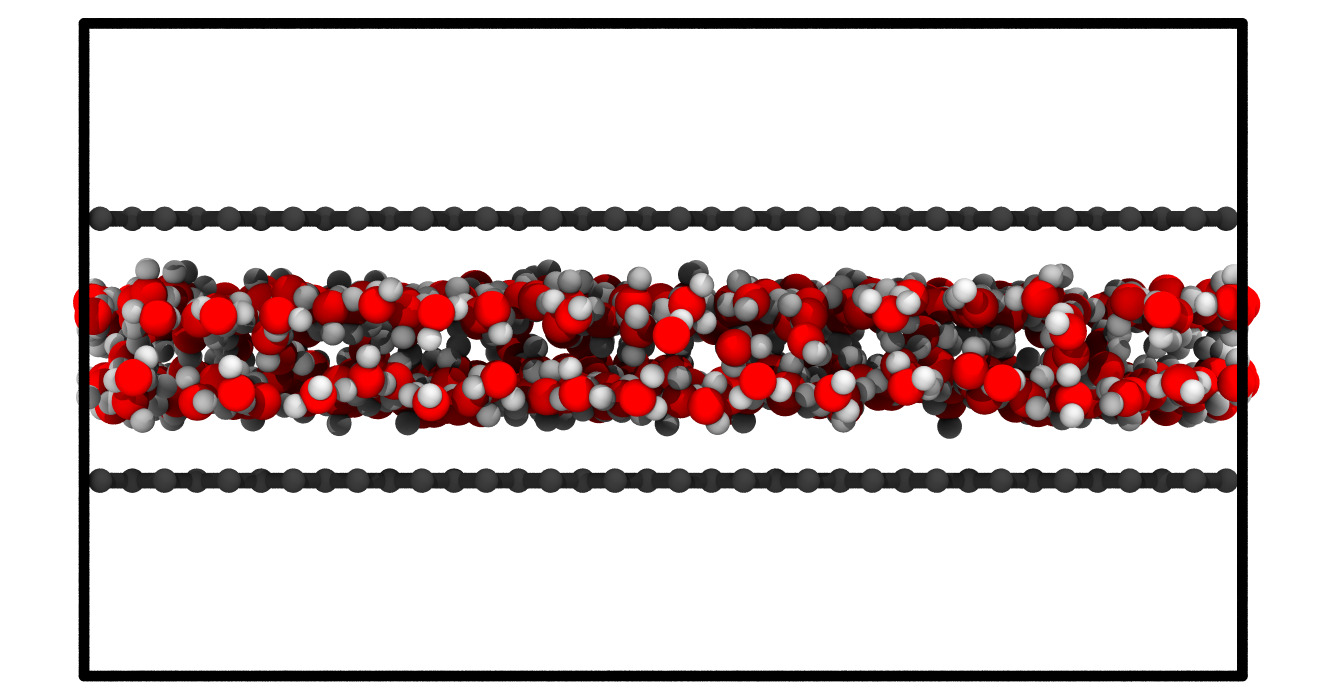}
\end{minipage}   \\* \midrule
\begin{tabular}[c]{@{}c@{}}Trilayer water confined between\\ parallel rigid GRA sheets\\  (44.460\;\AA\;$\times$\;47.058\;\AA\;$\times$\;28.400\;\AA)\end{tabular}  & 
\begin{tabular}[c]{@{}c@{}} $N_{\textrm{atoms}}=3474$\\ $N_{\textrm{C}}=1584$\\ $N_{\textrm{H}_2\textrm{O}}=630$\\ $N_{\textrm{umbrellas}}=31$\\ $t_{\textrm{eq}}=50$ ps\\ $t_{\textrm{sim/umbrella}}=100$ ps \\ $p\textrm{K}_{\textrm{w}}=14.48 \pm 0.22	
$\end{tabular}  & 
\begin{minipage}{0.25\textwidth}
      \includegraphics[width=\linewidth]{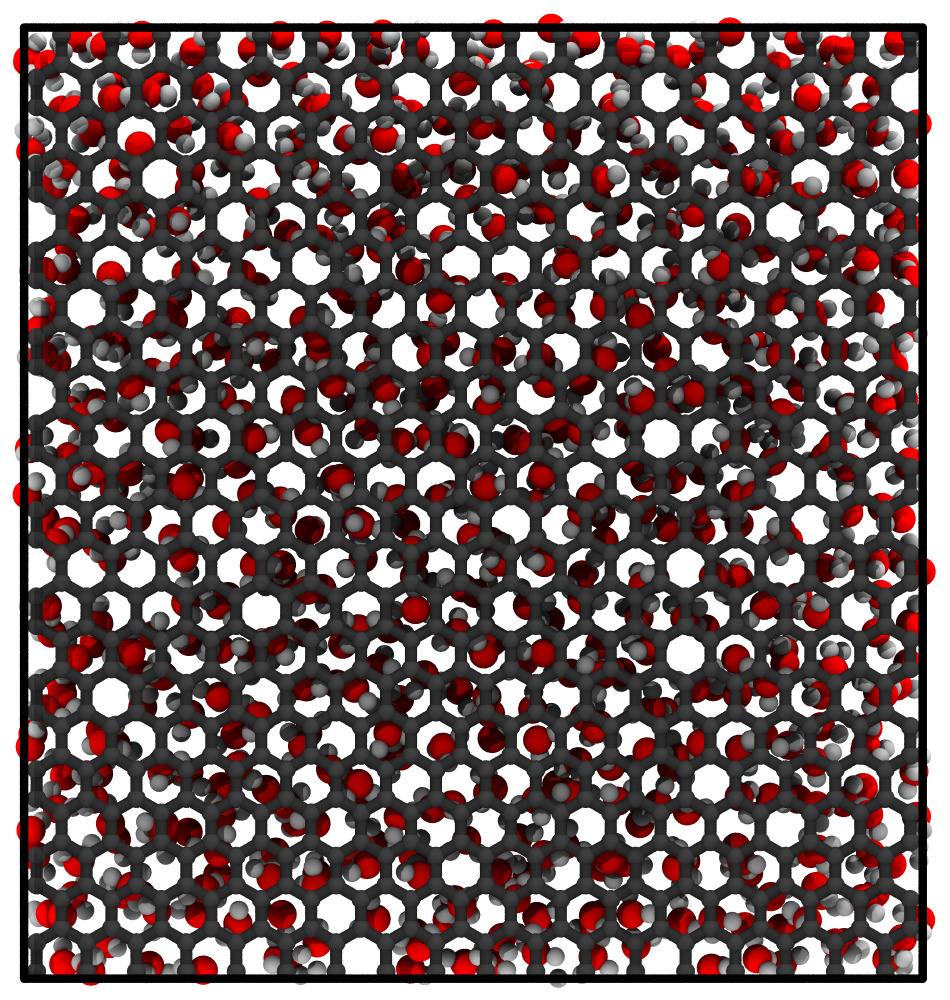}
      \includegraphics[width=\linewidth]{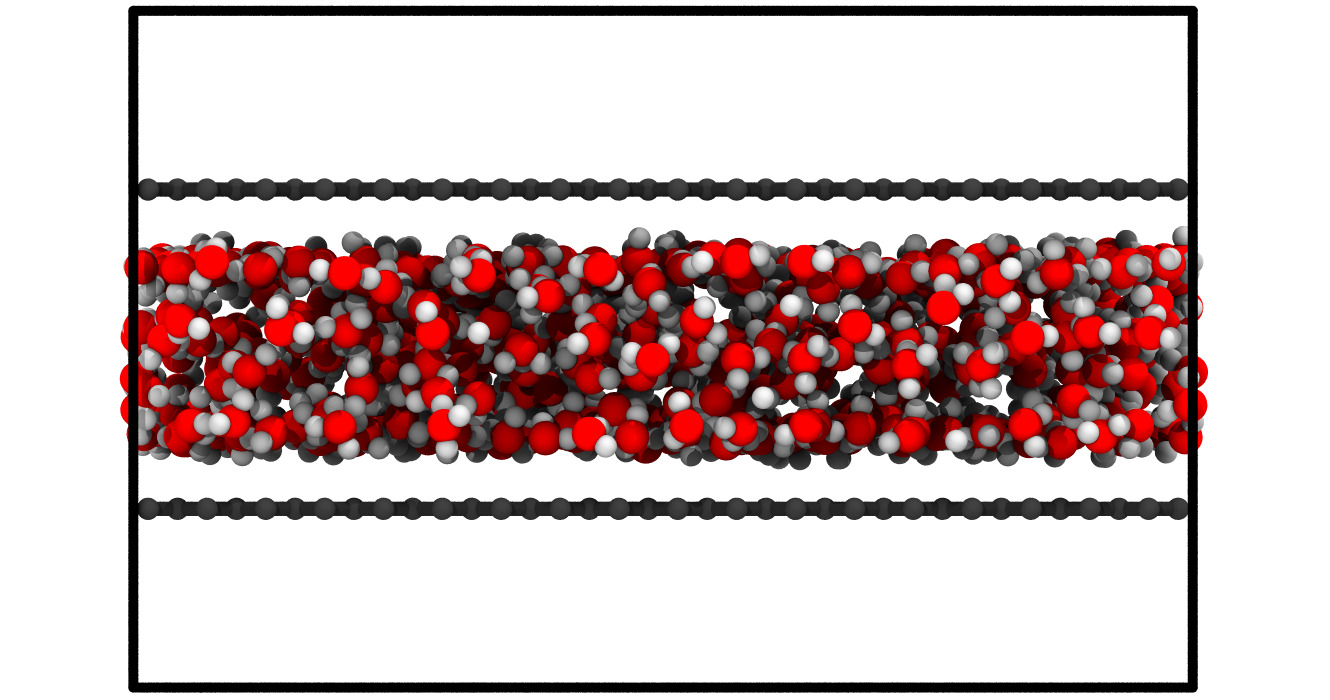}
\end{minipage}   \\* \bottomrule

\begin{tabular}[c]{@{}c@{}}Monolayer water confined between\\ parallel flexible GRA sheets\\ (44.460\;\AA\;$\times$\;47.058\;\AA\;$\times$\;21.700\;\AA)\end{tabular}  & 
\begin{tabular}[c]{@{}c@{}} $N_{\textrm{atoms}}=2214$\\ $N_{\textrm{C}}=1584$\\ $N_{\textrm{H}_2\textrm{O}}=210$\\ $N_{\textrm{umbrellas}}=31$\\ $t_{\textrm{eq}}=50$ ps\\ $t_{\textrm{sim/umbrella}}=100$ ps \\ $\rho_{\textrm{O}}^ {\textrm{2D}}=0.1003$ \#O/\AA$^{2}$\\ $p\textrm{K}_{\textrm{w}}=15.13 \pm 0.13	
$\end{tabular}  &
\begin{minipage}{0.25\textwidth}
      \includegraphics[width=\linewidth]{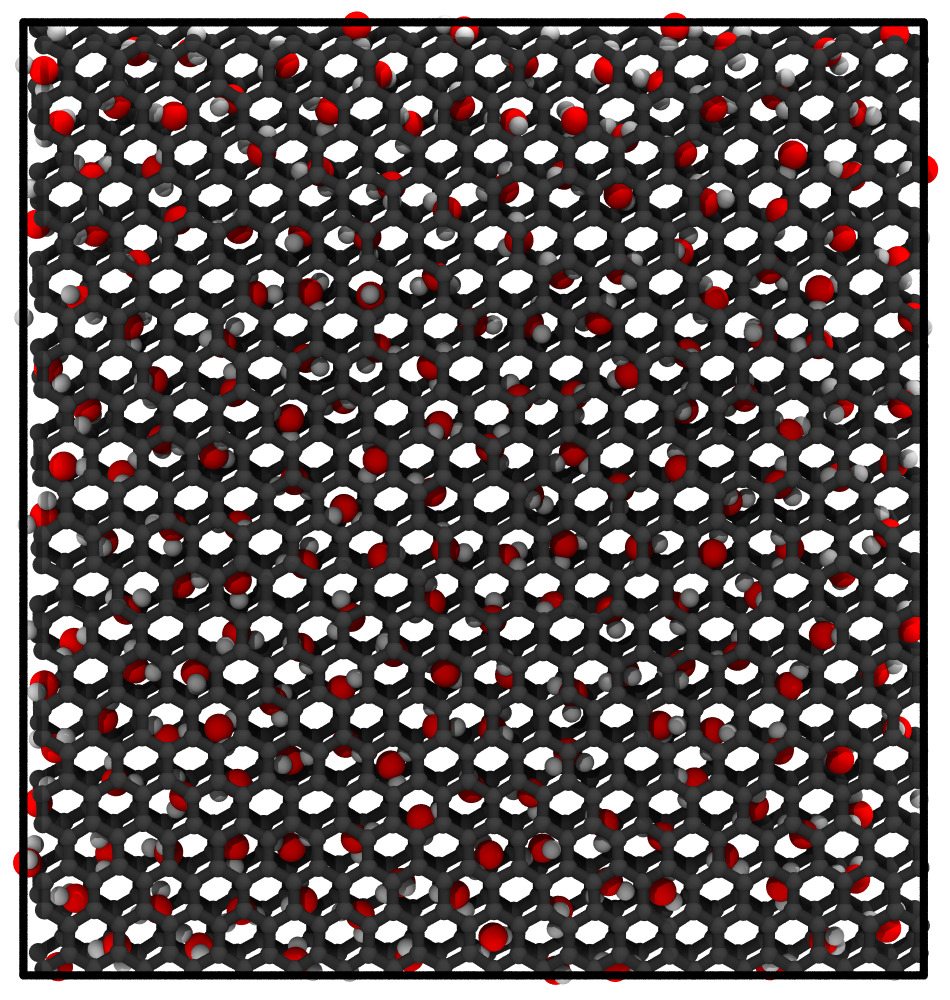}
      \includegraphics[width=\linewidth]{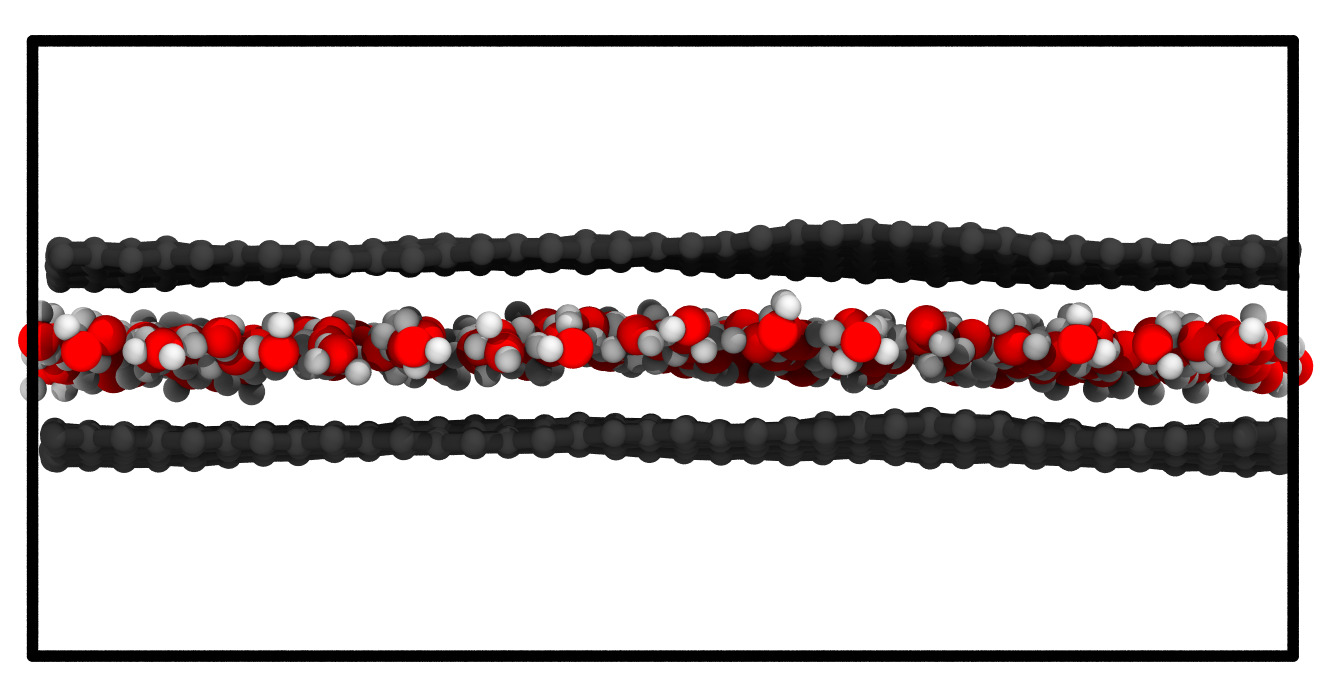}
\end{minipage}   \\* \midrule
\begin{tabular}[c]{@{}c@{}}Bilayer water confined between\\ parallel flexible GRA sheets\\ (44.460\;\AA\;$\times$\;47.058\;\AA\;$\times$\;25.050\;\AA)\end{tabular}  & 
\begin{tabular}[c]{@{}c@{}} $N_{\textrm{atoms}}=2844$\\ $N_{\textrm{C}}=1584$\\ $N_{\textrm{H}_2\textrm{O}}=420$\\ $N_{\textrm{umbrellas}}=31$\\ $t_{\textrm{eq}}=50$ ps\\ $t_{\textrm{sim/umbrella}}=100$ ps \\ $\rho_{\textrm{O}}^ {\textrm{2D}}=0.1003$ \#O/\AA$^{2}$\\ $p\textrm{K}_{\textrm{w}}=15.27 \pm 0.13	
$\end{tabular}  &
\begin{minipage}{0.25\textwidth}
      \includegraphics[width=\linewidth]{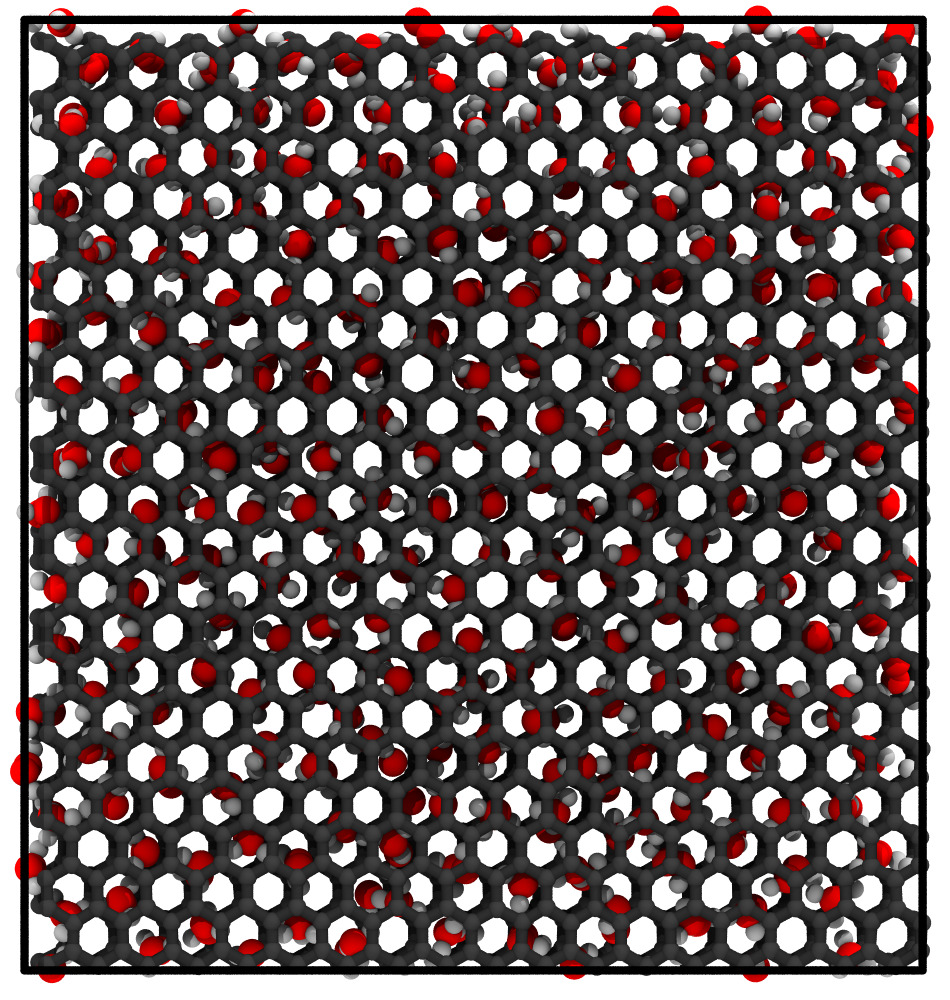}
      \includegraphics[width=\linewidth]{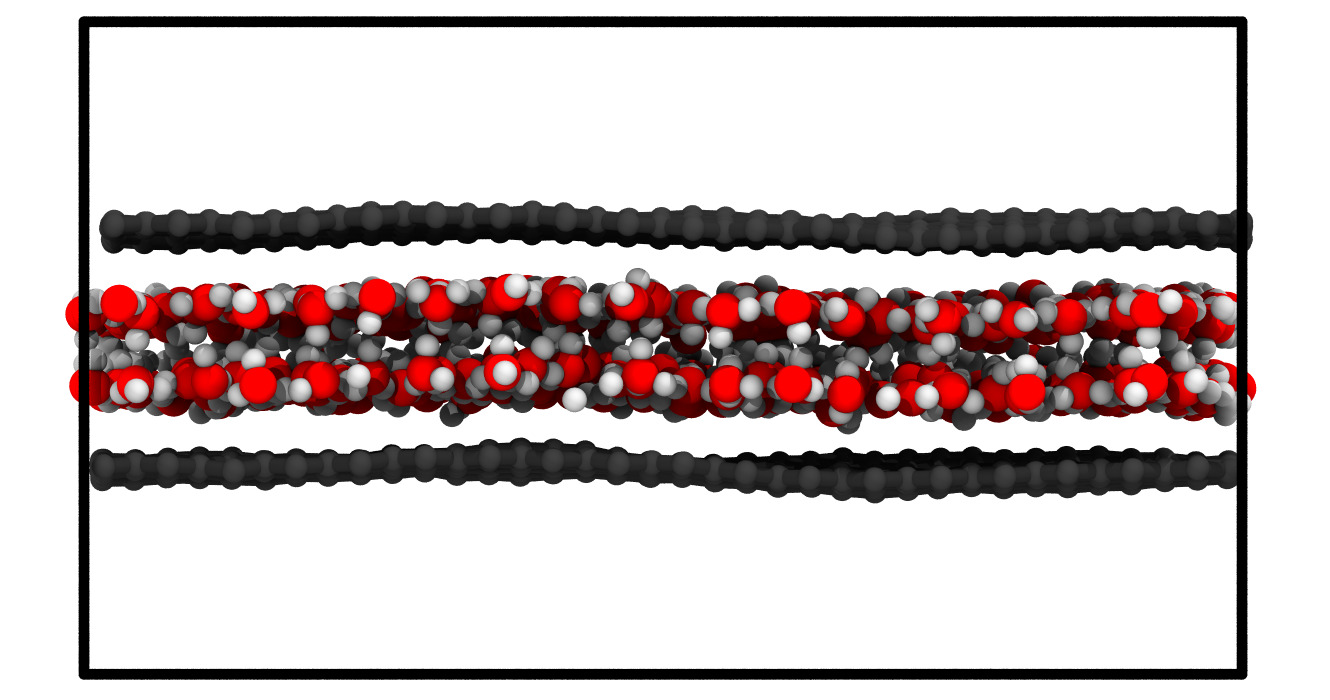}
\end{minipage}   \\* \midrule
\begin{tabular}[c]{@{}c@{}}Trilayer water confined between\\ parallel flexible GRA sheets\\  (44.460\;\AA\;$\times$\;47.058\;\AA\;$\times$\;28.400\;\AA)\end{tabular}  & 
\begin{tabular}[c]{@{}c@{}} $N_{\textrm{atoms}}=3474$\\ $N_{\textrm{C}}=1584$\\ $N_{\textrm{H}_2\textrm{O}}=630$\\ $N_{\textrm{umbrellas}}=31$\\ $t_{\textrm{eq}}=50$ ps\\ $t_{\textrm{sim/umbrella}}=100$ ps \\ $p\textrm{K}_{\textrm{w}}=14.03 \pm 0.23	
$\end{tabular}  & 
\begin{minipage}{0.25\textwidth}
      \includegraphics[width=\linewidth]{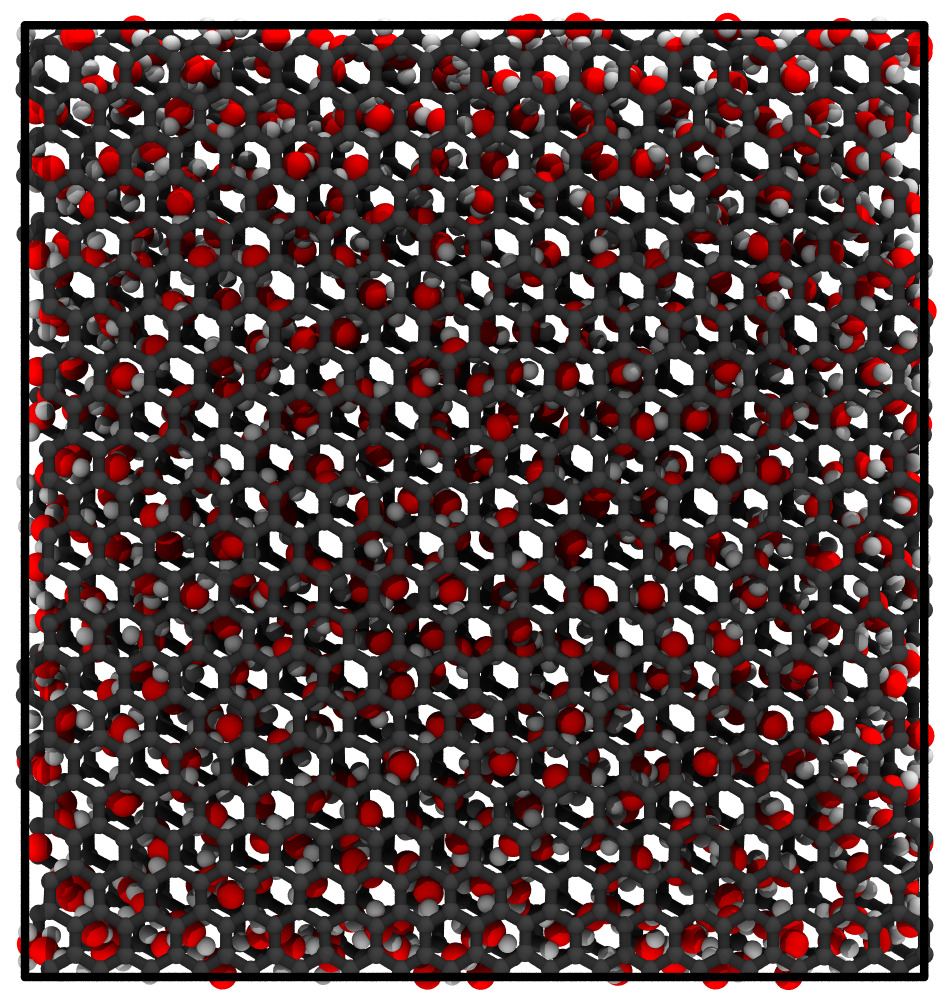}
      \includegraphics[width=\linewidth]{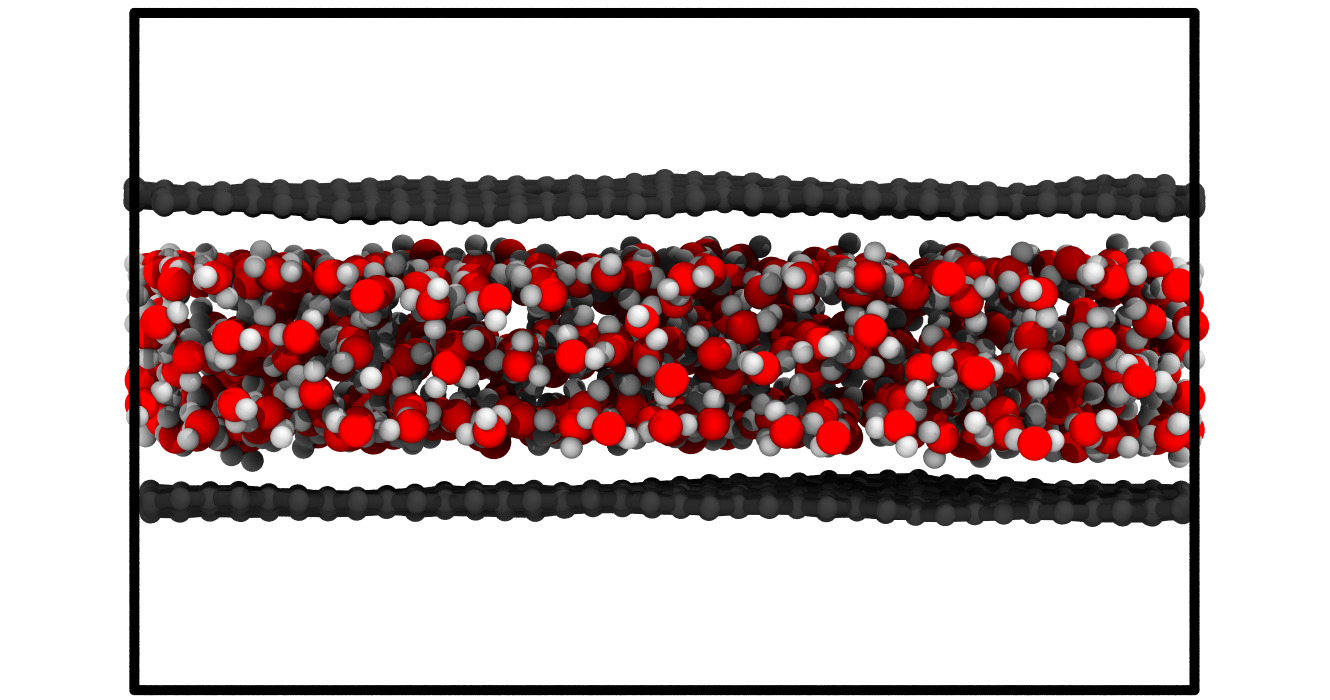}
\end{minipage}   \\* \bottomrule

\bottomrule
\multicolumn{3}{c}{
\parbox{0.95\textwidth}{
  \vspace{0.6cm}
  Table S3: Detailed overview of the systems reported in Figure 1c of the manuscript.
  For each system, we report the total number of atoms, $N_{\textrm{atoms}}$; the number of carbon atoms, $N_{\textrm{C}}$; the corresponding number of water molecules, $N_{\textrm{H}_2\textrm{O}}$; the number of umbrellas sampled $N_{\textrm{umbrellas}}$; the equilibration time, $t_{\textrm{eq}}$; the simulation production time per umbrella, $t_{\textrm{sim/umbrella}}$; its surface density, $\rho_{\textrm{O}}^ {\textrm{2D}}$; and its $p\textrm{K}_{\textrm{w}}$ value.
  As specified in the main text, the rigid 1L, 2L, and 3L setups correspond to slit widths of 6.70, 10.05, and 13.40 \AA, respectively. These values are commensurate with pore sizes that can be experimentally realized using vdW assembly \cite{radha_b_2016, kara_pairing_2024}. 
  In the bilayer setup, the number of water molecules was set to twice that of the monolayer case, based on the observation that two well-defined layers of water form at this width.
    A similar assumption was applied to the trilayer setup.
It is important to note, however, that these configurations are used for illustrative purposes.
As emphasized throughout the manuscript, rigorous comparisons require thermodynamic consistency across these systems.
For example, direct comparisons between rigid and flexible pores are complicated by the fact that flexible systems, even when exhibiting a similar average interlayer spacing, can accommodate a wider range of densities and layering motifs\cite{teresa_flexible_2018}.
  }}
  \label{tab:syst_gra}
\end{longtable}

\newpage

\begin{longtable}[c]{@{}c >{\centering\arraybackslash}p{6cm} >{\centering\arraybackslash}p{5cm}}
\toprule \toprule
\begin{tabular}[c]{@{}c@{}}\textbf{System}\\\textbf{(dimensions)}\end{tabular} & \textbf{Simulation details} & \textbf{Illustration} \\* \midrule
\endfirsthead
\endhead
\begin{tabular}[c]{@{}c@{}}Monolayer water confined between\\ parallel rigid hBN sheets\\ (43.490\;\AA\;$\times$\;45.198\;\AA\;$\times$\;21.700\;\AA)\end{tabular}  & 
\begin{tabular}[c]{@{}c@{}} $N_{\textrm{atoms}}=2070$\\ $N_{\textrm{B/N}}=1440$\\ $N_{\textrm{H}_2\textrm{O}}=210$\\ $N_{\textrm{umbrellas}}=31$\\ $t_{\textrm{eq}}=50$ ps\\ $t_{\textrm{sim/umbrella}}=100$ ps \\ $\rho_{\textrm{O}}^ {\textrm{2D}}=0.1003$ \#O/\AA$^{2}$\\ $p\textrm{K}_{\textrm{w}}=13.52 \pm 0.43	
$\end{tabular}  &
\begin{minipage}{0.25\textwidth}
      \includegraphics[width=\linewidth]{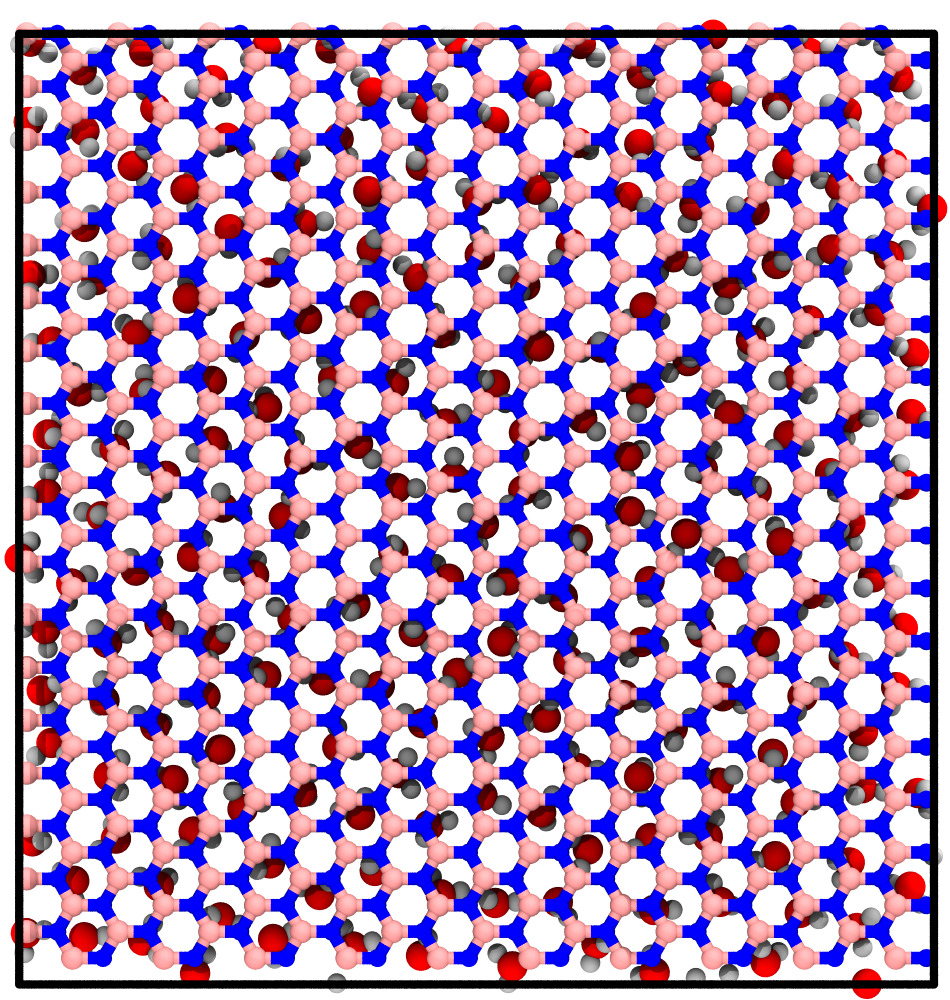}
      \includegraphics[width=\linewidth]{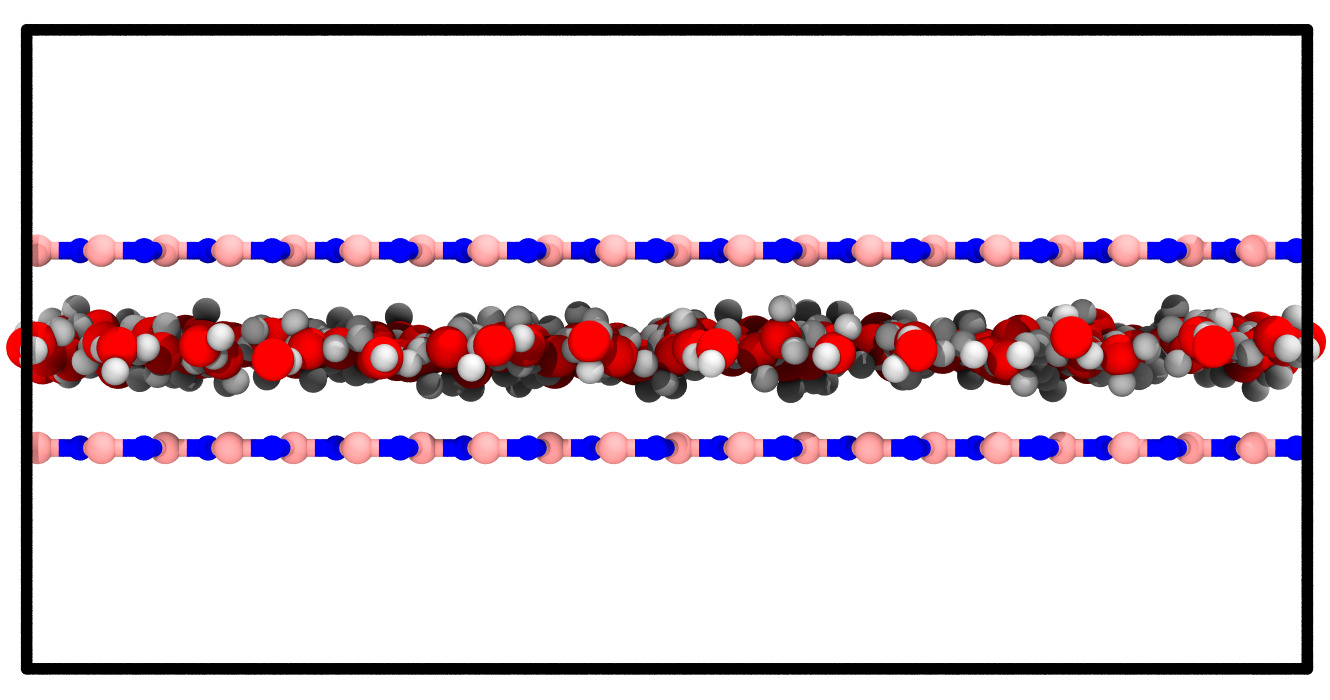}
\end{minipage}   \\* \midrule
\begin{tabular}[c]{@{}c@{}}Monolayer water confined between\\ parallel rigid hBN sheets\\  (43.490\;\AA\;$\times$\;45.198\;\AA\;$\times$\;21.700\;\AA)\end{tabular}  & 
\begin{tabular}[c]{@{}c@{}} $N_{\textrm{atoms}}=2106$\\ $N_{\textrm{B/N}}=1440$\\ $N_{\textrm{H}_2\textrm{O}}=222$\\ $N_{\textrm{umbrellas}}=31$\\ $t_{\textrm{eq}}=50$ ps\\ $t_{\textrm{sim/umbrella}}=100$ ps \\ $\rho_{\textrm{O}}^ {\textrm{2D}}=0.1062$ \#O/\AA$^{2}$\\ $p\textrm{K}_{\textrm{w}}=12.59 \pm 0.15	
$\end{tabular}  &
\begin{minipage}{0.25\textwidth}
      \includegraphics[width=\linewidth]{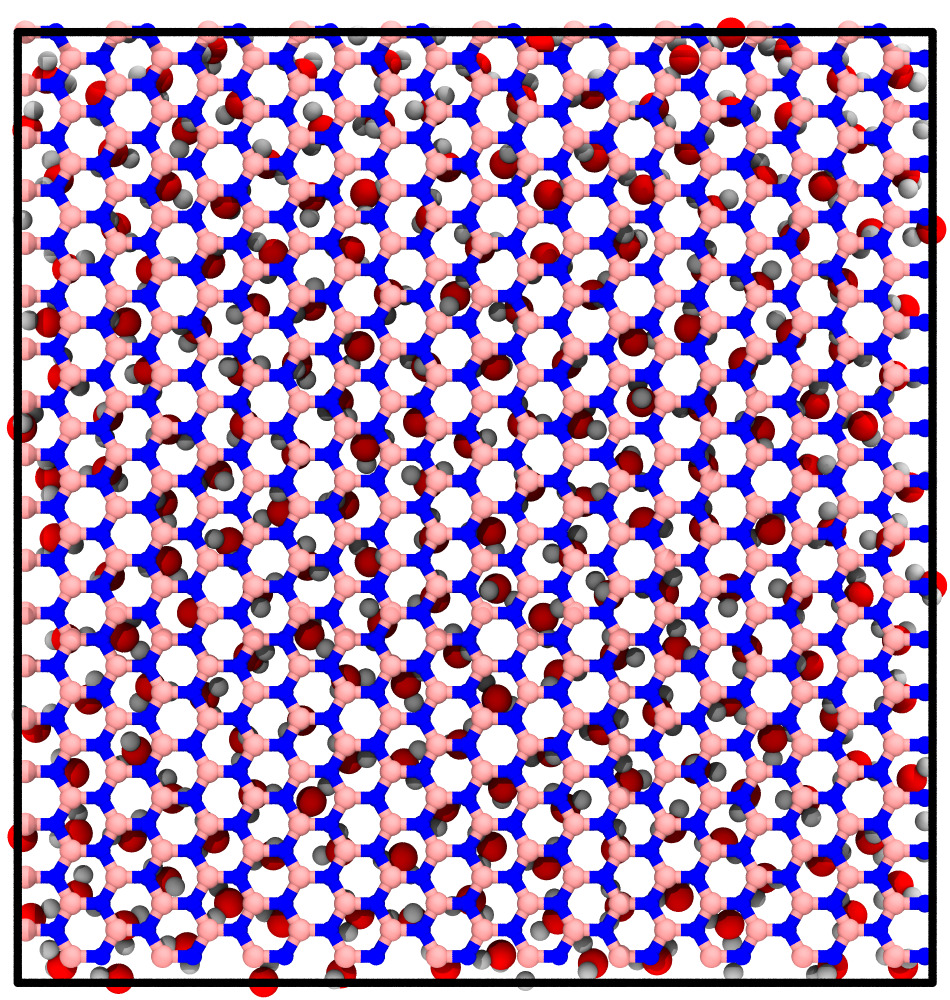}
      \includegraphics[width=\linewidth]{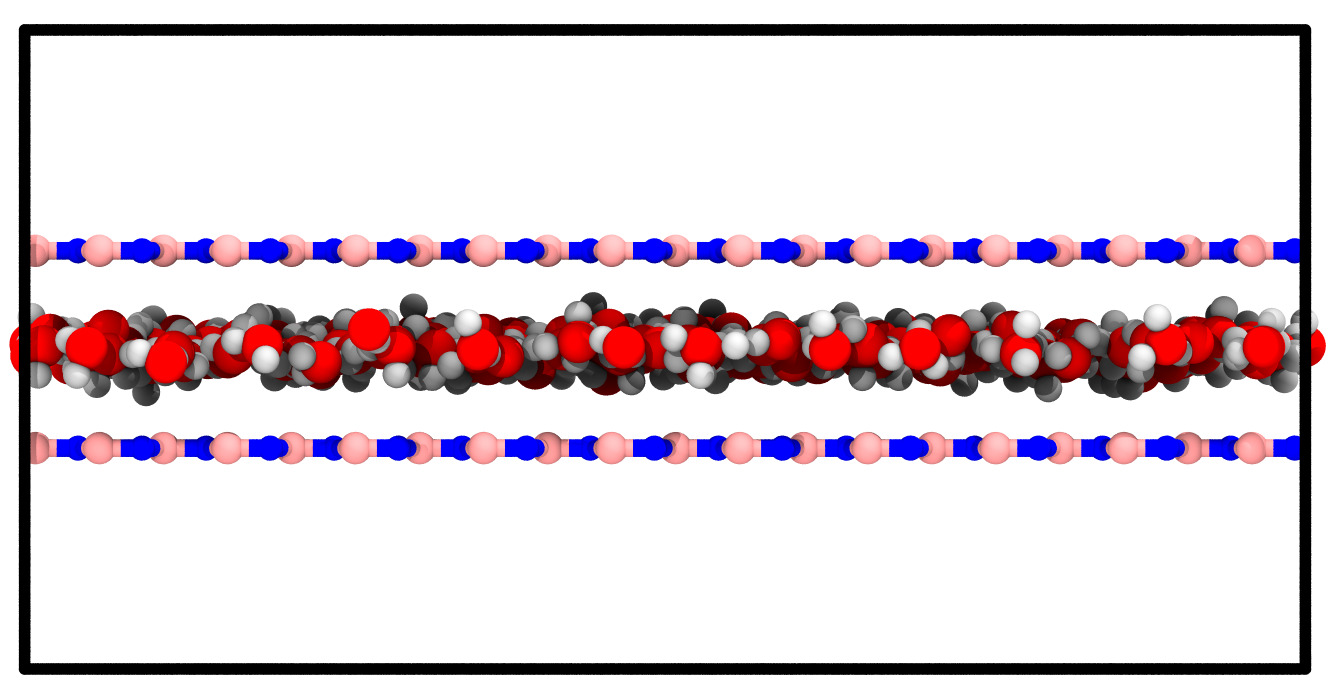}
\end{minipage}   \\* \midrule
\begin{tabular}[c]{@{}c@{}}Monolayer water confined between\\ parallel rigid hBN sheets\\  (43.490\;\AA\;$\times$\;45.198\;\AA\;$\times$\;21.700\;\AA)\end{tabular}  & 
\begin{tabular}[c]{@{}c@{}} $N_{\textrm{atoms}}=2145$\\ $N_{\textrm{B/N}}=1440$\\  $N_{\textrm{H}_2\textrm{O}}=235$\\ $N_{\textrm{umbrellas}}=31$\\ $t_{\textrm{eq}}=50$ ps\\ $t_{\textrm{sim/umbrella}}=100$ ps \\ $\rho_{\textrm{O}}^ {\textrm{2D}}=0.1121$ \#O/\AA$^{2}$\\ $p\textrm{K}_{\textrm{w}}=12.48 \pm 0.23	
$\end{tabular}  &
\begin{minipage}{0.25\textwidth}
      \includegraphics[width=\linewidth]{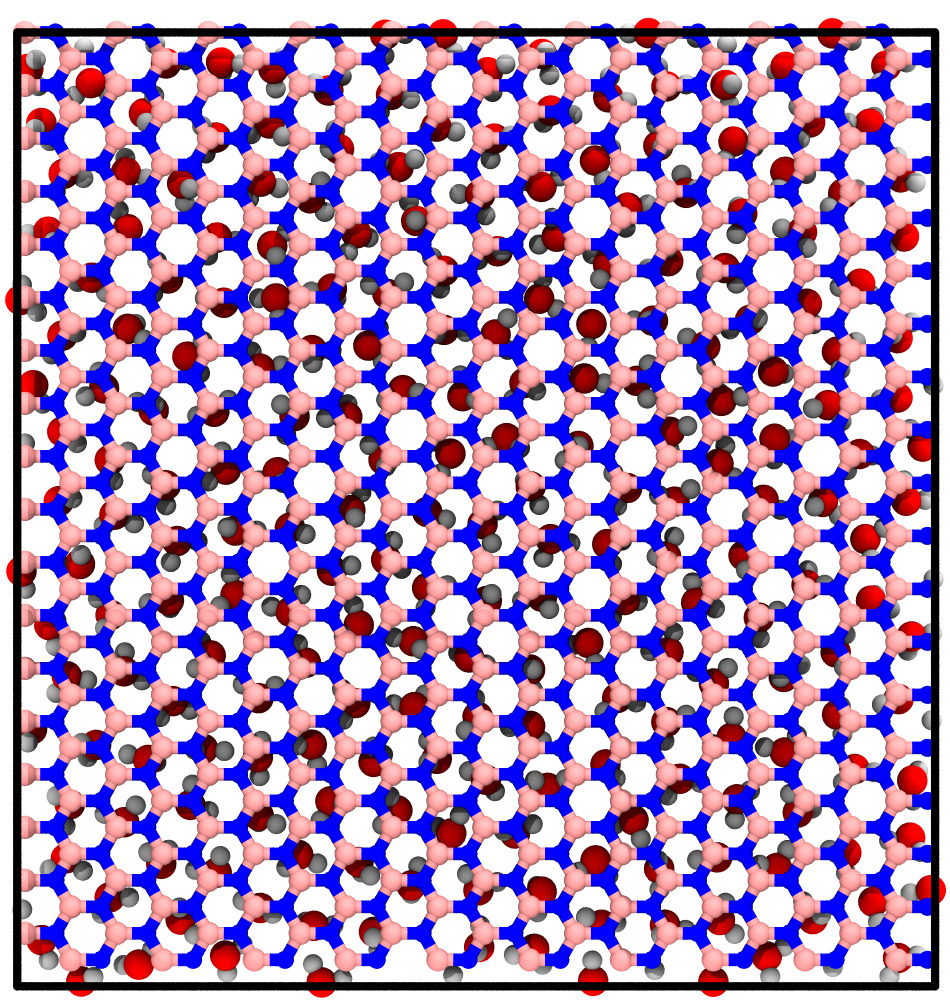}
      \includegraphics[width=\linewidth]{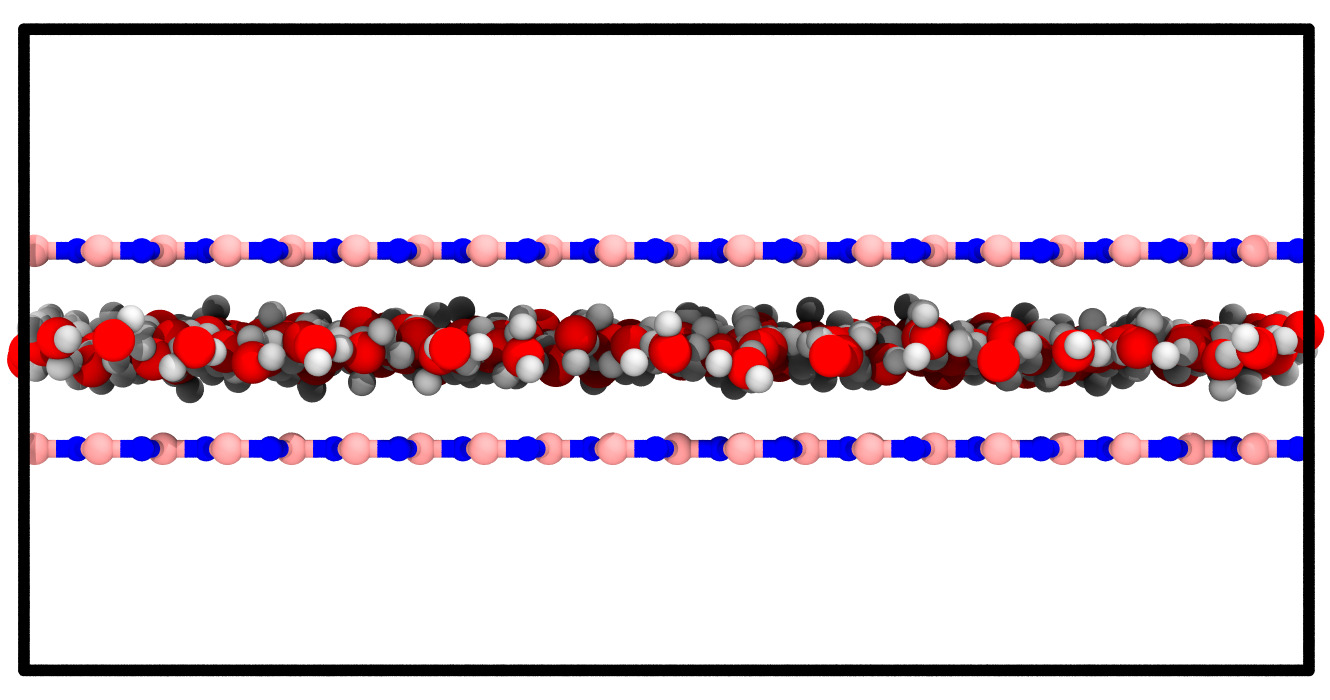}
\end{minipage}   \\* \midrule
\begin{tabular}[c]{@{}c@{}}Monolayer water confined between\\ parallel rigid hBN sheets\\  (43.490\;\AA\;$\times$\;45.198\;\AA\;$\times$\;21.700\;\AA)\end{tabular}  & 
\begin{tabular}[c]{@{}c@{}} $N_{\textrm{atoms}}=2181$\\ $N_{\textrm{B/N}}=1440$\\  $N_{\textrm{H}_2\textrm{O}}=247$\\ $N_{\textrm{umbrellas}}=31$\\ $t_{\textrm{eq}}=50$ ps\\ $t_{\textrm{sim/umbrella}}=100$ ps \\ $\rho_{\textrm{O}}^ {\textrm{2D}}=0.1180$ \#O/\AA$^{2}$\\ $p\textrm{K}_{\textrm{w}}=10.13 \pm 0.13	
$\end{tabular}  &
\begin{minipage}{0.25\textwidth}
      \includegraphics[width=\linewidth]{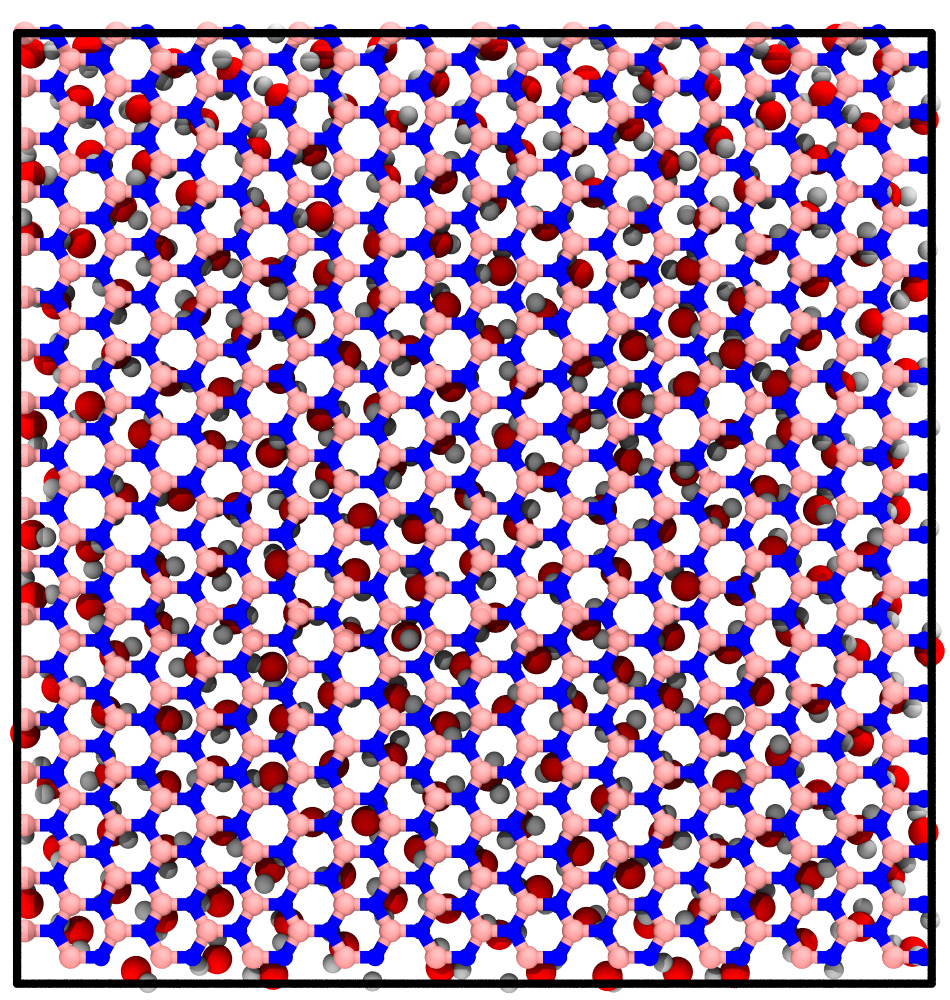}
      \includegraphics[width=\linewidth]{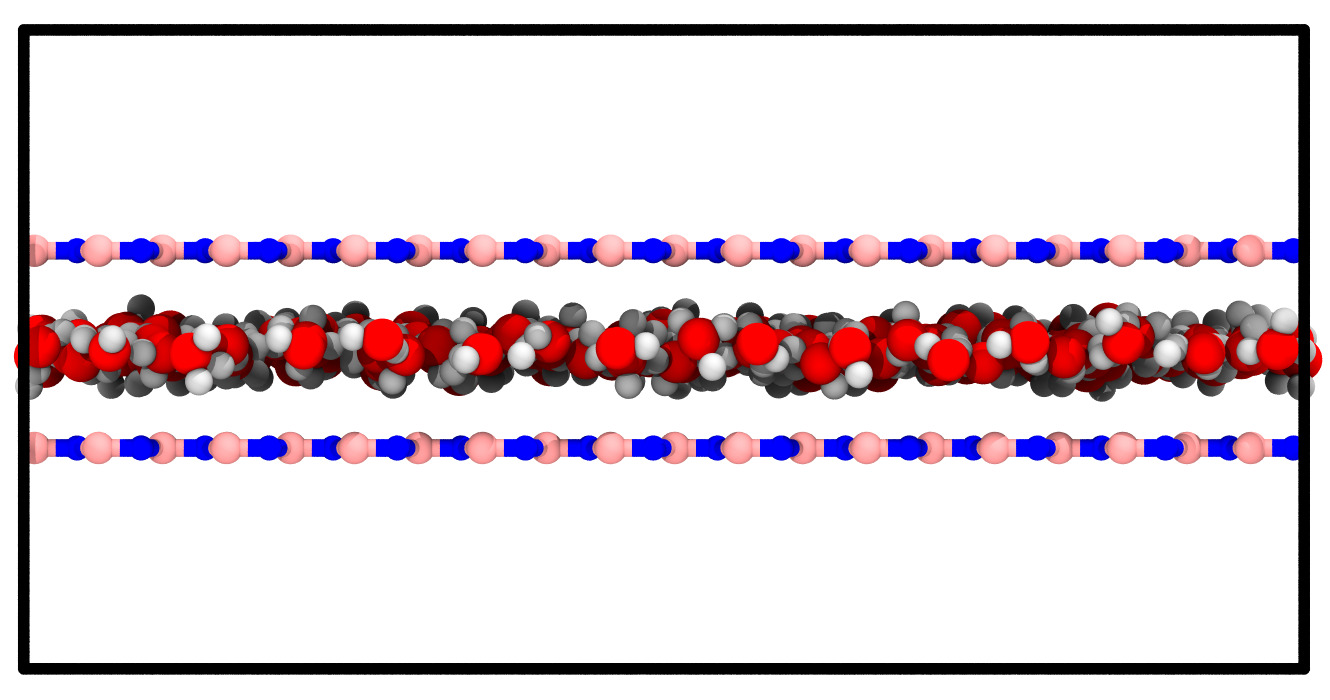}
\end{minipage}   \\* \midrule
\begin{tabular}[c]{@{}c@{}}Monolayer water confined between\\ parallel rigid hBN sheets\\ (43.490\;\AA\;$\times$\;45.198\;\AA\;$\times$\;21.700\;\AA)\end{tabular}  & 
\begin{tabular}[c]{@{}c@{}} $N_{\textrm{atoms}}=2217$\\ $N_{\textrm{B/N}}=1440$\\  $N_{\textrm{H}_2\textrm{O}}=259$\\ $N_{\textrm{umbrellas}}=31$\\ $t_{\textrm{eq}}=50$ ps\\ $t_{\textrm{sim/umbrella}}=100$ ps \\ $\rho_{\textrm{O}}^ {\textrm{2D}}=0.1239$ \#O/\AA$^{2}$\\ $p\textrm{K}_{\textrm{w}}=10.29 \pm 0.09	
$\end{tabular}  & 
\begin{minipage}{0.25\textwidth}
      \includegraphics[width=\linewidth]{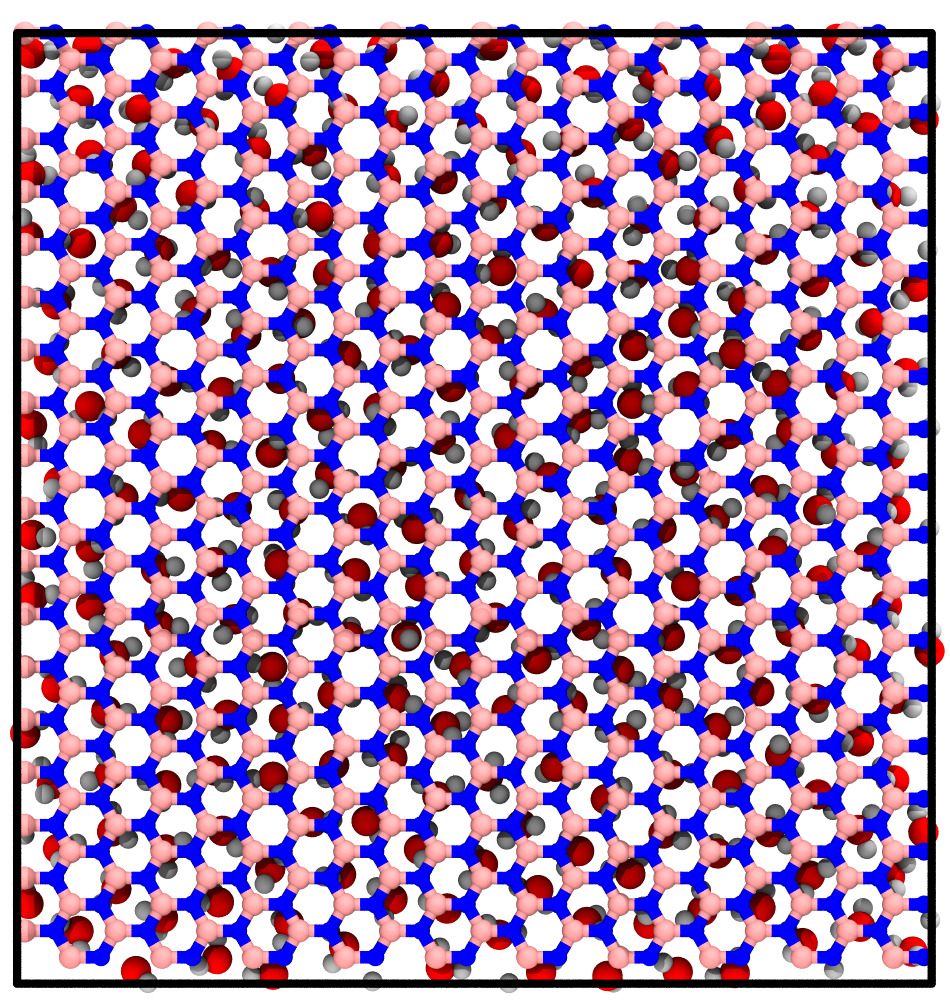}
      \includegraphics[width=\linewidth]{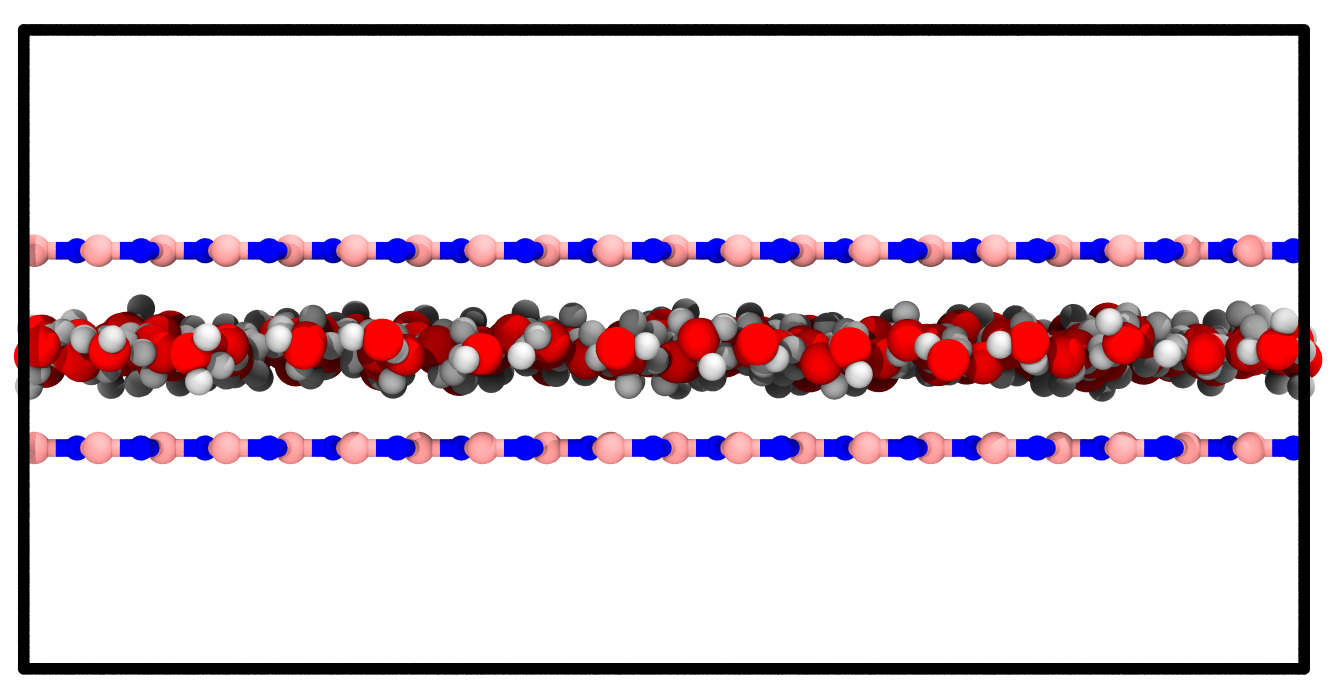}
\end{minipage}   \\* \midrule
\begin{tabular}[c]{@{}c@{}}Monolayer water confined between\\ parallel rigid hBN sheets\\ (43.490\;\AA\;$\times$\;45.198\;\AA\;$\times$\;21.700\;\AA)\end{tabular}  & 
\begin{tabular}[c]{@{}c@{}} $N_{\textrm{atoms}}=2256$\\ $N_{\textrm{B/N}}=1440$\\  $N_{\textrm{H}_2\textrm{O}}=272$\\ $N_{\textrm{umbrellas}}=31$\\ $t_{\textrm{eq}}=50$ ps\\ $t_{\textrm{sim/umbrella}}=100$ ps \\ $\rho_{\textrm{O}}^ {\textrm{2D}}= 0.1298$ \#O/\AA$^{2}$\\ $p\textrm{K}_{\textrm{w}}=9.02 \pm 0.15	
$\end{tabular}  & 
\begin{minipage}{0.25\textwidth}
      \includegraphics[width=\linewidth]{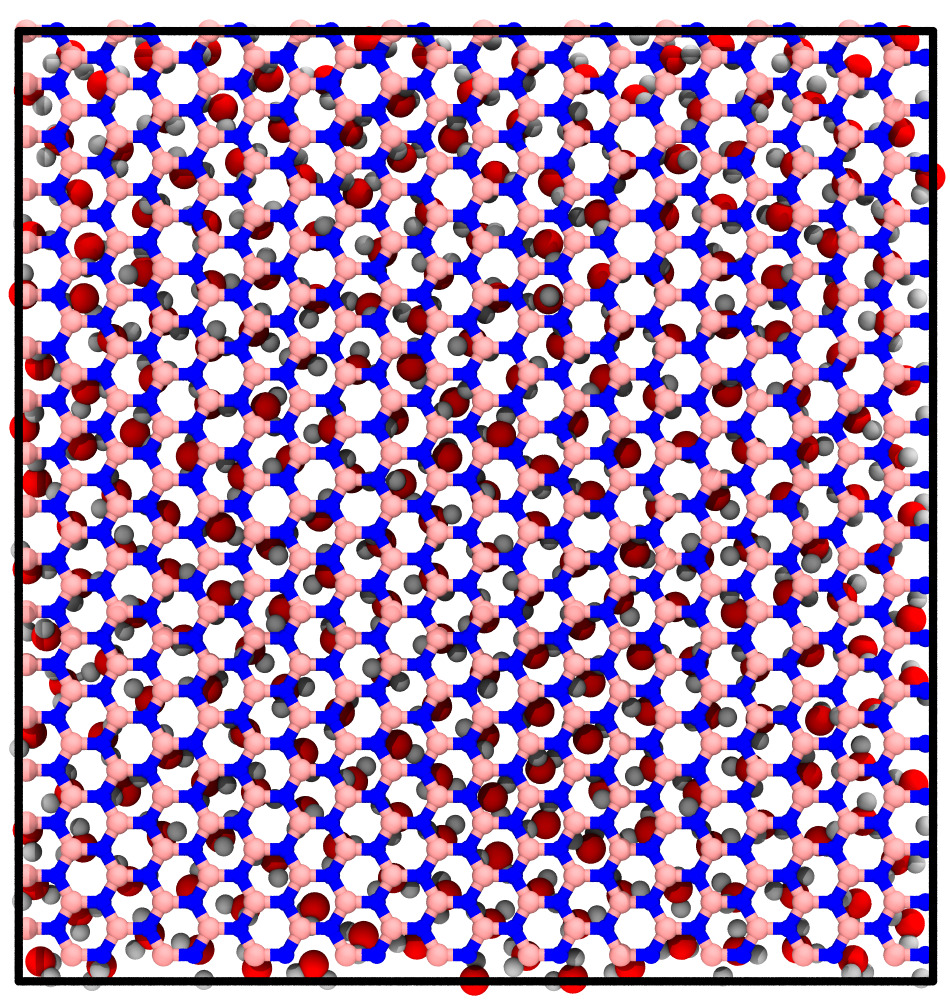}
      \includegraphics[width=\linewidth]{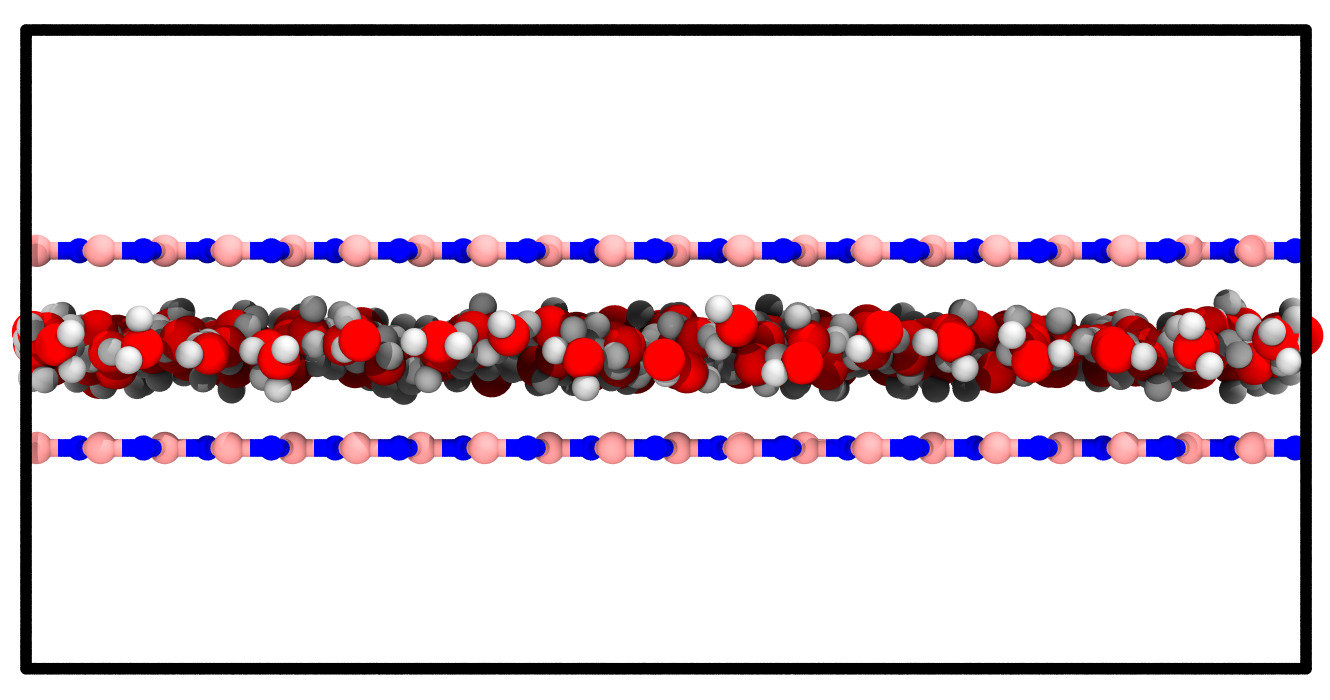}
\end{minipage}   \\* \bottomrule

\bottomrule
\multicolumn{3}{c}{
\parbox{0.95\textwidth}{
  \vspace{0.6cm}
  Table S4: Detailed overview of the monolayer confined water systems between parallel hexagonal boron nitride layers considered in this work.
  For each system, we report the total number of atoms, $N_{\textrm{atoms}}$; the number of boron and nitrogen atoms, $N_{\textrm{B/N}}$; the corresponding number of water molecules, $N_{\textrm{H}_2\textrm{O}}$; the number of umbrellas sampled $N_{\textrm{umbrellas}}$; the equilibration time, $t_{\textrm{eq}}$; the simulation production time per umbrella, $t_{\textrm{sim/umbrella}}$; its surface density, $\rho_{\textrm{O}}^ {\textrm{2D}}$; and its $p\textrm{K}_{\textrm{w}}$ value.}}
  \label{tab:syst_gra}
\end{longtable}

\newpage

\begin{longtable}[c]{@{}c >{\centering\arraybackslash}p{6cm} >{\centering\arraybackslash}p{5cm}}
\toprule \toprule
\label{tab:syst_drop}
\begin{tabular}[c]{@{}c@{}}\textbf{System}\\\textbf{(dimensions)}\end{tabular} & \textbf{Simulation details} & \textbf{Illustration} \\* \midrule
\endfirsthead
\endhead
\begin{tabular}[c]{@{}c@{}}GRA nanodroplet confined water\\ (79.040\;\AA\;$\times$\;81.282\;\AA\;$\times$\;40.000\;\AA)\end{tabular}  & 
\begin{tabular}[c]{@{}c@{}} $N_{\textrm{atoms}}=5260$\\ $N_{\textrm{C}}=4864$\\ $N_{\textrm{H}_2\textrm{O}}=132$\\ $N_{\textrm{umbrellas}}=31$\\ $t_{\textrm{eq}}=50$ ps\\ $t_{\textrm{sim/umbrella}}=100$ ps \\ $\rho_{\textrm{O}}^{\textrm{2D}}=0.1180\pm 0.0010$ \#O/\AA$^{2}$\end{tabular}  & 
\begin{minipage}{0.25\textwidth}
      \includegraphics[width=\linewidth]{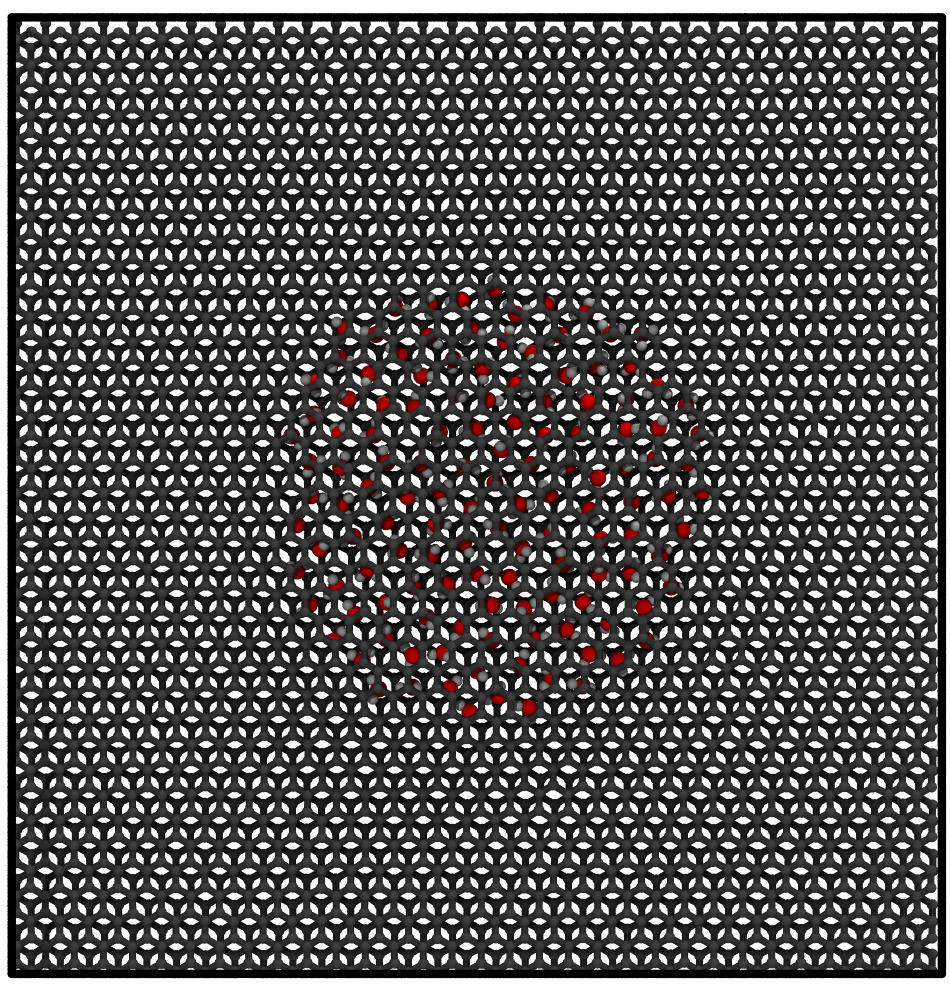}
      \includegraphics[width=\linewidth]{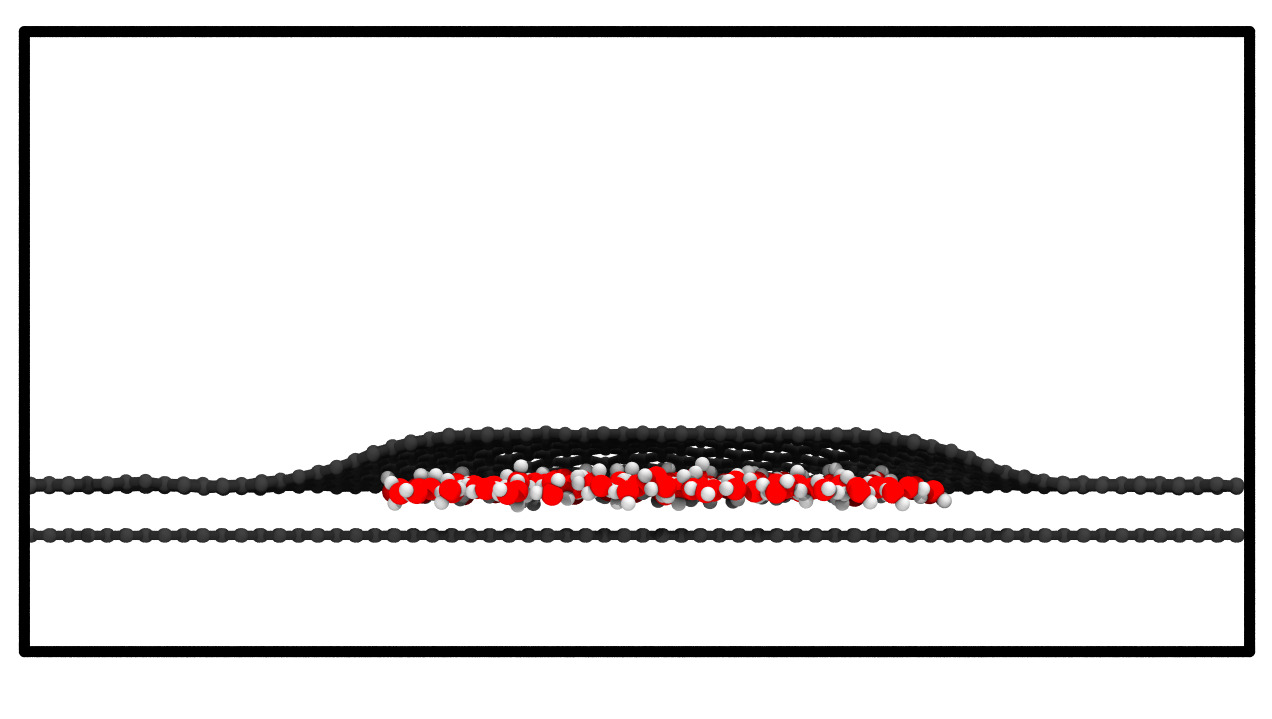}
\end{minipage}   \\* \bottomrule

\bottomrule
\multicolumn{3}{c}{
\parbox{0.95\textwidth}{
  \vspace{0.6cm}
  Table S5: Detailed overview of the GRA nanodroplet confined water system considered in this work.
  For the system, we report the total number of atoms, $N_{\textrm{atoms}}$; the number of carbon atoms, $N_{\textrm{C}}$; the corresponding number of water molecules, $N_{\textrm{H}_2\textrm{O}}$; the number of umbrellas sampled $N_{\textrm{umbrellas}}$; the equilibration time, $t_{\textrm{eq}}$; the simulation production time per umbrella, $t_{\textrm{sim/umbrella}}$; and its radial density (see Section \ref{sec:nanodrop_equi}), $\rho_{\textrm{O}}^{\textrm{2D}}$.}}
\end{longtable}

\newpage
\begin{longtable}[c]{@{}c >{\centering\arraybackslash}p{6cm} >{\centering\arraybackslash}p{5cm}}
\toprule \toprule
\label{tab:syst_drop}
\begin{tabular}[c]{@{}c@{}}\textbf{System}\\\textbf{(dimensions)}\end{tabular} & \textbf{Simulation details} & \textbf{Illustration} \\* \midrule
\endfirsthead
\endhead
\begin{tabular}[c]{@{}c@{}}hBN nanodroplet confined water\\ 
(78.282\;\AA\;$\times$\;80.352\;\AA\;$\times$\;40.000\;\AA)\end{tabular}  & 
\begin{tabular}[c]{@{}c@{}} $N_{\textrm{atoms}}=5004$\\ $N_{\textrm{B/N}}=4608$\\ $N_{\textrm{H}_2\textrm{O}}=132$\\ $N_{\textrm{umbrellas}}=31$\\ $t_{\textrm{eq}}=50$ ps\\ $t_{\textrm{sim/umbrella}}=100$ ps \\ $\rho_{\textrm{O}}^{\textrm{2D}}=0.1170 \pm 0.0010$ \#O/\AA$^{2}$\end{tabular}  & 
\begin{minipage}{0.25\textwidth}
      \includegraphics[width=\linewidth]{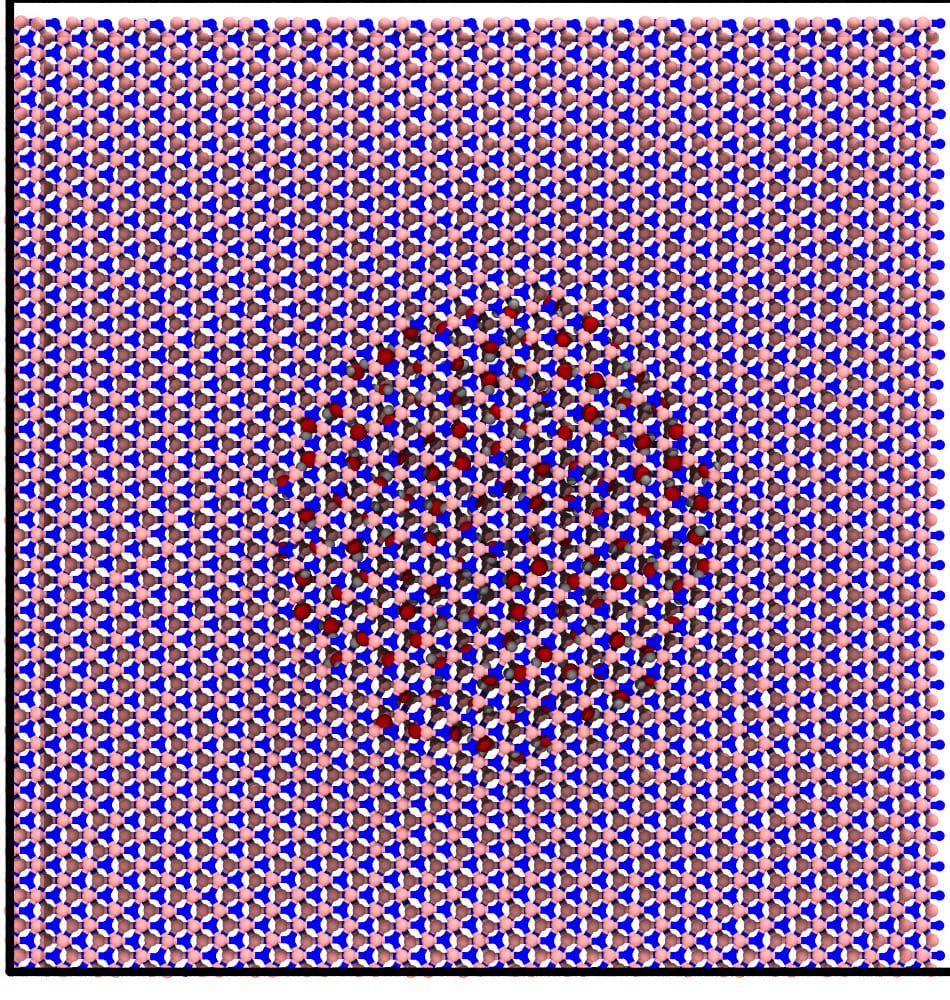}
      \includegraphics[width=\linewidth]{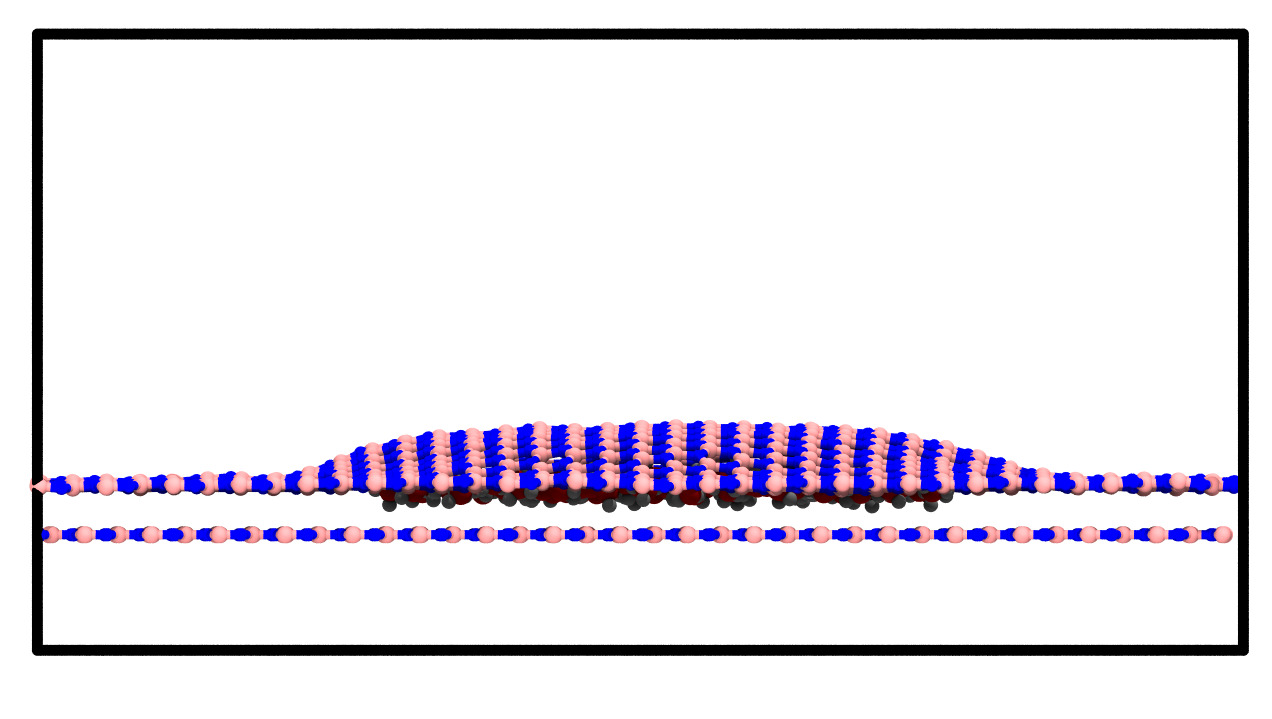}
\end{minipage}   \\* \bottomrule

\bottomrule
\multicolumn{3}{c}{
\parbox{0.95\textwidth}{
  \vspace{0.6cm}
  Table S6: Detailed overview of the hBN nanodroplet confined water system considered in this work.
  For the system, we report the total number of atoms, $N_{\textrm{atoms}}$; he number of boron and nitrogen atoms, $N_{\textrm{B/N}}$; the corresponding number of water molecules, $N_{\textrm{H}_2\textrm{O}}$; the number of umbrellas sampled $N_{\textrm{umbrellas}}$; the equilibration time, $t_{\textrm{eq}}$; the simulation production time per umbrella, $t_{\textrm{sim/umbrella}}$; and its radial density (see Section \ref{sec:nanodrop_equi}), $\rho_{\textrm{O}}^{\textrm{2D}}$.}}
\end{longtable}

\newpage
\begin{longtable}[c]{@{}c >{\centering\arraybackslash}p{6cm} >{\centering\arraybackslash}p{5cm}}
\toprule \toprule
\label{tab:syst_liq_reserv}
\begin{tabular}[c]{@{}c@{}}\textbf{System}\\\textbf{(avg. dimensions, since NPT)}\end{tabular} & \textbf{Simulation details} & \textbf{Illustration} \\* \midrule
\endfirsthead
\endhead
\begin{tabular}[c]{@{}c@{}}Parallel rigid GRA layers\\immersed in liquid water\\  (76.771\;\AA\;$\times$\;47.058\;\AA\;$\times$\;36.000\;\AA)\end{tabular} & 
\begin{tabular}[c]{@{}c@{}} $N_{\textrm{atoms}}=13827$\\ $N_{\textrm{C}}=1584$\\ $a=44.460$\;\AA\\ $b=47.058$\;\AA \\ $N_{\textrm{H}_2\textrm{O}}=4081$\\ $W=6.7$\;\AA\\$t_{\textrm{prod}}=300$ ps\\$t_{\textrm{eq}}=150$  ps\\ $\rho_{\textrm{O}}^{\textrm{2D}}=0.1097$ \#O/\AA$^{2}$   \end{tabular}  & 
\begin{minipage}{0.25\textwidth}
      \includegraphics[width=\linewidth]{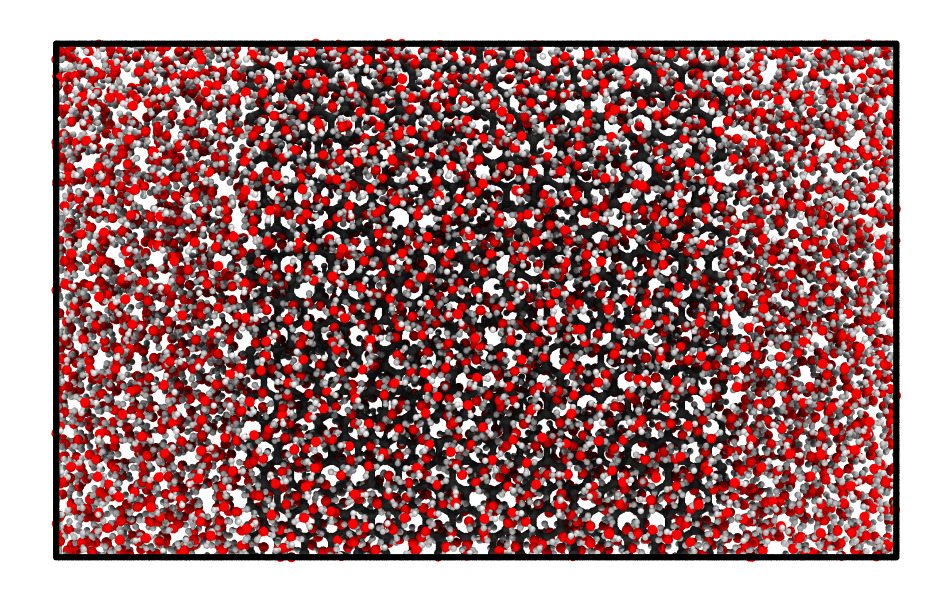}
      \includegraphics[width=\linewidth]{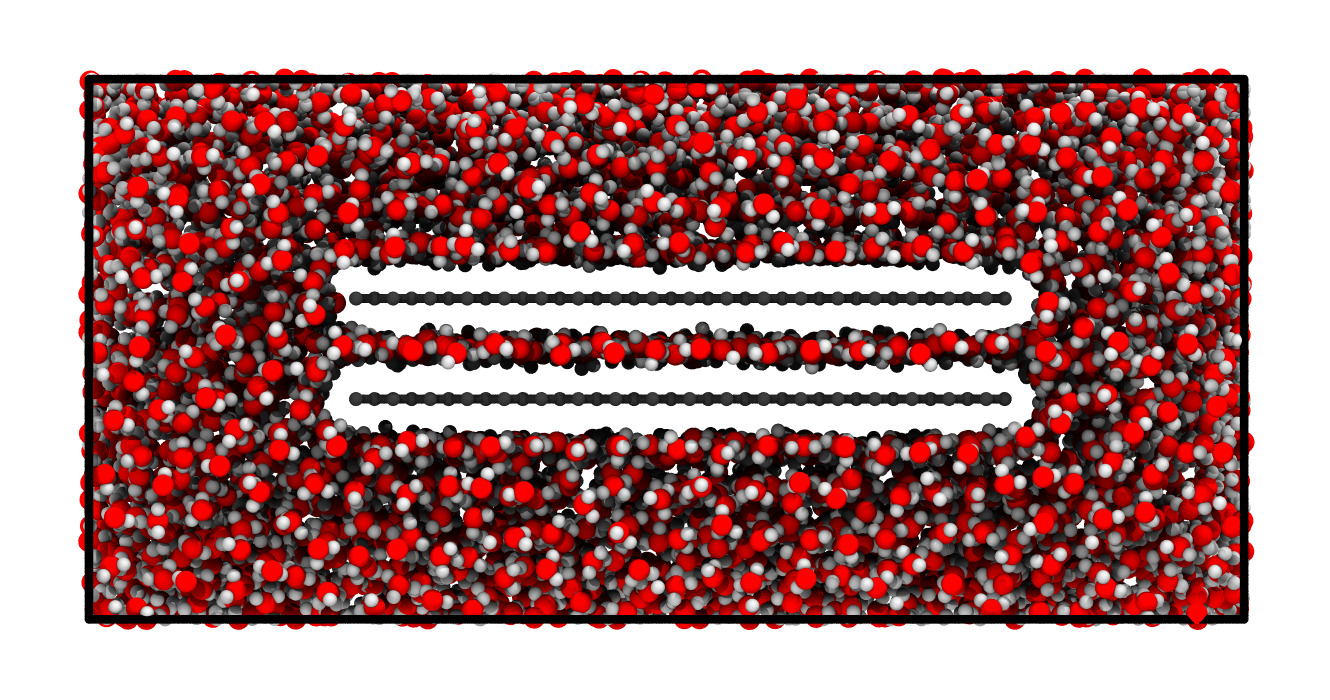}
\end{minipage}   \\* \midrule

\begin{tabular}[c]{@{}c@{}}Parallel rigid GRA layers\\immersed in liquid water\\  (86.762\;\AA\;$\times$\;55.614\;\AA\;$\times$\;36.000\;\AA)\end{tabular} & 
\begin{tabular}[c]{@{}c@{}} $N_{\textrm{atoms}}=18208$\\ $N_{\textrm{C}}=2288$\\ $a=54.340$\;\AA\\ $b=55.614$\;\AA \\ $N_{\textrm{H}_2\textrm{O}}=5308$\\ $W=6.7$\;\AA\\$t_{\textrm{prod}}=300$ ps\\$t_{\textrm{eq}}=150$  ps\\ $\rho_{\textrm{O}}^{\textrm{2D}}=0.1083$ \#O/\AA$^{2}$   \end{tabular}  & 
\begin{minipage}{0.25\textwidth}
      \includegraphics[width=\linewidth]{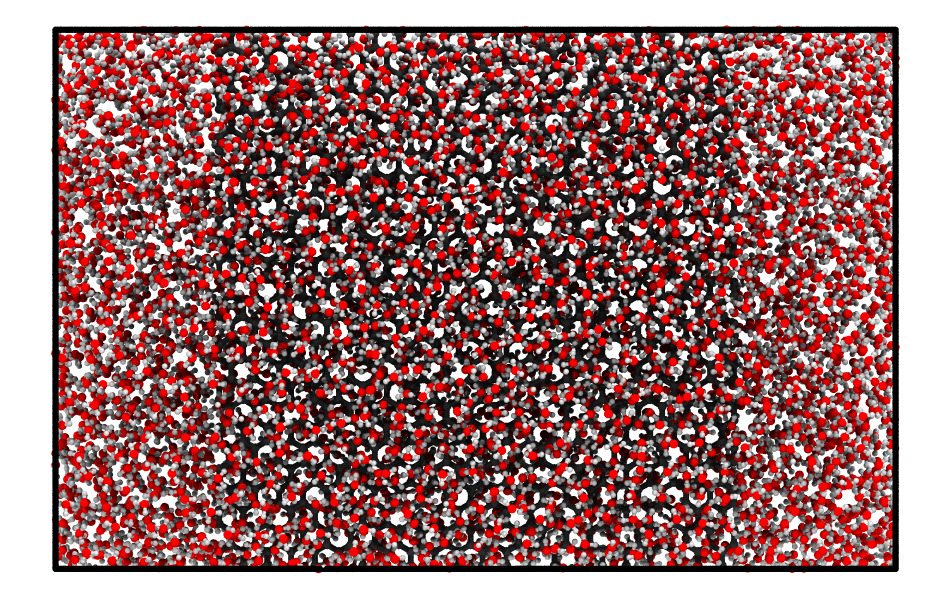}
      \includegraphics[width=\linewidth]{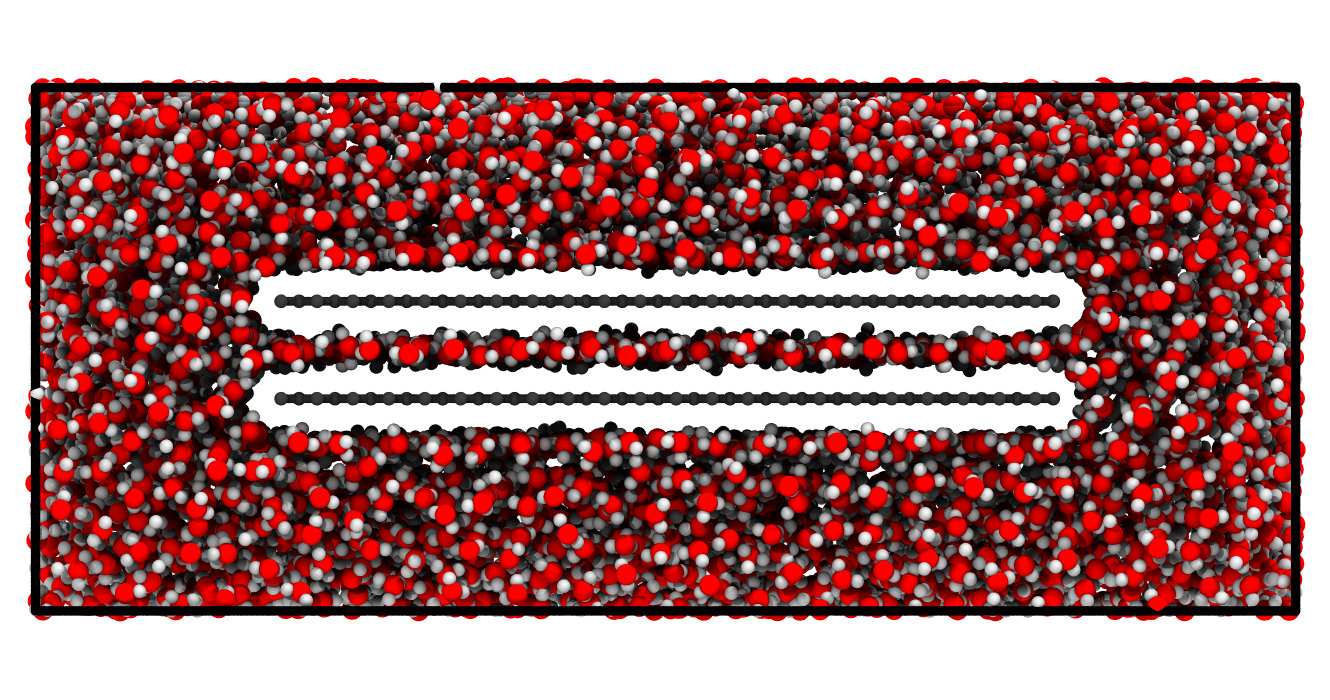}
\end{minipage}   \\* \midrule

\begin{tabular}[c]{@{}c@{}}Parallel rigid GRA layers\\immersed in liquid water\\  (96.694\;\AA\;$\times$\;64.170\;\AA\;$\times$\;36.000\;\AA)\end{tabular} & 
\begin{tabular}[c]{@{}c@{}} $N_{\textrm{atoms}}=23051$\\ $N_{\textrm{C}}=3120$\\ $a=64.220$\;\AA\\ $b=64.170$\;\AA \\ $N_{\textrm{H}_2\textrm{O}}=6645$\\ $W=6.7$\;\AA\\$t_{\textrm{prod}}=300$ ps\\$t_{\textrm{eq}}=150$  ps\\ $\rho_{\textrm{O}}^{\textrm{2D}}=0.1066$ \#O/\AA$^{2}$   \end{tabular}  & 
\begin{minipage}{0.25\textwidth}
      \includegraphics[width=\linewidth]{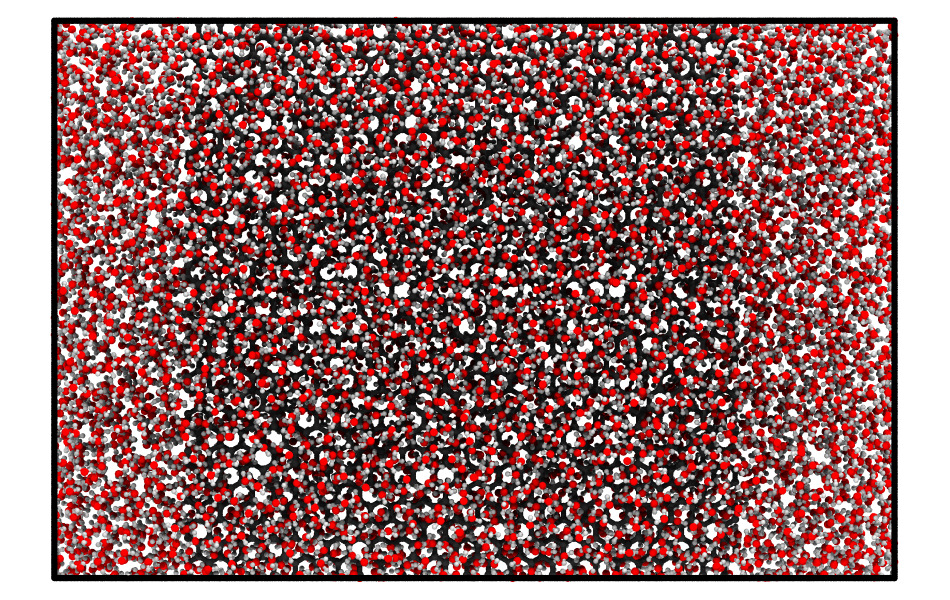}
      \includegraphics[width=\linewidth]{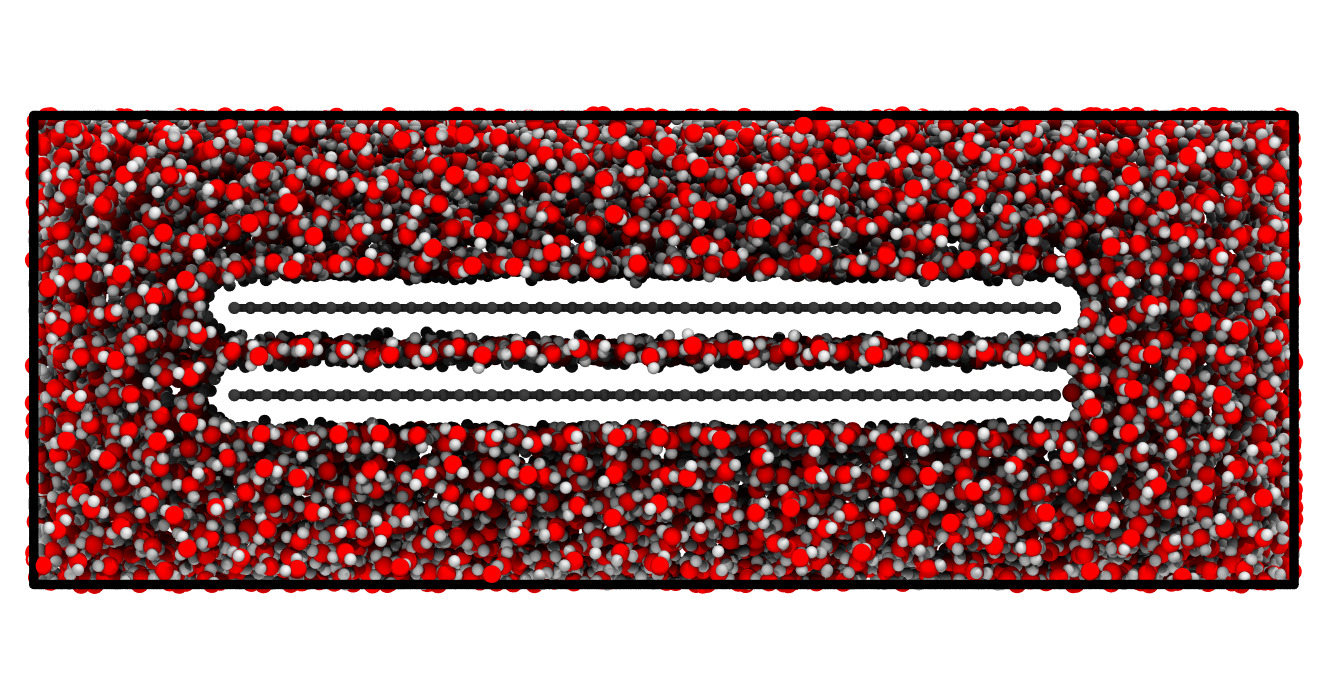}
\end{minipage}   \\* \midrule

\begin{tabular}[c]{@{}c@{}}Parallel rigid GRA layers\\immersed in liquid water\\  (105.742\;\AA\;$\times$\;77.004\;\AA\;$\times$\;36.000\;\AA)\end{tabular} & 
\begin{tabular}[c]{@{}c@{}} $N_{\textrm{atoms}}=29971$\\ $N_{\textrm{C}}=4320$\\ $a=74.100$\;\AA\\ $b=77.004$\;\AA \\ $N_{\textrm{H}_2\textrm{O}}=8553$\\ $W=6.7$\;\AA\\$t_{\textrm{prod}}=300$ ps\\$t_{\textrm{eq}}=150$  ps\\ $\rho_{\textrm{O}}^{\textrm{2D}}=0.1061$ \#O/\AA$^{2}$   \end{tabular}  & 
\begin{minipage}{0.25\textwidth}
      \includegraphics[width=\linewidth]{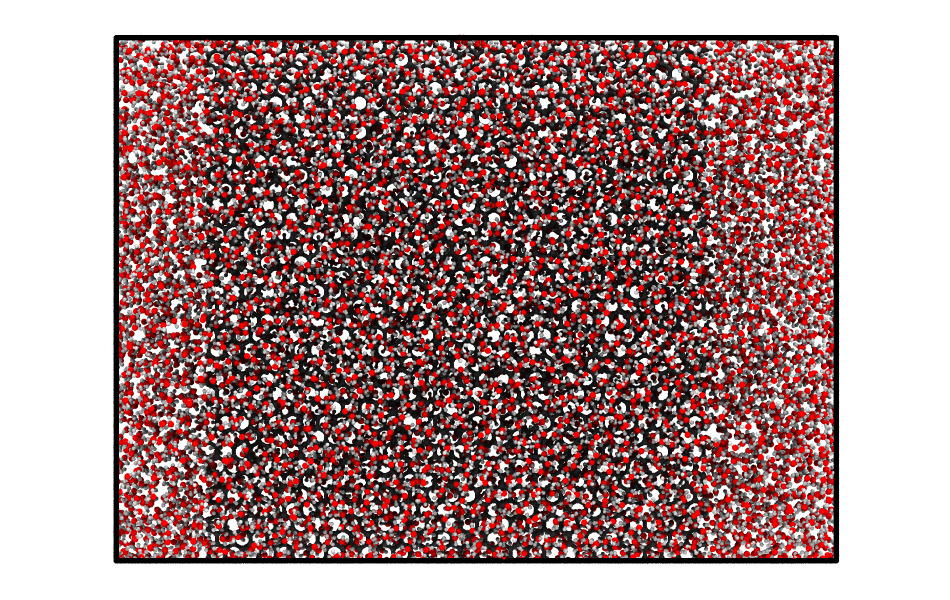}
      \includegraphics[width=\linewidth]{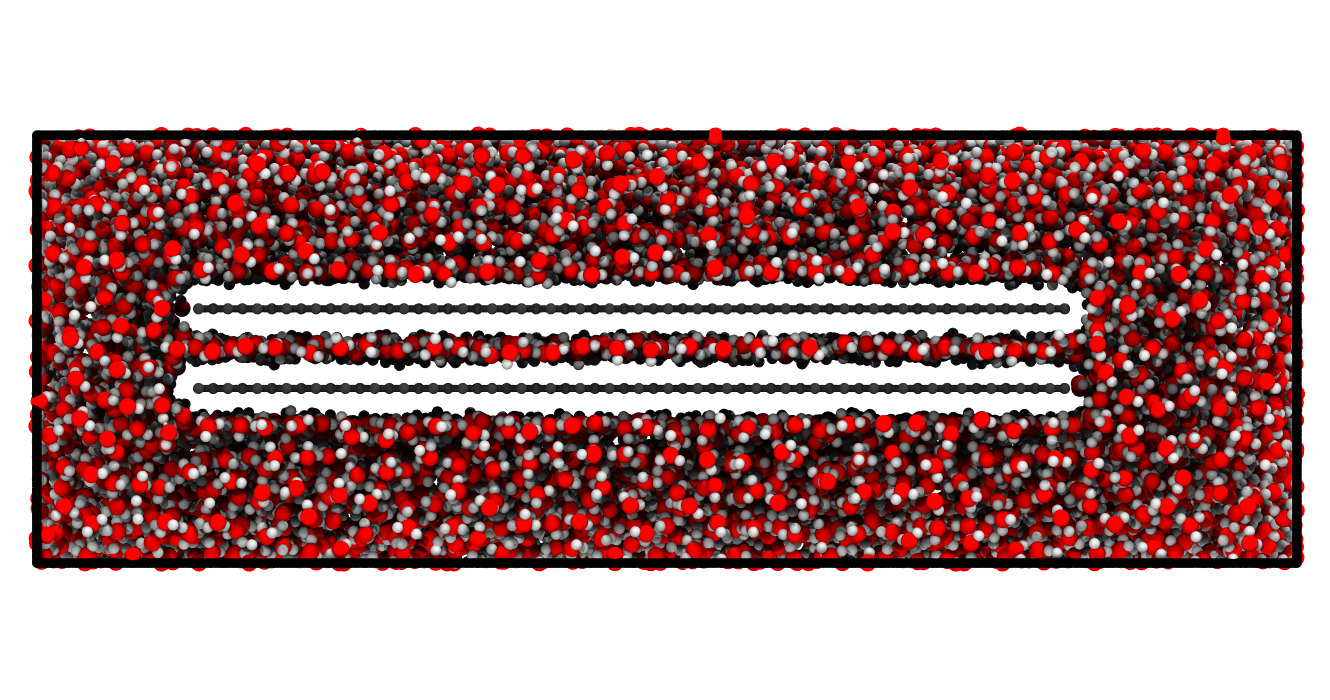}
\end{minipage}   \\* \bottomrule

\bottomrule
\multicolumn{3}{c}{
\parbox{0.95\textwidth}{
  \vspace{0.6cm}
  Table S7: Detailed overview of the parallel rigid GRA layer systems immersed in liquid water, used in this work to establish a chemical potential reference.
  For each system, we report the total number of atoms, $N_{\textrm{atoms}}$; the number of carbon atoms, $N_{\textrm{C}}$; the graphene in-plane lattice constant along the $x$-direction, $a$; the graphene in-plane lattice constant along the $y$-direction, $b$; the corresponding number of water molecules, $N_{\textrm{H}_2\textrm{O}}$; the interlayer distance between the graphene sheets, $W$; the equilibration time, $t_{\textrm{eq}}$; the simulation production time per umbrella, $t_{\textrm{sim/umbrella}}$; and the surface density within the central region of the slit, $\rho_{\textrm{O}}^{\textrm{2D}}$.}}
\end{longtable}

\newpage
\subsection{Simulation setup}
In this work, we conducted five different types of molecular dynamics (MD) simulations: (i) short MD simulations using the machine-learned potentials (MLPs) to generate additional training data for its development; (ii) \textit{ab initio} MD (AIMD) simulations to validate the MLP; (iii) constrained MD simulations using the developed MLP, which form the core of the results presented in this work; (iv) extended MD simulations using the developed MLP to obtain more accurate estimates of key observables; and (v) large-scale simulations using the MLP, where rigid slit pores are fully immersed in bulk liquid water, to determine the equilibrium surface density inside the central region of the pores.
All simulations used hydrogen atom masses.

\textbf{Short MD simulations}

To ensure that the MLP was trained on the range of configurations sampled during the production runs, we expanded the training dataset beyond our base data (from Refs.~\citenum{gra_wat_acid_2025} and \citenum{flt_nanotubes}).
Specifically, we included configurations that accurately describe graphene–graphene interactions in both AA and AB stackings, as well as structures spanning a range of water densities, from high to low.
For this purpose, we used the MLP to propagate dynamics via Langevin MD simulations in the NVT ensemble at 300 K, using a time step of 0.5 fs and a friction coefficient of 2.5 ps$^{-1}$.
These simulations were carried out using the ASE software \cite{ase_2017}.
A similar data generation and training procedure was used to develop a separate MLP tailored to water interacting with hBN in both slit pore and nanodroplet environments.

\textbf{AIMD simulations}

The AIMD simulations used to validate the MLP (see Section \ref{sec:mlp_valid}) were performed in the NVT ensemble using the CP2K/Quickstep code \cite{cp2k_2020}, interfaced with the i-PI program \cite{ipi3_2024}, with a time step of 1.0 fs.
The temperature was set to 300\;K, 330\;K, 360\;K, and 400\;K, maintained using a CSVR thermostat \cite{csvr_2007} with a 100 fs coupling constant.
We employed the revPBE generalized gradient approximation exchange-correlation functional \cite{revpbed3_1}, combined with the zero-damping version of Grimme’s D3 dispersion correction \cite{revpbed3_2}.
Atomic cores were represented using dual-space Goedecker–Teter–Hutter pseudopotentials \cite{gth_1996}, and a plane-wave cutoff of 400 Ry was applied.
The Kohn–Sham orbitals of oxygen and hydrogen atoms were expanded using the TZV2P basis set \cite{basis_set_2007}.

\newpage

\textbf{Constrained MD simulations}

The constrained MD simulations were performed using the ASE software \cite{ase_2017} with the PLUMED plugin \cite{plumed}.
Each umbrella window was propagated via Langevin MD simulations in the NVT ensemble at 300 K, using a time step of 0.5 fs and a friction coefficient of 2.5 ps$^{-1}$.
For each window, a 50\;ps equilibration period was followed by a 100\;ps production period, from which statistics of the umbrellas were obtained.

\textbf{Extended MD simulations}

To obtain further microscopic insights into the systems analysed, extended (unconstrained) MD simulations were performed using the ASE software \cite{ase_2017}.
For this, a 50\;ps equilibration period was followed by a 1\;ns production period, from which statistics of the properties of interest were obtained (mainly pressure, distance, and hydrogen bonding statistics).
The dynamics were propagated via Langevin MD simulations in the NVT ensemble at 300 K, using a time step of 0.5 fs and a friction coefficient of 2.5 ps$^{-1}$

\textbf{Large-scale MD simulations}

To determine the equilibrium surface density within the central region of slit pores --as required for setting a reference chemical potential (see Section \ref{sec:chem_pot})-- we performed large-scale molecular dynamics simulations using the Symmetrix library \cite{mace_off, symmetrix}, which interfaces directly with LAMMPS \cite{lammps_2022}.
The systems consisted of rigid slit pores fully immersed in bulk liquid water. Simulations were conducted in the NPT ensemble at 300 K with a 1 fs time step.
Each system was equilibrated for 150 ps, followed by a 300 ps production run, during which the surface density in the central pore region was measured.
Pressure control was applied only to the water atoms and restricted to the $x$-direction (perpendicular to the slit walls), to account for confinement and maintain periodicity along the $y$-axis.

\newpage
\section{Machine learning potential} \label{sec:mlp_valid}
\subsection{Model development}
The MLP was developed iteratively over several generations, beginning with training data from Refs.~\citenum{gra_wat_acid_2025} and \citenum{flt_nanotubes}, selected to provide broad coverage of the relevant physical regimes.
These datasets provide (i) self-dissociated water configurations across a wide range of densities under graphene confinement, including the ultra-confined limit, and (ii) water–graphene interfaces spanning from flat sheets to highly curved environments within carbon nanotubes of varying radii.
Together, they ensure that the model captures both the physics of water dissociation and the bending rigidity of graphene---an essential feature for accurately modeling nanodroplet formation.

To refine the description of graphene–graphene interactions, we introduced structures corresponding to the equilibrium interlayer distance and included variations across both AA and AB stacking configurations.
This was critical for capturing the van der Waals interactions that drive the graphene encapsulation of water.
Given the range of densities (and, effectively, pressures) sampled in this work, we enriched the dataset with both high- and low-density configurations.
We also incorporated water confined between AB-stacked graphene layers, as this stacking is key to the geometry of the encapsulated nanodroplet system shown in Figure~1a.
Finally, we extended the training data to include configurations with expanded graphene sheet dimensions to ensure the model's applicability to the large-scale systems considered here.
This comprehensive dataset allows the MLP to robustly describe the full range of thermodynamic and structural conditions explored in this work.

The MLP for hBN nanodroplets was developed following the same iterative strategy, with the addition of configurations from Ref.~\citenum{wang2025} to accurately capture water-hBN interactions specific to those environments.

\subsection{Model validation}
To evaluate the validity of the MLP for the systems studied in this work, we quantified the root-mean-square errors (RMSEs) in energies and forces predicted by the model.
Specifically, we randomly selected 300 snapshots from 100 ps MLP-based MD simulations that sample the water self-dissociation reaction pathway in either bulk or confined conditions, using values of $n_{\textrm{H}} = {1.00, 1.20, \ldots, 2.00}$.
For each configuration, we performed single-point DFT calculations at the same level of theory used to train the MLP(i.e., revPBE-D3).
This provides a direct and robust measure of the MLP’s accuracy, as it compares predictions for structures sampled from its potential energy surface against the reference \textit{ab initio} values.

\begin{figure}[htp!]
\centering
\includegraphics[width=0.6\textwidth]{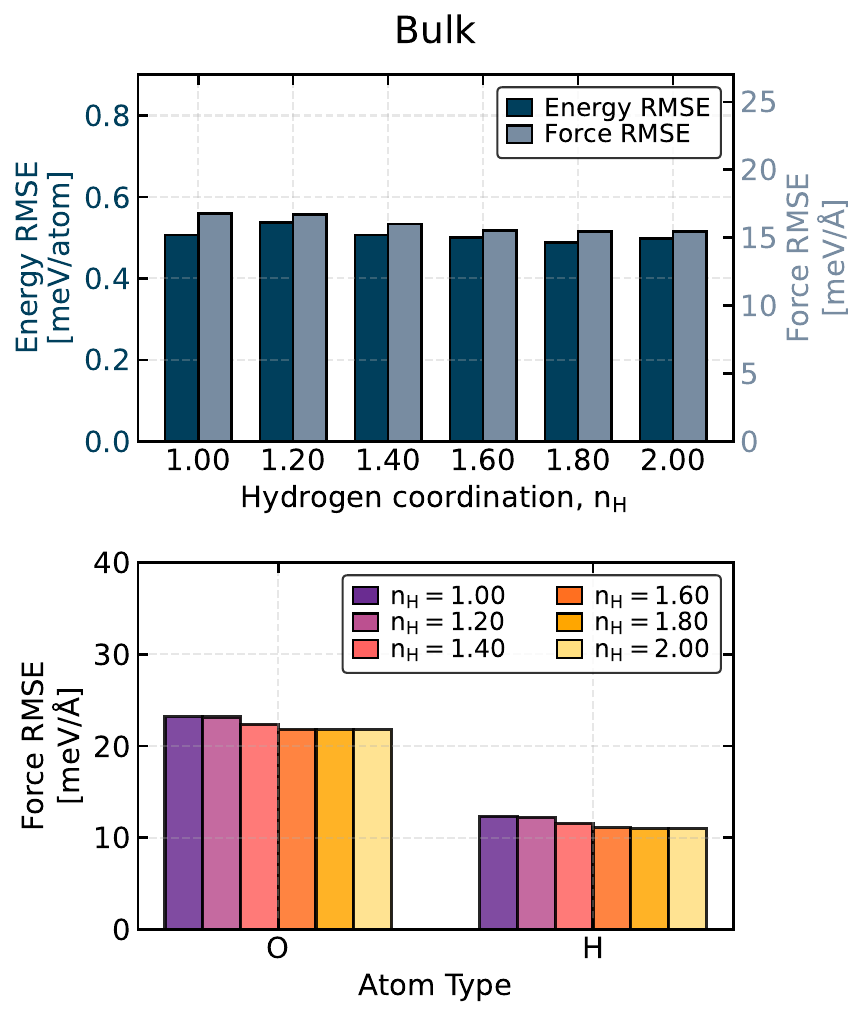}
    \caption{RMSE in energies and forces along the water self-dissociation reaction pathway in the bulk setup, comparing the MLP predictions to the underlying DFT reference.
    The force RMSE decomposed by atom type (O and H) along the reaction coordinate is also shown, highlighting the model's accuracy across all atomic species throughout the dissociation process.}
    \label{fig:bmark_bulk}
\end{figure}

\begin{figure}[htp!]
\centering
\includegraphics[width=0.7\textwidth]{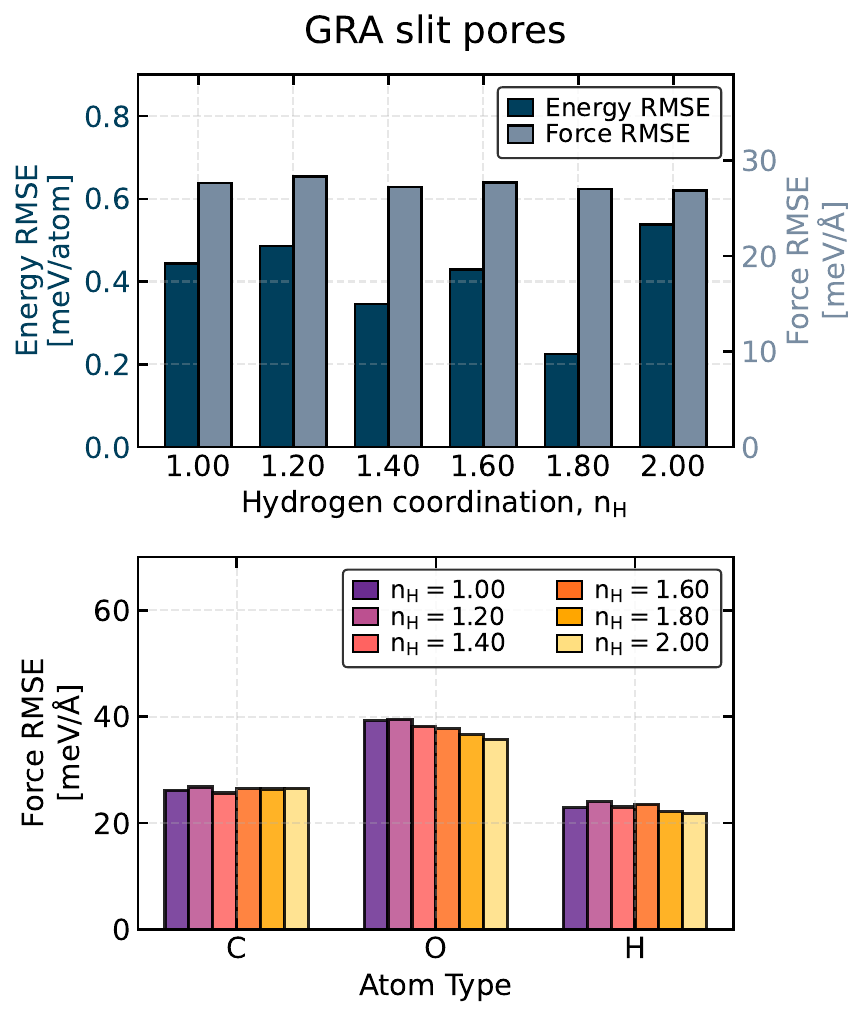}
    \caption{RMSE in energies and forces along the water self-dissociation reaction pathway in the GRA slit pores, comparing the MLP predictions to the underlying DFT reference.
    The force RMSE decomposed by atom type (C, O, and H) along the reaction coordinate is also shown, highlighting the model's accuracy across all atomic species throughout the dissociation process.}
    \label{fig:bmark_gra}
\end{figure}

\begin{figure}[htp!]
\centering
\includegraphics[width=0.7\textwidth]{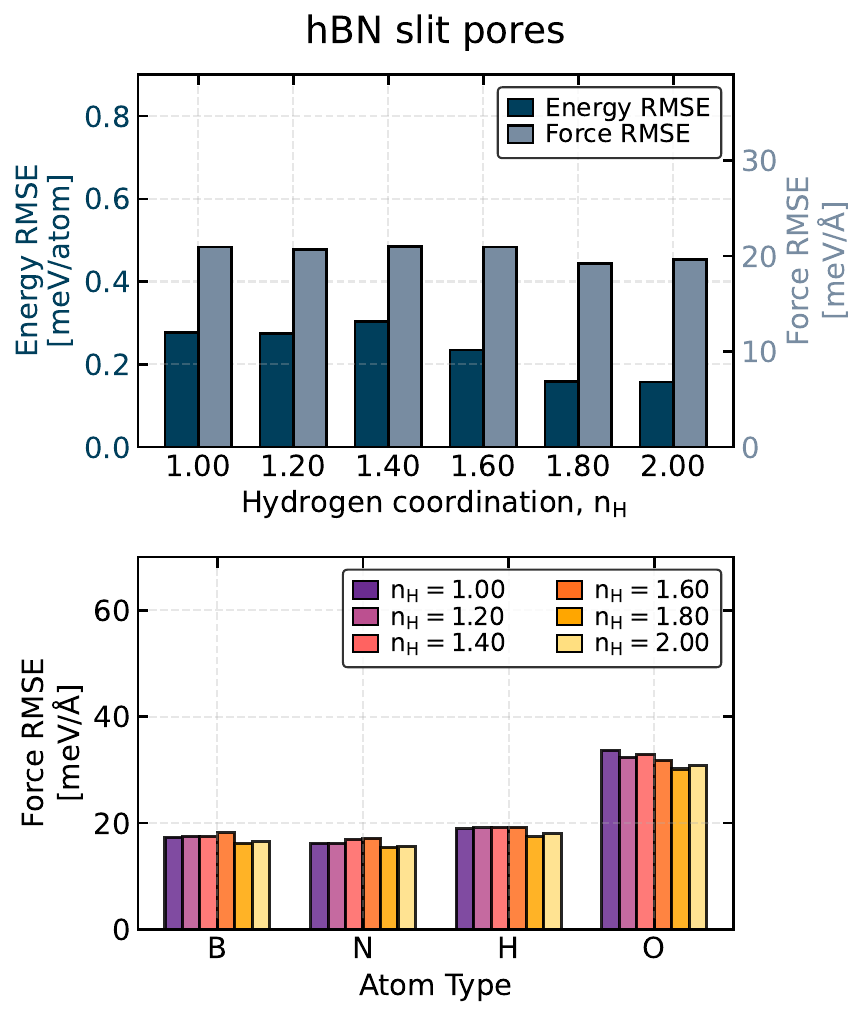}
    \caption{RMSE in energies and forces along the water self-dissociation reaction pathway in the hBN slit pores, comparing the MLP predictions to the underlying DFT reference.
    The force RMSE decomposed by atom type (B, N, O, and H) along the reaction coordinate is also shown, highlighting the model's accuracy across all atomic species throughout the dissociation process.}
    \label{fig:bmark_hbn}
\end{figure}

To reduce the computational cost of these electronic structure calculations, we used smaller simulation cells. For graphene, the sheet dimensions were set to $L_x = 12.350$\;\AA\; and $L_y = 12.834$\;\AA.
For hBN, we similarly employed reduced sheet dimensions of $L_x = 13.047$\;\AA\; and $L_y = 12.555$\;\AA.
As shown in Figs. \ref{fig:bmark_bulk}--\ref{fig:bmark_hbn}, the MLP closely reproduces both the energies and forces obtained from the underlying DFT method across all systems and reaction coordinate values.
This confirms the model’s reliability in capturing the key physicochemical features relevant to water self-dissociation in the environments considered.

\newpage

\newpage

\begin{figure}[htp!]
\centering
\includegraphics[width=\textwidth]{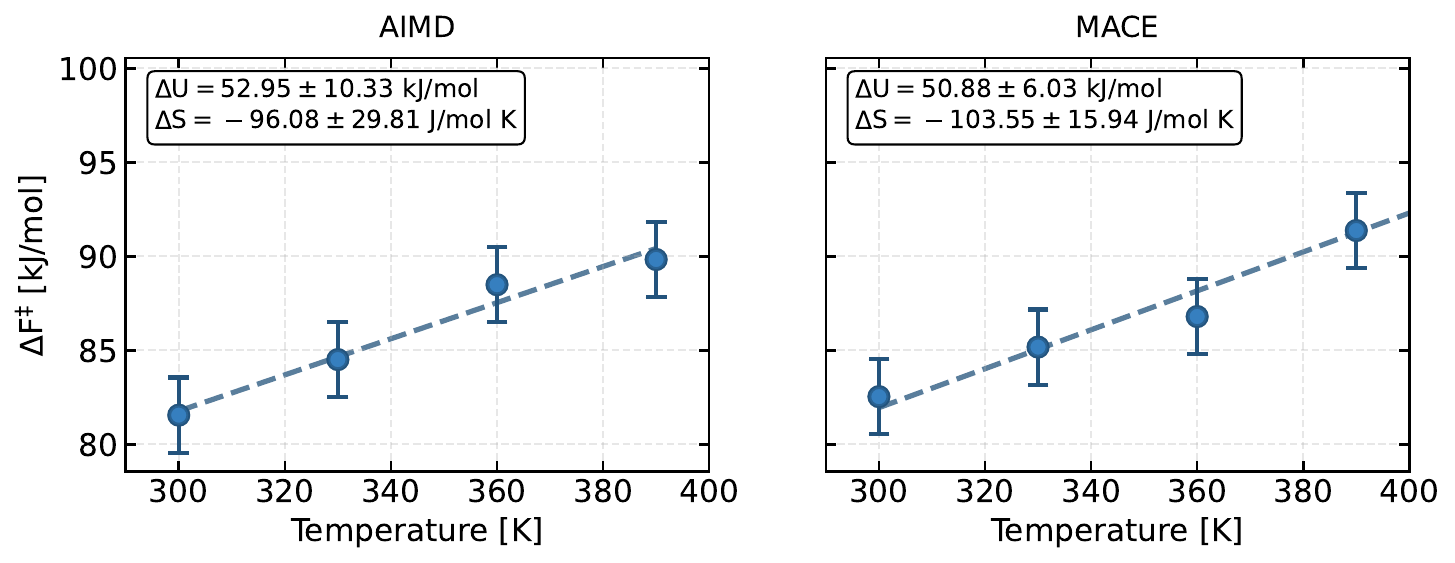}
    \caption{Comparison of temperature-dependent free energy barriers for water self-dissociation computed using AIMD\citenum{litman_efield_2025}  and MACE MLP simulations.
    The free energy barrier, $\Delta F^{\ddagger}$, is shown as a function of temperature $T$ for AIMD (left) and MACE (right), and fitted using the relation $\Delta F = \Delta U - T \Delta S$ to extract the enthalpic and entropic contributions.
    The close agreement between the two methods highlights the accuracy of the MLP in describing the dissociation process.}
    \label{fig:deltaf_temp_mlp}
\end{figure}

\newpage
To further validate the robustness of the MLP developed in this work, we assessed its ability to reproduce the temperature dependence of the water self-dissociation reaction.
While a focus of this study is on how water self-dissociation varies with density, accurately reproducing the thermodynamic response over a broader temperature range is a strong indicator of model transferability and physical fidelity.
As shown in Figure \ref{fig:deltaf_temp_mlp}, the free energy barriers computed using MACE closely match those obtained from AIMD across temperatures\cite{litman_efield_2025}, with excellent agreement in the extracted enthalpic and entropic contributions.
This consistency demonstrates that the MLP not only reproduces accurate energetics but also captures the underlying thermodynamic landscape of the dissociation process.

Finally, because a central focus of this work is the self-dissociation of water in material-encapsulated nanodroplets, it is essential that the model accurately captures the bending rigidity of the confining material, which determines how the material deforms to accommodate the droplet shape.
For graphene, the bending rigidity, $B_M$, can be obtained by fitting the energy per atom in single-wall carbon nanotubes (SWCNTs) of varying radii using the following expression\cite{bm_2013}: 
\begin{equation}
    E_{\textrm{atom}}^{\textrm{CNT}} = E_0 + S_0 B_{M} r^{-2}/2 \label{eq:bending}
\end{equation}
where $E_{\textrm{atom}}^{\textrm{CNT}}$ is the energy per atom in a SWCNT, $E_0$ is the energy per atom in a flat graphene, and $S_0=2.63$ \AA$^{2}$ is the planar footprint of a carbon atom in graphene.

\begin{figure}[htp!]
\centering
\includegraphics[width=0.55\textwidth]{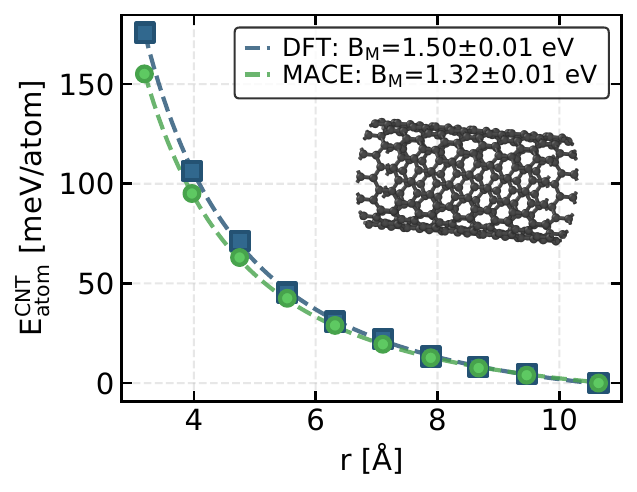}
    \caption{Energy per atom as a function of nanotube radius for SWCNTs rolled along zigzag directions.
    The dashed line indicates the fit used to extract the bending rigidity, $B_{M}$, of graphene.
    A representative SWCNT structure is shown as an inset.}
    \label{fig:bm_mlp}
\end{figure}

By computing $E_{\textrm{atom}}^{\textrm{CNT}}$ for nanotubes with different radii, $B_{M}$ can be obtained from the curvature dependence.
As shown in Figure \ref{fig:bm_mlp}, our MLP accurately reproduces this curvature dependence, in excellent agreement with the reference \textit{ab initio} values.

\begin{figure}[htp!]
\centering
\includegraphics[width=0.55\textwidth]{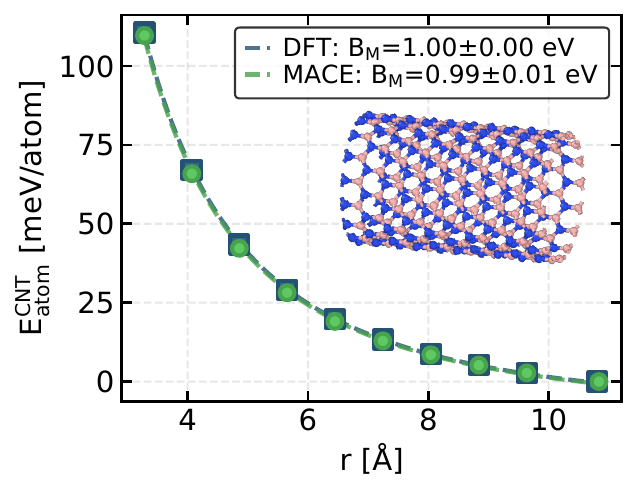}
    \caption{Energy per atom as a function of nanotube radius for SWCNTs rolled along zigzag directions.
    The dashed line indicates the fit used to extract the bending rigidity, $B_{M}$, of graphene.
    A representative SWCNT structure is shown as an inset.}
    \label{fig:bm_mlp_bn}
\end{figure}

A similar procedure is applied to hBN, where the bending rigidity is obtained from boron nitride nanotubes with varying curvature, yielding again excellent agreement with the reference data, as shown in Figure~\ref{fig:bm_mlp_bn}.

\newpage

Overall, the validations presented in this section demonstrate that the MLP developed in this work reliably captures the key physicochemical features governing water self-dissociation.
From energetic and force accuracy across a range of conditions to its ability to reproduce temperature-dependent thermodynamic quantities and structural properties of graphene, the MLP provides a robust and transferable framework for investigating confined water reactivity with first-principles fidelity.

\newpage
\section{Umbrella sampling}
To investigate how water self-dissociation is affected by confinement, we performed 1D umbrella sampling simulations using a reaction coordinate defined as the coordination number of a selected oxygen atom (O$^{*}$) with all hydrogen atoms in the system:
\begin{equation}
n_{\textrm{H}} = \sum^{N}_{i=1} \frac{1 - \left( r_i / R_0 \right)^{12} }{1 - \left( r_i / R_0 \right)^{24}}
\end{equation}
where $i$ iterates over all the hydrogens in the simulation box, $r_{i}$ is the distance between hydrogen $i$ and O$^{*}$, and $R_{0}=$1.38\;\AA\cite{sprik_pk_coord_2000}.

To sample configurations across the reaction coordinate, we applied a harmonic restraint around target values $n_{\textrm{H}}^{\prime}$ using a force constant of 200 kcal/mol per coordination unit squared.
Each umbrella window was centered between $n_{\textrm{H}}^{\prime} = 1.00$ and $2.20$ in steps of 0.04, for a total of 31 windows per system.
Each window was sampled for 100 ps.

\begin{figure}[htp!]
\centering
\includegraphics[width=0.6\textwidth]{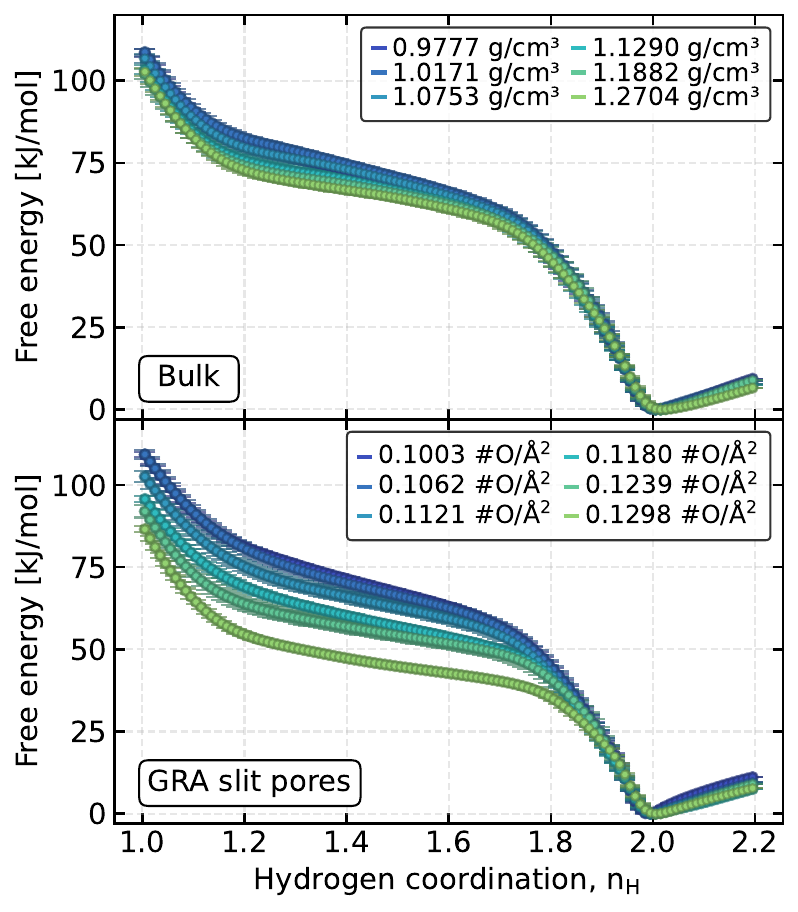}
    \caption{Free energy profiles as a function of the hydrogen coordination number, $n_{\textrm{H}}$, for the H$_2$O self-dissociation reaction across the different systems, obtained using umbrella sampling.}
    \label{fig:feps}
\end{figure}
\newpage
The resulting free energy profiles were reconstructed using thermodynamic integration\cite{umbrella_int_2005, umbrella_error_2006} (see Figure \ref{fig:feps}). The self-dissociation constant of water, $p\textrm{K}_{\textrm{w}}$, was then computed from the free energy difference $\Delta F^{\ddagger}$ between $n_\textrm{H} \approx 2.0$ and $n_\textrm{H} \approx1.2$ using the following expression:
\begin{equation}
    p\textrm{K}_{\textrm{w}} = \frac{\Delta F^{\ddagger}}{RT\ln(10)} \label{eq:pkw_relation}
\end{equation}
where $R$ is the molar gas constant, and $T$ is the temperature.

Because our goal is to understand how confinement modulates water dissociation, we emphasize relative differences in $p\textrm{K}_{\textrm{w}}$ across systems rather than precise absolute values.
This strategy ensures that our conclusions remain robust and transferable, as they are less sensitive to the choice of electronic structure method, neglect of nuclear quantum effects, or the specific sampling protocol.
While more complex approaches---such as multidimensional reaction coordinates---may offer increased absolute accuracy~\cite{tatsuya_2022, selloni_2023}, our systematic comparisons offer a reliable and meaningful picture of how confinement alters self-dissociation.
Finally, we verified that finite-size effects do not impact the reported trends.

\begin{figure}[htp!]
\centering\includegraphics[width=\textwidth]{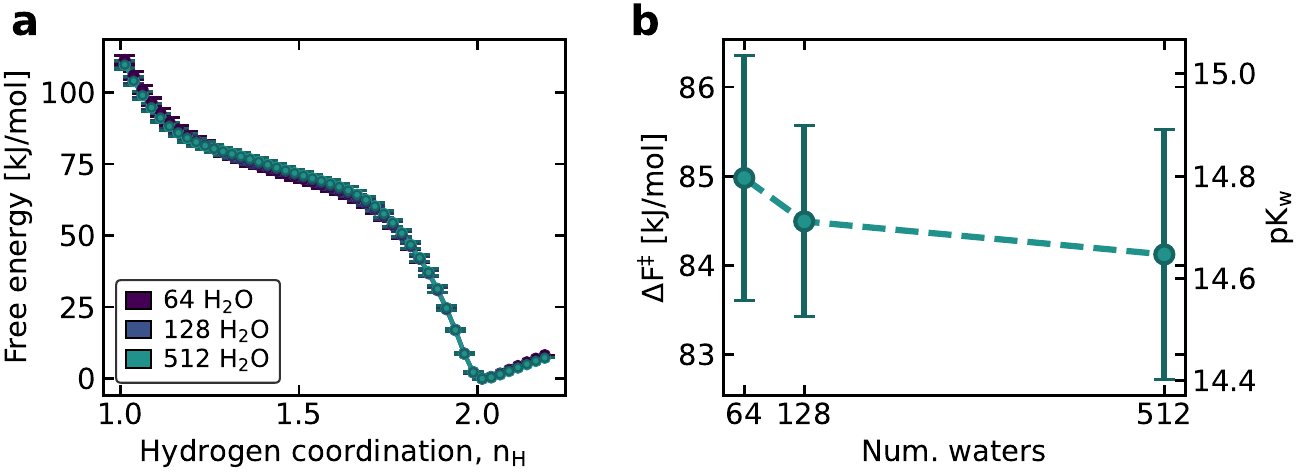}
    \caption{(a) Free energy profiles for the water self-dissociation reaction across systems with an increasing number of water molecules.
    (b) Finite size dependence of the computed free energy barrier $\Delta F^{\ddagger}$ (obtained as the free energy difference between $n_{\text{H}}\approx2$ an $n_{\text{H}}\approx1.2$) and self-dissociation constant of water $p\textrm{K}_{\textrm{w}}$.
    Convergence in these values is achieved for systems exceeding 128 water molecules per unit cell.}
    \label{fig:fs_effects_us}
\end{figure}

\newpage
\section{Consistent thermodynamic comparison between bulk and confined water} \label{sec:chem_pot}
To consistently compare how water self-dissociation changes with thermodynamic conditions in bulk and confined environments, we computed variations in the chemical potential.
This approach is necessary as other thermodynamic variables, such as pressure, are ill-defined in nanoconfined systems due to the ambiguity in specifying an effective volume within the slit pore.

\begin{figure}[htp!]
\centering\includegraphics[width=0.5\textwidth]{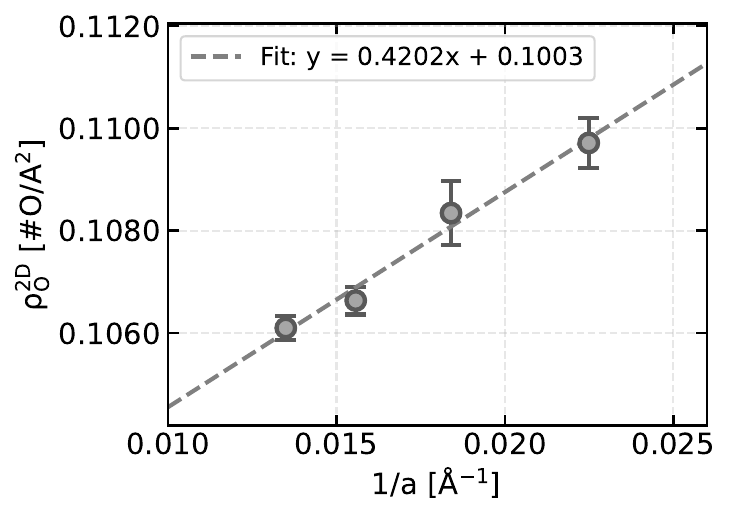}
    \caption{Surface density as a function of $1/a$, where $a$ is the graphene lattice constant along the $x$-axis (see Section S1). The extrapolated density value for $L \to \infty$ limit corresponds to 0.1003 \#O/\AA$^{2}$.}
    \label{fig:density_extrapolation}
\end{figure}

To quantify chemical potential changes under varying thermodynamic conditions, we first need to establish a reference chemical potential.
For bulk water, this reference corresponds to the chemical potential of a 1 bar liquid water box, a standard condition widely used in both experiments and simulations to represent standard ambient liquid water.
For nanoconfined water, the reference corresponds to the chemical potential at which the surface density within the central region of the slit pore reaches its equilibrium value, determined by extrapolating simulations across different system sizes.
To this end, we simulated systems with two parallel graphene layers of equal length, periodic along the $y$-axis, lateral dimensions of 44.460$\times$47.058\;\AA$^2$, 54.340$\times$55.614\;\AA$^2$, 64.220$\times$64.170\;\AA$^2$, and 74.100$\times$77.004\;\AA$^2$, all immersed in liquid water (see Section S1 for details).
For each system, we computed the oxygen surface density within the central region of the slit pore and extrapolated these results to the $L \to \infty$ limit to obtain the equilibrium density within the confined region, as shown in Figure~\ref{fig:density_extrapolation}.

\newpage
Having established consistent reference chemical potentials, we can now directly compare the behavior of bulk and nanoconfined water under varying thermodynamic conditions.
For large, homogeneous systems where the Gibbs free energy $G$ scales linearly with the number of moles $N$, the chemical potential simplifies to:
\[
\mu = \frac{G}{N} = G_m,
\]
where $G_m$ denotes the molar Gibbs free energy.
Similarly, the molar entropy and molar volume are given by:
\[
S_m = \frac{S}{N} \quad \text{and} \quad V_m = \frac{V}{N}.
\]
The differential form of the Gibbs free energy is:
\[
dG = -S \, dT + V \, dP,
\]
which, expressed per mole, becomes:
\[
d\mu = -S_m \, dT + V_m \, dP.
\]
Since our simulations are performed at constant temperature, this simplifies to:
\[
d\mu = V_m \, dP.
\]
Integrating, we obtain:
\begin{equation}
\mu(P') - \mu^{0}(P_0) = \int_{P_0}^{P'} V_m \, dP,
\label{eq:chem_bulk_1}
\end{equation}
where $\mu^{0}(P_0)$ is the reference chemical potential. For bulk water, we evaluate Eq.~\ref{eq:chem_bulk_1} using $P_0 = 1$~bar as the reference state.
In nanoconfined systems, however, as discussed earlier, specifying an effective confined volume introduces ambiguity, making direct application of Eq.~\ref{eq:chem_bulk_1} problematic.
To overcome this, we rewrite the expression as:
\[
\mu(P') - \mu^{0}(P_0) = \int_{P_0}^{P'} \frac{V}{N} \, dP.
\]
Noting that $dG = V \, dP$, we can express the chemical potential as:
\begin{equation}
\mu(G') - \mu^{0}(G_0) = \int_{G_0}^{G'} \frac{1}{N} \, dG.
\end{equation}
where $G$ is directly accessible from our enhanced sampling simulations.
To consistently compare bulk and confined systems, we set the reference states to be equivalent:
\[
\mu^{0}_{\mathrm{bulk}}(P_0 = 1 \, \mathrm{bar}) = \mu^{0}_{\mathrm{conf}}(G_0),
\]
where $G_0$ corresponds to the state with the surface density of confined water obtained in equilibrium.
With this, chemical potential variations under thermodynamic changes can be consistently compared between bulk and nanoconfined water.

\newpage
\section{Determination of equilibrium density in nanodroplet confinement} \label{sec:nanodrop_equi}

To ensure a meaningful comparison with bulk water, we designed the nanodroplet-confined systems such that the droplet core reaches its equilibrium density--defined as the density at which structural properties within the droplet's interior converge with respect to system size.
We then compared this state point to bulk water at its equilibrium density, thereby isolating the intrinsic effects of confinement without introducing artifacts from overpressurization.

In general, determining the equilibrium density in confinement is a nontrivial task.
The literature presents a variety of approaches---including piston-based equilibration, flexible boundary conditions, or matching the average pore density to bulk experimental values---but these methods often yield widely varying results.
Table \ref{tab:lit_1} provides a selection of representative examples, including both force field and first-principles molecular dynamics studies, illustrating the range of reported densities for systems mostly under graphene confinement.
This list is by no means comprehensive, as many additional studies exist.
Nevertheless, the variation in reported values highlights the challenge of comparing water self-dissociation across studies, particularly given that the dissociation constant of water is known to decrease with increasing pressure.

\begin{table}[htp!]
    \centering    
    \begin{tabular}[c]{c@{\hskip 0.2in}c@{\hskip 0.2in}c@{\hskip 0.2in}c@{\hskip 0.2in}c@{\hskip 0.2in}c@{\hskip 0.2in}c@{\hskip 0.2in}c@{\hskip 0.2in}c}
         \toprule
Ref.      & Method   & $n_{\textrm{H$_2$O}}$ & \begin{tabular}[c]{@{}c@{}}C-C\\distance [\AA]\end{tabular}  & $L_x$ [\AA] & $L_y$  [\AA] & $\rho_O^{\textrm{2D}}$ [\#O/\AA$^2$] \\
\midrule
\citenum{gra_wat_acid_2025}     & revPBE-D3  & 28  &  6.56  & 17.29  & 17.112  &0.0946\\
\citenum{Dufils_ChemSci_2024}         &  revPBE-D3 & 16 &6.63 & 12.30 & 12.78 & 0.1018\\
\citenum{Dufils_ChemSci_2024}         &   SPC/E & 147 & 6.63 & 38.34 & 36.89 & 0.1039\\
\citenum{kara_pairing_2024}       & revPBE-D3  & 248 &  6.70  & 44.46  & 47.058  &0.1185\\
\citenum{Das_JPCL_2023}                     & SPC/E         & 108  & 6.68 & 34.7484 & 34.3920 &0.0904\\
\citenum{Barragan_PCCP_2020}   &   SPC/E          &  27   &  6.63 & 17.12  & 17.30& 0.0912\\
\citenum{Ruiz-Barragan_JPCL_2019} &revPBE-D3      & 27    &   6.91 & 17.12  & 17.30&0.0912\\
\citenum{Duan_Langmuir_2022}& PBE &38 &6.98 & 19.71 & 21.37 &0.09021\\
\bottomrule
    \end{tabular}
    \caption{Monolayer water densities used in simulations of nanoslit-confined systems reported in various studies.
    Here, $n_{\textrm{H$_2$O}}$ denotes the number of water molecules, $L_x$ and $L_y$ refer to the graphene sheet dimensions in the $x$- and $y$-directions, respectively, and $\rho_O^{\textrm{2D}}$ represents the two-dimensional surface density,  defined as the number of oxygen atoms per unit area (O atoms/\AA$^{2}$).}
    \label{tab:lit_1}
\end{table}

To accurately determine the equilibrium density in nanodroplet confinement, we simulated a series of nanodroplets with varying dimensions and monitored the convergence of structural properties in the droplet core.
To reduce the computational cost of these large-scale simulations, we first employed classical force field methods using the i-PI program \cite{ipi3_2024} connected to the LAMMPS package \cite{lammps_2022}.
Water–water interactions were modeled using the TIP4P potential and water–carbon interactions with the AIREBO potential \cite{airebo_2000}.

We initially simulated a large nanodroplet system with dimensions 98.800\;\AA\;$\times$\;98.394\;\AA\; to identify the density at which structural properties converge, as detailed in Table \ref{tab:ff_bmark_1}.
The equilibrium density was determined by computing the cumulative number of oxygen atoms, $N_{\textrm{O}}(r)$, as a function of radial distance $r$ from the droplet center.
The 2D oxygen density, $\rho_{\textrm{O}}^{\textrm{2D}}(r)$, was then obtained from the average slope of multiple linear fits to this curve:

\begin{equation}
    \rho_{\textrm{O}}^{\textrm{2D}}(r) = \frac{1}{2 \pi r } \frac{d N_{\textrm{O}}(r)}{dr}.
\end{equation}

This method provides a smooth, binning-free estimate of the density, minimizing artifacts from interfacial fluctuations and statistical noise.

\begin{figure}[htp!]
\centering\includegraphics[width=\textwidth]{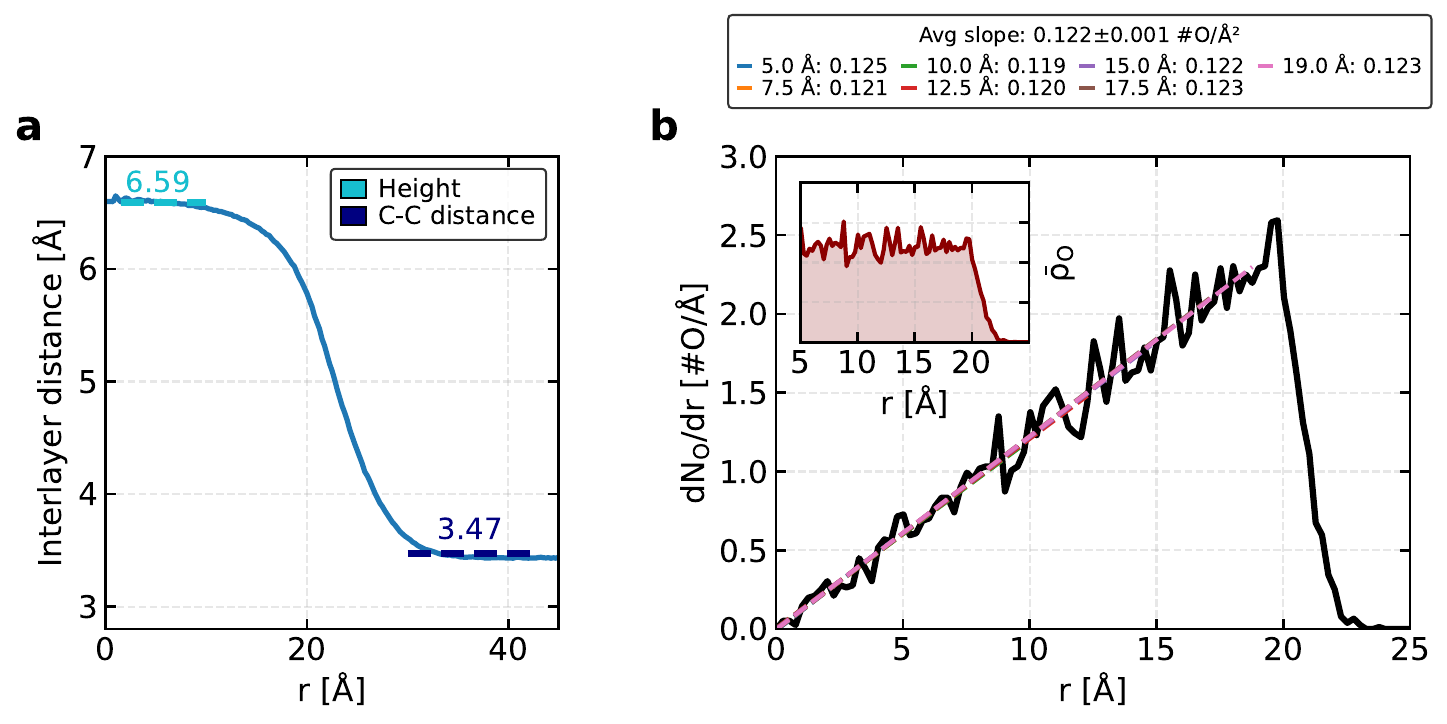}
    \caption{(a) Interlayer distance as a function of the nanodroplet radius, obtained using the classical force field described.
    The droplet height is determined from the central region, where water molecules are present, while the C–C distance is computed from the outer region, where water is absent.
    (b) Cumulative number of oxygen atoms as a function of the nanodroplet radius, obtained using the classical force field described.
    The 2D oxygen density, $\rho_{\textrm{O}}^{\textrm{2D}}$, is determined from the average slope of multiple linear fits.
    Each fit region and its corresponding slope are reported in the legend.
    The inset shows the normalized radial density profile of oxygen atoms, $\bar{\rho}_{\textrm{O}}$.}
    \label{fig:ff_nanodrop_equi}
\end{figure}

Following this procedure, we determined the equilibrium density for a system with sheet dimensions of 98.800\;\AA\;$\times$\;98.394\;\AA.
As shown in Table \ref{tab:ff_bmark_1}, convergence of the core structural properties is achieved when 132 water molecules are confined within the droplet.

\begin{table}[htp!]
    \centering    
    \begin{tabular}[c]{c@{\hskip 0.2in}c@{\hskip 0.2in}c@{\hskip 0.2in}c@{\hskip 0.2in}c@{\hskip 0.2in}c@{\hskip 0.2in}c@{\hskip 0.2in}c@{\hskip 0.2in}c}
         \toprule
Sheet dimensions (\AA$^2$) & $n_{\textrm{H$_2$O}}$ &   \begin{tabular}[c]{@{}c@{}}Droplet\\height [\AA]\end{tabular} & \begin{tabular}[c]{@{}c@{}}C-C\\distance [\AA]\end{tabular} &  \begin{tabular}[c]{@{}c@{}}Droplet\\radius [\AA]\end{tabular}  &  $\rho_{\textrm{O}}^{\textrm{2D}}$ [\#O/\AA$^2$]    \\
\midrule
98.800 $\times$ 98.394   & 33 & 6.31 & 3.43  &11.1     &  0.110 $\pm$0.010       \\
98.800 $\times$ 98.394   & 66 & 6.46 &3.44  & 14.9  &  0.122 $\pm$0.010  \\
98.800 $\times$ 98.394  & 99  & 6.55 &3.43  &17.9  & 0.124 $\pm$0.005  \\
98.800 $\times$ 98.394   & 132 &6.61  & 3.43 &20.4   & 0.120 $\pm$0.006   \\
98.800 $\times$ 98.394   & 165 & 6.64 &  3.45   & 22.71  & 0.120 $\pm$0.002  \\
98.800 $\times$ 98.394   & 198 & 6.66 & 3.43  & 24.74  & 0.121 $\pm$0.002  \\
98.800 $\times$ 98.394   & 231 &  6.65 & 3.44  & 26.59 & 0.122 $\pm$0.002  \\
\bottomrule
    \end{tabular}
    \caption{Convergence of structural properties for a system with dimensions of 98.800\;\AA\;$\times$\;98.394\;\AA, evaluated as a function of the number of water molecules, $n_{\textrm{H$_2$O}}$. Core structural features converge when 132 water molecules are present within the droplet.}
    \label{tab:ff_bmark_1}
\end{table}  

Since our goal is to ultimately simulate this nanodroplet using the MACE MLP developed in this work---which is significantly more computationally demanding than classical force fields---we investigated whether the system size could be reduced without compromising equilibrium properties.
By fixing the number of water molecules at 132 and varying the graphene sheet dimensions, we found that a system size of 79.040\;\AA\;$\times$\;81.282\;\AA\; is sufficient to preserve the target density in the droplet core (see Table \ref{tab:ff_bmark_2}).

\begin{table}[hbt!]
    \centering
    \begin{tabular}[c]{c@{\hskip 0.2in}c@{\hskip 0.2in}c@{\hskip 0.2in}c@{\hskip 0.2in}c@{\hskip 0.2in}c@{\hskip 0.2in}c@{\hskip 0.2in}c@{\hskip 0.2in}c}
         \toprule
Sheet dimensions (\AA$^2$) & $n_{\textrm{H$_2$O}}$ &   \begin{tabular}[c]{@{}c@{}}Droplet\\height [\AA]\end{tabular} & \begin{tabular}[c]{@{}c@{}}C-C\\distance [\AA]\end{tabular} &  \begin{tabular}[c]{@{}c@{}}Droplet\\radius [\AA]\end{tabular}  &  $\rho_{\textrm{O}}^{\textrm{2D}}$ [\#O/\AA$^2$]    \\
\midrule
59.280 $\times$ 59.892    & 132   &  6.58 & 3.46 &20.5  & 0.121 $\pm$0.002  \\  
79.040 $\times$ 81.282     & 132  &6.61  & 3.44  &20.5  & 0.120 $\pm$0.002   \\
98.800 $\times$ 98.394    & 132 & 6.61  & 3.43 &20.4   & 0.120 $\pm$0.006   \\

\bottomrule
    \end{tabular}
  \caption{Convergence of structural properties for a system with 132 water molecules, evaluated as a function of the sheet dimensions. Core structural features converge with sheet dimensions of 79.040 $\times$ 81.282.}
    \label{tab:ff_bmark_2}
\end{table}  

\begin{figure}[htp!]
\centering\includegraphics[width=\textwidth]{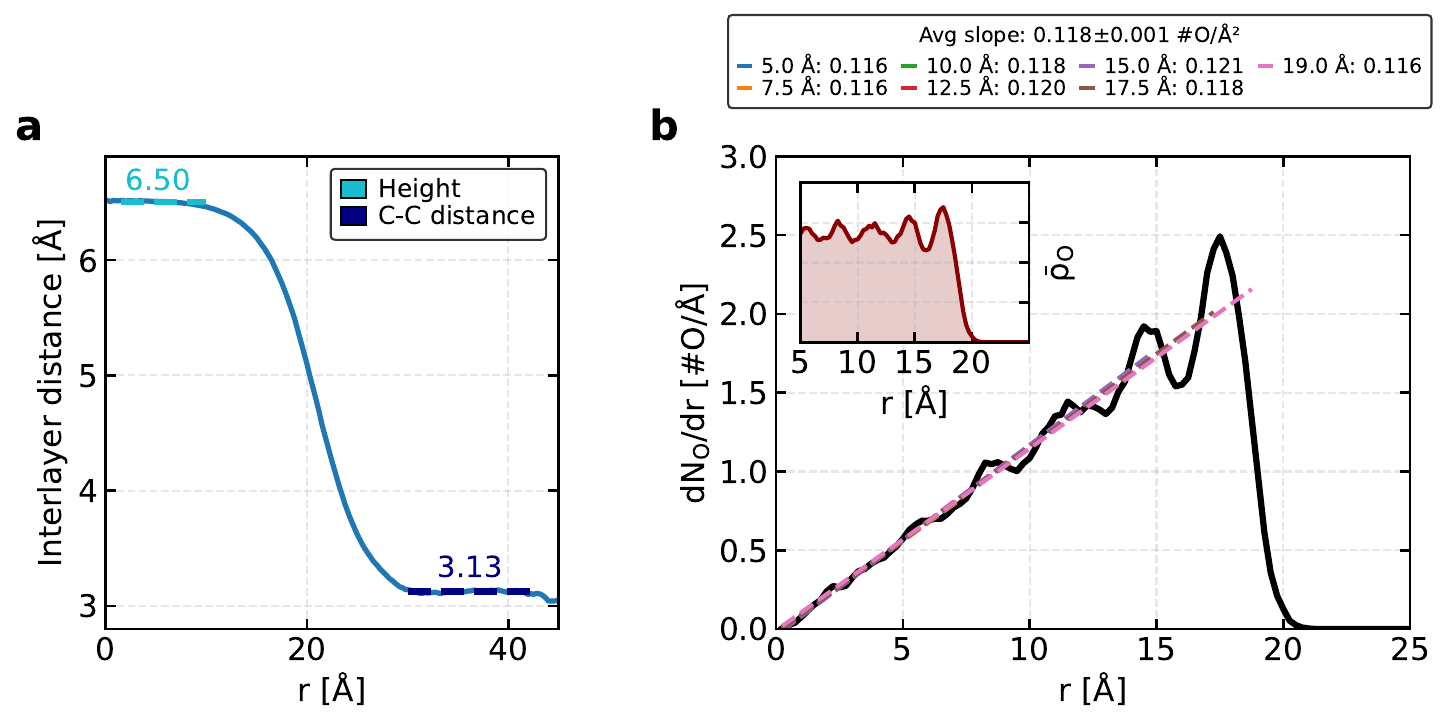}
    \caption{(a) Interlayer distance as a function of the nanodroplet radius, obtained using the MLP developed in this work.
    The droplet height is determined from the central region, where water molecules are present, while the C–C distance is computed from the outer region, where water is absent.
    (b) Cumulative number of oxygen atoms as a function of the nanodroplet radius, obtained using the MLP developed in this work. The 2D oxygen density, $\rho_{\textrm{O}}^{\textrm{2D}}$, is determined from the average slope of multiple linear fits.
    Each fit region and its corresponding slope are reported in the legend.
    The inset shows the normalized radial density profile of oxygen atoms, $\bar{\rho}_{\textrm{O}}$.}
    \label{fig:mace_nanodrop_equi}
\end{figure}

Based on these results, the nanodroplet system used for simulations with the MLP developed in this work consisted of 132 water molecules confined between graphene sheets with dimensions of 79.040\;\AA\;$\times$\;81.282\;\AA.
To better reflect experimental conditions, the bottom graphene sheet was held fixed, mimicking graphene on a substrate, covered by an additional flexible graphene sheet.
Following the same procedure described earlier, we determined the equilibrium density for this system using the MLP. As we see in Figure \ref{fig:mace_nanodrop_equi}, the value obtained corresponds to $\rho_{\textrm{O}}^{\textrm{2D}}=0.118$ \#O/\AA$^2$, which is the one reported in the main part of the manuscript and in Section \ref{sec:md_sims}.

\newpage
\section{Water self-dissociation in bulk and nanodroplet-confined systems}

\begin{figure}[htp!]
    \centering\includegraphics[width=\textwidth]{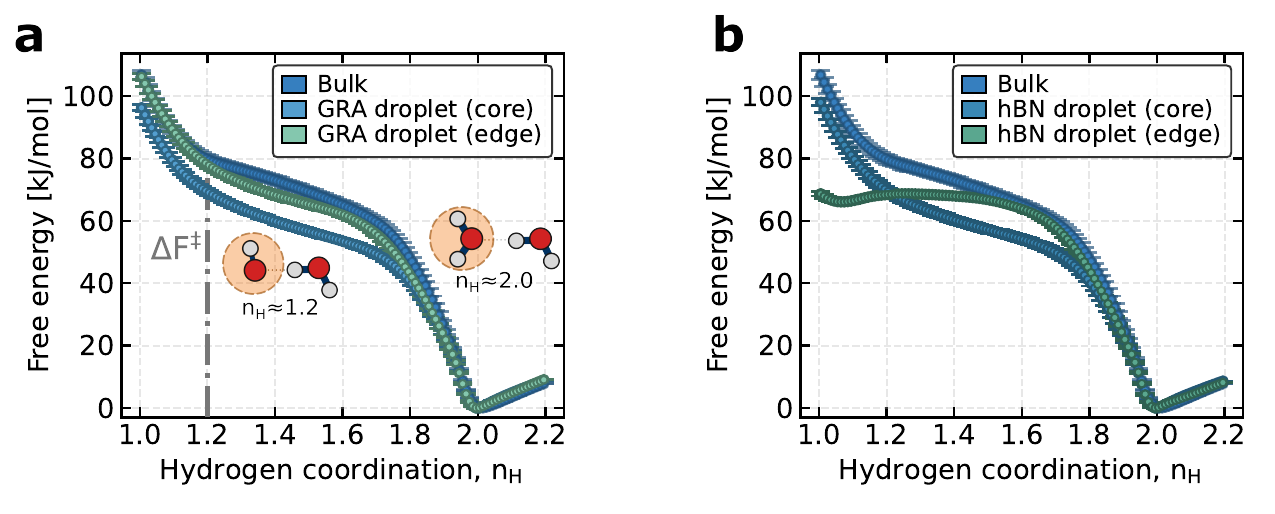}
    \caption{Free energy profiles for the water self-dissociation reaction across the different systems. The free energy barrier $\Delta F^{\ddagger}$, used to compute $p\textrm{K}{\textrm{w}}$, is defined as the free energy difference between $n_{\text{H}} \approx 2$ and $n_{\text{H}} \approx 1.2$. For dissociation events occurring at the edge of the hBN nanodroplet, we instead use $n_{\text{H}} \approx 1.05$.}
    \label{fig:feps_droplets}
\end{figure}

The $p\textrm{K}{\textrm{w}}$ values for water confined within graphene and hBN nanodroplets are obtained from the free energy profiles of the self-dissociation reaction shown in Figure \ref{fig:feps_droplets}, using the relationship defined in Equation \ref{eq:pkw_relation}.

\newpage

\begin{figure}[htp!]
    \centering\includegraphics[width=\textwidth]{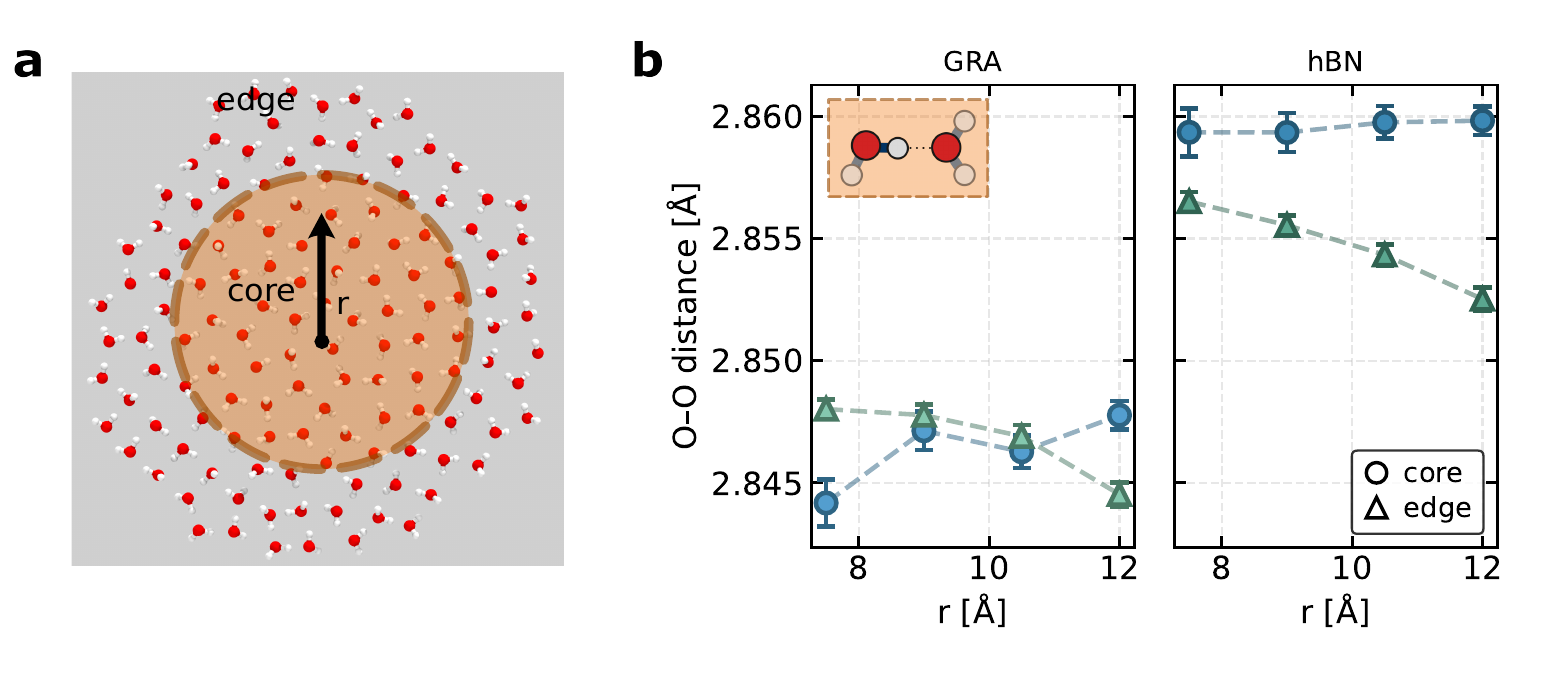}
    \caption{(a) Schematic illustration of the core–edge classification within a nanodroplet, where O–H$\cdots$O pairs located within a cutoff radius $r$ from the center of mass are labeled as ``core'' and those beyond as ``edge''.
    (b) O–O distances in O–H$\cdots$O pairs as a function of the radial distance $r$ to the droplet’s center of mass for water confined in graphene and hBN nanodroplets.
    The accompanying schematic illustrates a representative O–H$\cdots$O pair.
}
    \label{fig:nanodrop_mats_distances}
\end{figure}

To compare the O–O distances in O–H$\cdots$O pairs, which serve as key indicators of water self-dissociation, we evaluated these distances separately at the droplet core and near the edge, depending on the pair’s location within the nanodroplet.
Specifically, we analyzed O–H$\cdots$O pairs along a radial distance $r$ from the droplet’s center of mass.
Pairs located within a defined cutoff radius were classified as ``core'', while those beyond this threshold were labeled as ``edge''.
Figure 4 in the main manuscript shows a series of representative O–O distances for these two regions, using $r = 10.5$\;\AA{} as the core-edge boundary.
In Figure \ref{fig:nanodrop_mats_distances}, we show that these distances remain relatively insensitive to the specific cutoff value used to define the core and edge zones of the nanodroplet.

\newpage